\PassOptionsToPackage{table}{xcolor} %
\documentclass[sigconf,authorversion]{acmart}

\usepackage{acmart-taps}
\usepackage{multirow}

\usepackage{dirtytalk}

\usepackage[utf8]{inputenc} %
\usepackage[T1]{fontenc}

\usepackage{framed}
\usepackage{graphicx}
\usepackage{subcaption}
\usepackage{float}

\usepackage{enumitem}

\usepackage{makecell}

\expandafter\let\csname c@tblerows\endcsname\rownum

\newcolumntype{L}[1]{>{\raggedright\arraybackslash}p{#1}} %
\newcolumntype{C}[1]{>{\centering\arraybackslash}p{#1}} %
\newcolumntype{R}[1]{>{\raggedleft\arraybackslash}p{#1}} %
\newcolumntype{S}{>{\raggedleft\arraybackslash}X}

\usepackage{graphics}
\usepackage{caption}

\usepackage{fontawesome}

\usepackage{verbatim}

\newcommand{\contrastone}{Contrast\,\textsubscript{C/T}\,}
\newcommand{\contrasttwo}{Contrast\,\textsubscript{L/M}\,}
\setlist[description]{leftmargin=\parindent,labelindent=\parindent}

\usepackage{amsthm} %

\makeatletter

\newcommand{\boldparagraph}[1]{%
  \par\addvspace{0.4ex plus 0.8ex minus 0.2ex}%
  {\noindent\normalfont\normalsize\bfseries{#1}.\enspace}%
  \ignorespaces %
}
\makeatother

\usepackage[acronym]{glossaries}

\newacronym{ai}{AI}{artificial intelligence}
\newacronym{genai}{GenAI}{generative AI}
\newacronym{aiact}{AI Act}{Artificial Intelligence Act}
\newacronym{dsa}{DSA}{Digital Services Act}
\newacronym{c2pa}{C2PA}{Coalition for Content Provenance and Authenticity}
\newacronym{aigc}{AIGC}{AI-generated content}
\newacronym{aigi}{AIGI}{AI-generated image}
\newacronym{vlop}{VLOP}{very large online platform}
\newacronym{vlose}{VLOSE}{very large online search engine}
\newacronym{iptc}{IPTC}{International Press Telecommunications Council}
\newacronym{pai}{PAI}{Partnership on AI}
\newacronym{pki}{PKI}{public key infrastructure}
\newacronym{cac}{CAC}{Cyberspace Administration of China}
\newacronym{nist}{NIST}{National Institute of Standards and Technology}
\newacronym{aic}{AIC}{Akaike information criterion}
\newacronym{lmm}{LMM}{linear mixed model}
\newacronym{glmm}{GLMM}{generalized linear mixed model}
\newacronym{sdt}{SDT}{signal detection theory}
\newacronym{me}{ME}{main effect}
\newacronym{ia}{IE}{interaction effect}

\newlength{\maxstimuliwidth}
\newlength{\maxstimuliheight}
\copyrightyear{2026}
\acmYear{2026}
\setcopyright{cc}
\setcctype{by}
\acmConference[CHI '26]{Proceedings of the 2026 CHI Conference on Human Factors in Computing Systems}{April 13--17, 2026}{Barcelona, Spain}
\acmBooktitle{Proceedings of the 2026 CHI Conference on Human Factors in Computing Systems (CHI '26), April 13--17, 2026, Barcelona, Spain}
\acmPrice{}
\acmDOI{10.1145/3772318.3791006}
\acmISBN{979-8-4007-2278-3/2026/04}

\begin{document}

\title{"That's another doom I haven't thought about": A User Study on AI Labels as a Safeguard Against Image-Based Misinformation}

\author{Sandra Höltervennhoff}
\authornote{Equal contribution.}
\orcid{0000-0003-4284-0473}
\affiliation{\institution{CISPA Helmholtz Center for Information Security}
\city{Hannover}
\country{Germany}}
\affiliation{\institution{Leibniz University Hannover}
\city{Hannover}
\country{Germany}}
\email{sandra.hoeltervennhoff@cispa.de}

\author{Jonas Ricker}
\authornotemark[1]
\orcid{0000-0002-7186-3634}
\affiliation{\institution{Ruhr University Bochum}
\city{Bochum}
\country{Germany}}
\email{jonas.ricker@rub.de}

\author{Maike M. Raphael}
\orcid{0009-0007-2196-1889}
\affiliation{\institution{CISPA Helmholtz Center for Information Security}
\city{Saarbr\"{u}cken}
\country{Germany}}
\email{maike.raphael@cispa.de}

\author{Charlotte Schwedes}
\orcid{0000-0002-2554-208X}
\affiliation{\institution{CISPA Helmholtz Center for Information Security}
\city{Saarbr\"{u}cken}
\country{Germany}}
\email{schwedes@cispa.de}

\author{Rebecca Weil}
\orcid{0000-0002-3725-7948}
\affiliation{\institution{CISPA Helmholtz Center for Information Security}
\city{Saarbr\"{u}cken}
\country{Germany}}
\email{weil@cispa.de}

\author{Asja Fischer}
\orcid{0000-0002-1916-7033}
\affiliation{\institution{Ruhr University Bochum}
\city{Bochum}
\country{Germany}}
\email{asja.fischer@rub.de}

\author{Thorsten Holz}
\orcid{0000-0002-2783-1264}
\affiliation{\institution{Max Planck Institute for Security and Privacy}
\city{Bochum}
\country{Germany}}
\email{thorsten.holz@mpi-sp.org}

\author{Lea Sch\"{o}nherr}
\orcid{0000-0003-3779-7781}
\affiliation{\institution{CISPA Helmholtz Center for Information Security}
\city{Saarbr\"{u}cken}
\country{Germany}}
\email{schoenherr@cispa.de}

\author{Sascha Fahl}
\orcid{0000-0002-5644-3316}
\affiliation{\institution{CISPA Helmholtz Center for Information Security}
\city{Hannover}
\country{Germany}}
\affiliation{\institution{Leibniz University Hannover}
\city{Hannover}
\country{Germany}}
\email{sascha.fahl@cispa.de}

\begin{abstract}
As generative AI is increasingly contributing to the spread of deceptively realistic misinformation, lawmakers have introduced regulations requiring the disclosure of AI-generated content.
However, it is unclear if labels reduce the risk of users falling for AI-generated misinformation.
To address this research gap, we study the effect of labels on users' perception and the implications of mislabeling, focusing on AI-generated images.
We first explored users' opinions and expectations of labels using five focus groups.
Although participants were wary of practical implementations, they considered labeling helpful in identifying AI-generated images and avoiding deception.
Second, we conducted a survey with 1\,354 participants to assess how labels affect users' ability to recognize misinformation.
While labels reduced participants' belief in false claims supported by AI-generated images, we found evidence of \emph{overreliance}, leading to unintended side effects:
Participants were more susceptible to false claims accompanied by human-made images, and were more hesitant to believe true claims illustrated with labeled AI-generated images.
\end{abstract}

\ccsdesc[500]{Security and privacy~Usability in security and privacy}
\ccsdesc[300]{Information systems~Social networks}
\ccsdesc[300]{Human-centered computing~Empirical studies in HCI}

\keywords{AI-Generated Content, Misinformation, AI Labels, Social Media}

\maketitle
\renewcommand{\shortauthors}{S. Höltervennhoff, J. Ricker, M. M. Raphael, C. Schwedes, R. Weil, A. Fischer, T. Holz, L. Schönherr, and S. Fahl}

\section{Introduction}
Since the beginning of the so-called ``AI boom''~\cite{griffithNewAreaAI2023}, \gls{ai} has permeated many aspects of our digital lives. 
While a few years ago, state-of-the-art \gls{genai} methods required specialized knowledge and extensive computational resources, tools like ChatGPT, Midjourney, ElevenLabs, and Sora enable laypeople to transform simple prompts into text, images, audio, and videos that are almost indistinguishable from human-made content~\cite{frankARepresentativeStudy2024}. 
Besides countless productive and creative applications of \gls{genai}, there is significant potential for misuse, 
like deepfake scams~\cite{vakulovDeepfakeScamsAre,ChenFinanceWorker2024,LepidoFerrariExec2024} or the generation of non-consensual intimate content~\cite{FBISextortion2023,CongerExplicitDeepfake2024}.
Another major threat of \gls{aigc} is the spread of misinformation\footnotemark[1]\footnotetext[1]{We use the term ``misinformation'' throughout this work to emphasize the falsehood of information rather than the intent with which it is created or shared. Since AI labels can only inform \textit{how} an image was created, not \textit{why}, they are relevant for both intentionally created misleading content (i.e., disinformation) as well as inadvertently shared misleading content (i.e., misinformation).}~\cite{ryan-mosleyHowGenerativeAI,dufour2024ammeba}. 
This became increasingly apparent during the 2024 U.S.\ presidential election~\cite{jingnanAIgeneratedImagesHave2024,ai_in_elections,ccdh_report}, when, for instance, Donald Trump shared fabricated images supposedly showing Taylor Swift's support for him~\cite{ibrahimTaylorSwiftEndorsed2024}.
Similarly, a generated image of an explosion near the Pentagon caused a dip in the stock market~\cite{claytonFakeAIgeneratedImage2023}, highlighting the potential of \gls{aigc} to cause outrage and manipulate public opinion.
While image creation or manipulation are not new phenomena, \gls{genai} significantly lowers the skill, time, and effort required to create highly realistic images of almost anything, compared to traditional tools like Photoshop.
Experts fear that the sheer number of images paired with their rapid distribution might overwhelm fact-checkers and cause the public to be (too) skeptical of information in general~\cite{gretelkahnWillAIgeneratedImages2023}.

Given that a significant proportion of Internet users obtain at least part of their news from social media (54\% of adults in the U.S.~\cite{social_media_and_news}), legislators worldwide are considering regulatory measures to protect users against AI-generated misinformation.
An essential goal is the transparent disclosure of \gls{aigc} through the use of \emph{labels}. 
The European Parliament passed both the \gls{dsa}~\cite{other/DSA} and \gls{aiact}~\cite{other/AI_Act}, obligating large online platforms, search engines, and providers of \gls{genai} to disclose AI-generated and manipulated content.
One of the EU's goals is to prevent \say{systemic risks that may arise from the dissemination of content that has been artificially generated or manipulated, in particular the risk of the actual or foreseeable negative effects on democratic processes, civic discourse and electoral processes, including through disinformation}~\cite{other/AI_Act}.
Under the presidency of Joe Biden, the U.S.\ government attempted to implement similar rules.
However, the Trump administration has revoked most safety measures to not hinder the development of \gls{genai}~\cite{shepardsonTrumpRevokesBiden2025,WhiteHouseAIActionPlan,WhiteHouseAIAntiWoke}.
Recently, as part of a greater campaign to target harmful online content like misinformation, the Chinese government published a set of rules on how AI providers and content-sharing platforms must label \gls{aigc}~\cite{danChinaReleasesNew2025,publishedChinaWillEnforce2025}.

While legislators have high expectations that labeling will help to mitigate some of the societal risks posed by \gls{genai}, for users, such AI labels function primarily as a transparency mechanism.
They do not indicate whether the content accurately reflects reality or is fabricated, but only whether it was generated using AI.
Therefore, research is urgently needed to investigate if labels can serve their intended purpose in the fight against misinformation.
To close this research gap, we investigate whether AI labels can meet the expectations placed on them as a tool against misinformation, or whether they fall short of these expectations and even have negative side effects.
Visual content, such as images or videos, plays a major role in the spread of misinformation~\cite{VisualMisinformationFacebookYang2023, dufour2024ammeba,NewmanHowImages2024}. 
In this work, we focus on AI labels for images.

In a first study, we conducted five focus groups to collect qualitative insights about users' opinion towards AI labels in the context of misinformation.
We considered factors that influence users' acceptance, comprehension, and trust in labels and identified potential problems that could hinder adoption.
While past qualitative research focused mainly on the design of labels, to the best of our knowledge, we are the first to explore users' perception.
\begin{framed}
\noindent\textbf{RQ1: What are users' opinions, expectations, and concerns about AI labeling?}
\end{framed}

Secondly, we measured the actual effects of AI labels on users' perception through a pre-registered online survey ($N=1\,354$), in which users rated the veracity of news posts accompanied by images.
While previous work~\cite{wittenbergLabelingAIgeneratedMedia2024} showed that AI labels can reduce users' belief in misleading \glspl{aigi}, the survey did not include other types of stimuli, providing only a restricted view on the implications of labeling.
To investigate potential (side) effects, our stimuli set varied in terms of image origin (human-made/AI-generated) and claim veracity (true/false). 
Utilizing this design, we are able to answer if users are simply \emph{relying} on the presence of labels or if labels encourage users to \emph{think about the veracity} of a claim.
\begin{framed}
\noindent\textbf{RQ2: How does AI labeling affect users' perception of true and false claims with human-made and AI-generated images?} 
\end{framed}
Lastly, we examined the impact of mislabeling. 
At least within the near future, labeling mechanisms will not be without errors, resulting in unlabeled \glspl{aigi} and falsely labeled human-made images. 
In our focus groups, we investigated how mislabeling might change users' opinions and trust towards labels.
Moreover, in our survey, we measured its impact on participants' judgments of claims. %
\begin{framed}
\noindent\textbf{RQ3: What are the consequences of mislabeling, and how does it affect users' trust in AI labels?}
\end{framed}

By combining the two studies, we gain a comprehensive understanding of the merits of AI labels in combating misinformation.
We are not only able to give voice to users' approval and doubt about labels, but also measure the effects on misinformation in a large-scale controlled experiment.
Our work contributes to understanding the inherent value of AI labels in combating misinformation, explores unexpected risks, and ultimately aims to investigate measures for deploying labels for \glspl{aigi} that are sensible.

\section{Background and Related Work}
We first introduce current labeling mechanisms and summarize how popular social media platforms label \gls{aigc}.
Moreover, we present existing work on AI labeling and the related topic of misinformation warnings.

\subsection{AI Labeling Mechanisms} %
\label{sec:background:mechanisms}
At the time of writing, most social media platforms shift the responsibility for labeling \gls{aigc} to their users.
Facebook, Instagram, TikTok, YouTube, and LinkedIn require the disclosure of realistic-appearing, synthetic content that could mislead viewers~\cite{LabelAIContentInstagram,TikTokCommunityGuidelines,YouTubeDisclosingUse,linkedinFalseMisleadingContent}.
\gls{aigc} created using the platforms' own tools is usually automatically labeled~\cite{LabelAIContentInstagram,TikTokAboutAIGC,YouTubeDisclosingUse}.

To identify \gls{aigc} without relying on users, several platforms (e.g., Facebook, TikTok, and LinkedIn) have adopted the \gls{c2pa}~\cite{c2pa} metadata standard~\cite{nickcleggLabelingAIgeneratedImages2024,TikTokC2PA,LinkedInContentCredentials}.
Upon creation, information such as the used \gls{genai} tool, author, and editing history is added to the file in a cryptographically signed data structure.
Notably, \gls{c2pa} is not only intended to disclose \gls{aigc}, but also to prove that content is authentic~\cite{youtubeBuildingTrustYouTube}.
However, \gls{c2pa} metadata can be easily removed (e.g., by taking a screenshot of an image) or may be stripped if content is shared through platforms that do not support the standard.

An alternative approach to proactively tag \gls{aigc} is watermarking.
While visible watermarks are an established means of preventing the unauthorized use of copyrighted material, such as stock photos, deep neural networks can embed information directly into an image's pixels~\cite{Zhu2018hidden,tancik2020stegastamp}.
More recent approaches~\cite{fernandezStableSignatureRooting2023,wenTreeringsWatermarksInvisible2023,yangGaussianShadingProvable2024,ciRingIDRethinkingTreering2025} perform invisible watermarking during the generation process, such that all produced content can be detected and attributed to the respective model. 
However, to the best of our knowledge, such methods are not yet used in practice.

Another active research topic is the passive detection of \gls{aigc}.
\gls{aigi} detectors, often machine learning models, exploit imperceptible artifacts or inherent properties of the generation process to distinguish real from generated content~\cite{wangCNNgeneratedImagesAre2020,ojhaUniversalFakeImage2023,drct-24,rickerAEROBLADETrainingfreeDetection2024}.
An ongoing challenge is the generalization to unseen models and the robustness to (adversarial) perturbations~\cite{mavali2025adversarialrobustnessaigeneratedimage,saberiRobustnessAIimageDetectors2024,abdullahAnalysisRecentAdvances2024}.
To date, there is no clear evidence that social media platforms employ passive detection methods.
However, several websites and tools (e.g., \href{https://hivemoderation.com/ai-generated-content-detection}{Hive} and \href{https://isgen.ai}{isgen.ai}) exist where users can upload content and receive a score indicating the likelihood that it is AI-generated.

For our focus groups, we concentrate on three mechanisms:
\emph{self-disclosure} and \emph{metadata}, which are already used by social media platforms, and \emph{detection}, due to the growing number of available tools.

\subsection{Research on Labels for AI-Generated Content}
\citet{epsteinWhatLabelShould2023} first investigated the understanding of textual labels for \gls{aigc}.
While terms such as ``AI Generated'' or ``AI Manipulated'' were correctly associated with \gls{aigc}, participants considered ``Deepfake'' or ``Manipulated'' content to be intentionally misleading. 
\citet{LabelingSyntheticContent2025} explored AI label designs regarding four dimensions: sentiment, color and iconography, position, and detail level. 
They compared the effect of ten different designs. All labels made participants believe the content was AI-generated or edited, however, the trust in the label depended on the design. Moreover, labels did not affect the engagement level. %
\citet{SignalOfProvenance2025} conducted semi-structured interviews with sighted and visually impaired participants. The latter struggled to utilize visual cues that identified AI content, especially because of poor or inconsistent design decisions.

\citet{toffTheyCouldJust2025} evaluated how labeling affects trust in journalistic content. Participants found news less trustworthy if it was labeled as AI-generated. 
\citet{limTheEffectOf} investigated how participants react to AI-generated health prevention messages if labeled as such.
Disclosing the source had a negative impact on participants' assessment by a small but significant amount. 
Additionally, two similar studies~\cite{ternovskiDeepfakeWarningsPolitical2021, lewisDeepfakeDetectionContent2023} on the detectability of deepfake videos have shown that warnings did increase skepticism towards shown videos. 
However, due to the inability to reliably distinguish between fake and authentic videos, this effect existed regardless of whether the video was a deepfake or not.
\citet{raeEffectsPerceivedAI2024} studied whether labels for text matter in a future where content created by AI cannot be distinguished from human-made content. 
They found that participants had more negative feelings towards creators when they believed AI was involved and were less satisfied.

Focusing on misinformation, \citet{altayPeopleAreSceptical2024} investigated AI labels for news headlines and accounted for human-made and \gls{aigc}, as well as for true news and misinformation.
They found that labels reduce trust in headlines, even if they are true or authentic, as participants believed that the whole text was written by AI.
\citet{ImpactofArtificialIntelligenceLi2024} conducted a similar study, investigating the influence of labels on text, considering accurate and inaccurate, as well as for-profit and not-for-profit articles.
They found no significant effect for labels on perceived accuracy, credibility, or sharing intentions.
Targeting images instead of text,\citet{wittenbergLabelingAIgeneratedMedia2024} investigated the effect of different label variants for \glspl{aigi}.
The authors found that labels generally reduced belief in \glspl{aigi}.
However, their stimuli set contained only images that were AI-generated \emph{and} misleading.
By design, the experiment could not investigate effects of labels on benign posts with \glspl{aigi} or misinformation accompanied by human-made images.

In summary, while most prior work focused on labels for AI-generated text, we address a research gap in understanding how AI labels function for images.
Investigating how such labels affect the perception of visual misinformation is critical, as it has been shown that images increase the attention that messages receive as well as their perceived truthfulness~\cite{NewmanHowImages2024,newmanNonprobativePhotographsWords2012}.
Moreover, extending previous work~\cite{wittenbergLabelingAIgeneratedMedia2024}, our study design not only measures the warning effect for misleading \gls{aigc}, but also allows us to uncover side effects and the impact of mislabeling.

\begin{table*}
    \caption{Demographics of our 18 valid focus group participants. Our sample included 6 participants from the U.S.\ and 12 from the EU. The latter were from Portugal (3), France (2), Slovenia (2), Denmark (1), Germany (1), Greece (1), Italy (1), and Spain (1).}
    \Description{Table with demographics of the valid focus group participants. It includes gender, region, education, age, ATI-S, and AIAS-4. The demographics are divided into EU and U.S. participants, the column labeled ‘N’ covers both regions.}
    \label{tab:focus_group_demographics}
    \centering
    \small
    \begin{tabular}{lrrr@{\qquad}lrrr}
    \toprule
    \textbf{Gender} & \textbf{EU} & \textbf{U.S.} & \textbf{N} & \textbf{Age} & \textbf{EU} & \textbf{U.S.} & \textbf{N} \\
        \quad Female & 5 & 2 & 7 & \quad 18--29 & 4 & 2 & 6 \\
        \quad Male & 7 & 3 & 10 & \quad 30--49 & 7 & 3 & 10 \\
        \quad Non-binary & - & 1 & 1 & \quad 50--69 & 1 & 1 & 2 \\
    \textbf{Education} &&&& \textbf{ATI-S~\cite{wesselATISUltrashortScale2019}} \\
        \quad Secondary school & - & 1 & 1 & \quad $0 \le x \le 2$ & 1 & 1 & 2 \\
        \quad University/College w/o degree & 3 & 3 & 6 & \quad $2 < x \le 4$ & 6 & - & 6 \\ 
        \quad Associate degree & 1 & 1 & 2 & \quad $4 < x \le 6$ & 6 & 4 & 10  \\
        \quad Bachelor's degree & 4 & - & 4 & \textbf{AIAS-4~\cite{grassiniDevelopmentValidationAI2023}} \\
        \quad Master's degree & 4 & 1 & 5 & \quad $0 \le x \le 2$ & - & 1 & 1 \\
        &&&& \quad $2 < x \le 4$ & 1 & - & 1\\
        &&&& \quad $4 < x \le 6$ & 11 & 5 & 16 \\
    \bottomrule
    \end{tabular}
\end{table*}

\subsection{Research on Misinformation Warnings}
Reviewing previous research, \citet{martelMisinformationWarningLabels2023} found that warnings, presented alongside misinformation, can be used as an effective tool to combat deceptive media.
Investigating the effect of warning labels on Twitter, \citet{papakyriakopoulosTwitterLabels} found that, overall, labels did not impact the interaction with posts, but that contextual or well-explained warnings could reduce it. %
To increase the effectiveness of misinformation warnings, previous work investigated different designs:
While \citet{kaiserAdaptingSecurityWarnings2021} compared contextual and interstitial warnings, 
\citet{sharevskiContextRedFlag2022} investigated contextual and iconographic designs. 

However, misinformation warnings do have side effects.
\citet{hoes2024prominent} investigated the effectiveness of three misinformation intervention strategies.
The strategies reduced participants' belief in misinformation, but also made participants more suspicious of authentic information.
Adding to that, \citet{pennycook_implied_truth} found that misinformation labels can lead to an \textit{implied truth effect}, meaning that, in the presence of labels, users trust unlabeled content more, as they assume that it passed a fact check.
\citet{hameleersNothingDeepfake2023} studied the consequences if misinformation labels are maliciously assigned and found that they can reduce the credibility of authentic content.

This literature uncovers the effects and unwanted side effects of misinformation warnings, further inspiring us to explore the side effects of AI labels and how users would interact with them.
While AI labels have the potential to warn against (AI-generated) misinformation, their benefits and risks are not yet well understood.

\section{Study 1: Focus Groups on Users' Expectations and Concerns About AI Labels}
In our first study, we investigated how users perceive AI labels in the context of misinformation, focusing on their expectations and concerns (RQ1).
We explored factors boosting or hindering users' trust and, consequently, the adoption of AI labels, paying special attention to users' opinions on mislabeling (RQ3).
While related work \cite{epsteinWhatLabelShould2023, LabelingSyntheticContent2025, SignalOfProvenance2025} mainly investigates the implications of label design, our study focused on underlying aspects such as labeling rules and mechanisms.
When reporting the methodology and findings of this study and our survey (see Section~\ref{sec:survey}), we follow the transparency guidelines of \citet{klemmer2025transparency}.

\subsection{Method}
We conducted five semi-structured online focus groups between December 2024 and February 2025 to gain insight into people's fundamental thinking about AI labeling. The results served as a basis for our second study (see Section~\ref{sec:survey}).
We chose focus groups to gain first qualitative insights on perceptions of AI labels in the context of misinformation. 
Since AI labels are a relatively novel topic, we aimed to understand the fundamental principles that influence their acceptance.
For such ``introductory session[s]''~\cite{rosenbaum2002focusgroupshci}, focus groups have been found to be especially suitable.
Participants are able to explore and challenge their opinions while
being stimulated by cues and anecdotes of others, which is harder to reach in interviews~\cite{kitzinger2006}.
Utilizing focus groups, we could observe participants' agreement, disagreement, as well as reactions to novel perspectives.
Moreover, focus groups can facilitate the interactive development of ideas beyond individual opinions~\cite{rosenbaum2002focusgroupshci}.
This makes them valuable for topics that have a societal impact~\cite[p.~2-14]{krueger2014focus} such as AI-generated misinformation and AI labeling. 
For AI labels to be effective, especially in mitigating the spread of misinformation on social media platforms, users must agree on some kind of shared understanding, even if the concrete perception depends on each individual.
As recommended for online focus groups~\cite{abrams2017online} and in line with previous work in the HCI and security community~\cite{Agha2024TrickyVs, Chen2024Exploration, Davis2024Fashioning}, we formed rather small groups of three to five participants.
Besides fostering interaction, smaller focus groups can be more appropriate if participants are expected to show strong involvement or feelings, as misinformation is a contentious topic~\cite[p.~82]{krueger2014focus}.

\boldparagraph{Procedure}
To reach a diverse and international group of participants, we conducted focus groups online using Zoom and in English. 
The focus groups lasted an average of 74~minutes.
The discussions were guided by a set of initial questions (see Appendix~\ref{app:focus_groups:guide}).
Participants viewed a slide deck that included the current question and exemplary images or visualizations. 
We included both harmless and misleading images to not give the impression that labels only apply to misinformation.
The slide deck is available as supplementary material.
We tested our focus group guide with two pilots.
We occasionally adjusted questions or added new ones between focus groups if novel perspectives came up.
After five focus groups we reached thematic saturation and therefore stopped recruiting.

\boldparagraph{Recruitment}
We recruited participants through Prolific. %
We focused on adult participants from the EU and the U.S., as these regions had the most advanced \gls{genai} legislation at the start of the focus groups, increasing the likelihood that participants had already come into contact with labels.
Based on the answers to a short pre-screening questionnaire, we formed groups considering age, gender, country of residence, social media usage, and attitude toward technology interaction (ATI-S~\cite{wesselATISUltrashortScale2019}) and AI (AIAS-4~\cite{grassiniDevelopmentValidationAI2023}). 
In total, we conducted six focus groups. 
For one group, only two participants showed up, and they also had insufficient English skills, which is why we excluded their results from our analysis. 
In the remaining five focus groups, three to five participants took part, with $N=18$ participants in total.
Participants were compensated \pounds 23.75 for an estimated duration of 90~minutes to account for unexpected events.
We provide the aggregated demographics of our participants in Table~\ref{tab:focus_group_demographics}. 

\boldparagraph{Analysis}
All focus groups were transcribed by a GDPR-compliant transcription service. Afterward, we removed personally identifiable information and coded the focus groups using ATLAS.ti, utilizing an open coding approach and thematic analysis.
Each focus group was independently coded by two researchers, who afterward discussed the codes and agreed on a shared coding.
A total of three researchers were involved in the whole process.
As these discussions, including the resolving of conflicts, were crucial for forming our final codebook, we did not calculate the inter-rater reliability, which is in line with previous work~\cite{mcdonaldReliability2019, conf/ccs/klemmer24, BouwmanCupofTI2020}.
We provide the final codebook in Appendix~\ref{app:focus_groups:codebook}.
Through affinity mapping, we condensed our codebook into relevant themes. %

\subsection{Results}
\label{sec:focus_group_results}
To present the findings of our focus groups, we first describe participants' previous experiences and expectations towards AI labels, followed by their concerns, and, finally, their worries about mislabeling. Alongside these findings, we also describe the main questions that prompted the participants' responses.
We use quantifiers to describe how many participants gave certain answers 
(``few'': 2--5, ``some'': 6--9, ``many'': 10--13, ``most'': 14--17).

\boldparagraph{Experiences With \glspl{aigi}}
To guide our participants into the topic, we first established a common understanding of \gls{genai}, in particular regarding the generation of images, without mentioning potential risks.
Many participants already came across \glspl{aigi} on social media.
Some of our participants generated images themselves, often for personal use, but a few utilized AI for their professional endeavors. 
Our participants' overall attitude towards \glspl{aigi} was ambivalent.
A few stated that they would not engage with content if they knew it was AI-generated, indicating that they are not interested in \gls{aigc}.
But disapproval was also content-sensitive:
Some participants stated that AI could lead to negative perceptions if disclosed, e.g., in news, ads, or in part art.
On the contrary, art and humorous \glspl{aigi} were also examples where AI could lead to a positive perception.
When asked about potential problems of \glspl{aigi}, all focus groups identified misinformation as one pressing issue, followed by the potential of \glspl{aigi} to be used by criminals, e.g., for blackmailing, pornographic material, or scams.

\boldparagraph{AI Labels Help to Avoid Deception}
For the next part, we focused on the risk that \glspl{aigi} can be used to create misinformation. We then introduced the idea of AI labels as a potential countermeasure.
For most participants, AI labels had not played a significant role in their online activity so far.
While some had already experienced AI labels on social media, %
and a few were already confronted with labeling systems when uploading their content, some had never even heard of them.

However, many participants liked the concept of AI labels. 
They generally considered them helpful for distinguishing real from AI-generated images and, as a consequence, useful to uncover AI-generated misinformation.
As one participant summarized their merit:
\begin{quote}
    \textit{``I think they are great. So you don't have to question yourself whether something is real or not. Especially if you're not like very tech-savvy''} - FG1\_P4.
\end{quote}
However, participants also mentioned additional factors, e.g., AI labels could help contrast 
\textit{``genuine work''} (FG4\_P15) 
of human artists from AI work or adjust unrealistic expectations on reality.
A few participants found that labels will be a necessity in the future to still be able to distinguish \glspl{aigi} from human-made images, since they feared that \glspl{aigi} will become even more realistic than they are today.

Participants were especially concerned about the vulnerability of loved ones with lower media literacy, e.g., elderly people: 
\begin{quote}
    \textit{``{[...] and I'm afraid that someday [my grandmother] will be misleading [sic] by some sort of page [...] and [...] it will end up very bad}''} - FG3\_P8.
\end{quote}
Therefore, they considered labels especially valuable for these groups. %

To better distinguish AI from reality,
some participants found that labeling \glspl{aigi} should be mandatory.
Some also stated that, as a side effect, labels could increase the general awareness of AI.
It was hypothesized that the effect could even persist if not all \glspl{aigi} are correctly labeled. 
Critically, half of our participants stated that if an image is labeled as AI-generated, they would perceive the content more negatively.
Others stated that disclosing AI usage could make a site or company appear less credible, either because users could feel misled by \glspl{aigi} or because they would rather like to support human artists.

We also asked our participants what labeling mechanisms they could think of.
Many participants named detection, often assuming that it would involve AI: 
\textit{``Which tool can be used to detect AI more [sic] than AI?''} (FG1\_P1).
Some participants were aware of visible and invisible watermarks.
Others considered the option to disclose the use of AI when uploading media to a social network or to implement a mechanism similar to X’s ``Community Notes''.
Interestingly, only few participants mentioned the embedding of metadata into the file, which is already employed by several platforms.
Our results indicate that, despite the beginning adoption of standards like \gls{c2pa}, the concept of metadata as a transparency mechanism is not yet well known.
Instead, users might overestimate the real-world deployment of AI detectors.

\boldparagraph{AI Labels Are Full Of Pitfalls}
Most participants did not consider AI labels a perfect solution to counter misinformation, raising both overarching concerns and reservations regarding individual labeling mechanisms.
We first present two general concerns.

\paragraph{Unclear Standardization} %
Many participants were concerned about standardization, finding it hard to decide which \glspl{aigi} should or should not be labeled and what labels should look like. 
A few discussed whether images should be labeled depending on their context, differentiating between innocent and disturbing or contentious images:
\begin{quote}
    \textit{``{If it's a dog, if it's a pie, if it's a car. Okay, I don't care. But if it's a politician, if it's a global event, [...] I think it should definitely be labeled}''} - FG2\_P7.
\end{quote}
But they found it hard to decide which content would fit into this category and were concerned about edge cases, e.g., images that might only upset a small subset of people.
Therefore, most of our participants decided that all \glspl{aigi} should be labeled.
Participants anticipated negative implications if only some were labeled while others were not.
When discussing labeling rules for images that are only partly generated or edited using AI, the decision was even harder. %
A few participants considered labels unnecessary if AI was only used for minor edits, like background enhancements or filters, as ``conventional'' filters have been used for years without a label.  
Contrary, some participants found that all edits involving AI need a label, as it is hard to quantify if the meaning of an image has changed, even if the manipulation is subtle. %
To account for differences between entirely generated and edited images, some participants proposed to use different labels, e.g., ``enhanced with AI'' or ``edited with AI''.  %
However, one participant wondered if this would make the labeling system too complex.
Some participants stressed the importance of consistency regarding similar rules across different social media platforms and countries.

\paragraph{Abuse of Power}
Half of our participants were concerned about the power a platform or authority would have if in charge of regulating and enforcing a labeling system.
They suspected that companies might push their own agenda:
\begin{quote}
    \textit{``{I don't know if I necessarily trust a platform to do the right thing because I've heard of many instances where they're like, oh we're gonna try to do the right thing and [...] they don't}''} - FG2\_P6.
\end{quote}
As lacking trust in authorities was a major theme, we added a question about who should be responsible for AI labeling after the second focus group.
A few participants felt that the platforms should take responsibility.
A few mentioned community efforts leaning on already existing community notes for misinformation. 
Other answers included central organizations, providers of AI services, and content creators themselves.
Some participants even saw the necessity to pour AI labeling into law, mistrusting voluntary commitment.

\medskip
Shifting to the technical aspect of AI labeling, we explained three relevant mechanisms (self-disclosure, detection, and metadata). Based on this information, our participants identified three main concerns:

\paragraph{Dishonest Users}
Most of our participants questioned other users' honesty and were uncomfortable with mechanisms that solely rely on it.
This issue was especially discussed for self-disclosure, since users could easily lie about using AI.
As one participant put it: 
\textit{``{Self-disclosure is like probably the least trustworthy because [...] it would be almost impossible to tell if someone is being honest}''} (FG5\_P16).
But participants were also concerned about the intentional removal of metadata.
Other participants assumed that criminals could simply use (custom) \gls{genai} tools that do not insert metadata, thus bypassing detection.

\paragraph{Reliability}
Some participants were concerned about platforms using detectors to label \glspl{aigi}, which were perceived as particularly non-transparent. 
Participants had general doubts about the reliability of AI detectors, partly due to experiences with other \gls{genai} tools: 
\textit{``But AI assessing AI, it's probably not reliable. [...] like ChatGPT is not always reliable''} (FG1\_P2).
A recurring theme was that the performance of such a detector would depend on the data it is trained on, allowing the responsible party to influence what is labeled and what not. 
This uncertainty regarding how AI will behave or evolve caused discomfort among a few participants: 
\textit{``I don't like thinking about this, this is scary''} (FG4\_P15).

\paragraph{Usability Issues}
Some of our participants identified usability issues.
A recurring theme was that \gls{aigc} might be accidentally not labeled, e.g., considering self-disclosure, a user might forget to add a label.
A few participants even encountered difficulties themselves, suggesting that users are not well-informed about the labeling mechanism and its effect: 
\begin{quote}
    \textit{``For Instagram [...] there is an option to say it's [...] AI content, which I have tried but I don't know how to operate it maybe. And I'm like, okay there is no big difference whatever I try to do with that option''} - FG1\_P3.
\end{quote}
A few participants also stated usability issues for metadata, as it can be unintentionally removed, e.g., when uploading images to platforms not supporting the respective standard.
A concern regarding detectors was that, if they output a probability of an image being AI-generated, this might be difficult to interpret.
However, one focus group found that including such information in labels would make them more informative, allowing users to make their own decision.

\medskip
To let our participants reach an informed verdict about the presented mechanisms, we laid out their main advantages and disadvantages.
The concluding assessment of our participants regarding labeling mechanisms was mixed.
Metadata was considered the most favorable approach by many.
A few participants stated the advantage that metadata does neither depend on users nor AI since metadata is directly added to an image upon creation. 
But, some participants were genuinely surprised at how easily metadata can be removed and had thought that it would require more technical expertise.
Most participants chose their favorite mechanism by exclusion and concentrated on the disadvantages of the individual mechanisms.
In this regard, self-disclosure was considered the least reliable.
While several participants were relatively confident in their favorite mechanism, individual participants struggled to pick a trustworthy one: 
\textit{``I have actually zero confidence in any of these methods being able to identify \glspl{aigi}''} (FG2\_P5).
Some participants suggested using a combination of mechanisms to compensate for the weaknesses of individual ones, or to have a fallback in case a mechanism is bypassed.

\boldparagraph{Mislabeling Might Erode Trust in Labels}
\label{sec:focus_group_results:mislabeling}
The final part of our focus groups addressed mislabeling.
Some participants worried about mislabeling when discussing the problems of labels and labeling mechanisms.
A few of those already came across a mislabeled image on social media. 
Participants mainly mentioned unlabeled \glspl{aigi}, but one focus group also identified the problem of wrongly labeled human-made images.
To ensure that all participants had the same level of knowledge, we explained both false negatives (unlabeled \glspl{aigi}) and false positives (labeled human-made images).
After this introduction, most participants found mislabeling to be problematic. 
Participants feared that it could lead to a distortion of reality or incite fear if crimes or disasters are fabricated.
In this respect, a few worried that users would over-rely on the presence or absence of labels.
However, the level of concern differed between the two types of labeling errors. 
Half of the participants found unlabeled \glspl{aigi} more concerning due to their potential to misinform and cause confusion or fear. 
In contrast, the implications of mislabeled human-made images were considered not as severe. 
A few participants reasoned that identifying false positives is easier due to common knowledge or the existence of other images of, e.g., the same event, making them 
\textit{``easier to authenticate''} (FG4\_P14).
Nevertheless, one group agreed that mislabeled historic photos could make people question past events, like 9/11 or the Holocaust: 
\begin{quote}
    \textit{``That's another doom I haven't thought about until now. Unraveling the implication on history books or politics. Well, that's [a] huge mess''} - FG2\_P6.
\end{quote}
Another concern was the potential reputational damage, e.g., politicians being perceived as dishonest or artists being falsely accused of not creating original work. 
One focus group highlighted the necessity of an appealing system for content creators, if labels are assigned incorrectly, to not harm innocuous content creators.
Noteworthy, a few participants were surprised by the possibility of false positives, which suggests that the dangers of those might be less present or tangible.

Some participants considered both mislabeling cases to be equally dangerous, with one participant stating 
\textit{``It's just as bad. We have to be able to tell what's reality and what's not and it's just as bad to me''} (FG3\_P11).
Some participants noted that the consequences strongly depend on the image and the context.
Problematic examples of mislabeling involved politicians or celebrities. 
However, mislabeling harmless content was regarded less problematic, e.g., one participant previously noticed a mislabeled video of a dancing person and did not expect negative consequences.

We finally asked participants how mislabeling would affect their trust in the labeling system. 
While many stated that observing mislabeled images would make them lose confidence in the label, they had different views on what degree of mislabeling is acceptable. 
Some could tolerate the occasional mislabeling of images, e.g., because they were inherently skeptical of social media content anyway.
Others would not, fearing that every case of mislabeling could lead to great harm:
\textit{``[As] we're formulating our opinions and ideas off of the content we're receiving there has to be like zero error''} (FG3\_P10).
Interestingly, a few participants stated they would lose trust more quickly if ``obvious'' \glspl{aigi} are mislabeled, hinting towards a misconception regarding the functioning of labeling mechanisms: 
\textit{``[If] you can clearly tell something is like created by AI and it's not labeled [...] it's kind of like, okay, is this really working?''} (FG3\_P9).
Beyond the labels themselves, a few participants found that mislabeling could damage the trust towards the post's source, e.g., a newspaper posting an image alongside a headline: 
\textit{``Errors do happen, but if they happen multiple times you start questioning about it''} (FG2\_P7).
Especially if otherwise credible institutions shared mislabeled \glspl{aigi}, this would strongly erode participants' trust in them.

Overall, participants were still in favor of AI labels after discussing the possibility of mislabeling. 
Many emphasized that labeling is still helpful or should be mandatory. 
Only a few participants concluded that labeling may be insufficient to combat misinformation and a few emphasized that labeling strategies must be more sophisticated. 
However, a few participants found that platforms should at least attempt to use labels, since even imperfect labels are still better than doing nothing to counter misinformation and would draw attention to the presence of \glspl{aigi}.

\boldparagraph{Key Findings}
\label{sec:focus_group_results:takeaways}
Investigating \textbf{RQ1} (Users’ opinions, expectations, and concerns about AI labeling) in the context of misinformation, we found that participants, 
despite having little prior experience with AI labels, initially considered them a useful tool to differentiate real from generated images.
However, this positive view is clouded by several concerns:
Participants questioned the rules about what should be labeled and what not, and were suspicious of platforms abusing their power by selectively enforcing labels. %
Moreover, they questioned the reliability of mechanisms that depend on others' honesty, worried about ambiguous outputs of detectors, and identified usability issues.
Regarding \textbf{RQ3} (Consequences of mislabeling and effect on users' trust) we found that at first not all participants were aware of the possibility of mislabeling.
However, after learning about it, it was considered a significant threat to the success of AI labeling.
While participants found both instances of mislabeling problematic, the implications of unlabeled \glspl{aigi} were often rated worse than wrongly labeled human-made images.
Our results strongly suggest that users lose trust in the labels if they encounter mislabeling. 
While some participants would be more lenient, others reported that they would already lose trust if mislabeling happens seldomly.
Despite their existing flaws, many participants would still welcome AI labels as a tool to combat misinformation.
However, to increase users' trust towards labels they must be thoughtfully designed and implemented.

\section{Study 2: Survey on Effects and Side Effects of AI Labels}\label{sec:survey}
Through our second study, we measured how the presence (or absence) of AI labels influences users' belief in accurate and misleading social media posts (RQ2).
While the majority of focus group participants considered labels a useful tool against deception, they also identified mislabeling as a critical problem.
Therefore, our study does not only assess the effect of AI labels, but also examines how mislabeling affects users' judgments (RQ3).

\begin{table}
    \centering
    \small
    \Description{This table summarizes the design of our main experiment. Each column represents a factor (Image, Claim, and Group), where Group is split into Control, Labeling, and Mislabeling. Each table cell contains the number of labeled and unlabeled images that participants rated.}
    \caption{Summary table of our 2 $\mathbf{\times}$ 2 $\mathbf{\times}$ 3 mixed factorial design. Each participant rated 24 posts, divided into four subsets with six images each. The subsets varied regarding the origin of the image (human, AI) and the veracity of the claim (true, false). In the control group, all images were unlabeled. In the labeling group, all AI-generated images were labeled. In the mislabeling group, two out of six images per subset were mislabeled, i.e., human-made images labeled as ``AI-generated'' and AI-generated images without a label (highlighted in \textbf{bold}).}
    \label{tab:experiment_design}
    \begin{tabular}{@{}lll@{\quad}l@{\quad}l@{}}
        \toprule
        \multirow{2}{*}[-2pt]{\textbf{Image}} & \multirow{2}{*}[-2pt]{\textbf{Claim}} & \multicolumn{3}{c}{\textbf{Group}} \\
        \cmidrule{3-5}
        & & \textbf{Control} & \textbf{Labeling} & \textbf{Mislabeling} \\
        \midrule
        \multirow{2}{*}{Human} & True & 6 unlabeled & 6 unlabeled & 4 unlabeled / \textbf{2 labeled} \\
        & False & 6 unlabeled & 6 unlabeled & 4 unlabeled / \textbf{2 labeled} \\
        \midrule
        \multirow{2}{*}{AI} & True & 6 unlabeled & 6 labeled & 4 labeled / \textbf{2 unlabeled} \\
        & False & 6 unlabeled & 6 labeled & 4 labeled / \textbf{2 unlabeled} \\
        \bottomrule
    \end{tabular}
\end{table}

\subsection{Method}
We conducted an online survey with $N=1\,354$ valid participants in April 2025.
We chose a survey setting to quantitatively investigate the effects of AI labels with a diverse set of participants.
The pre-registration, describing our hypotheses and analysis plan, is available at 
\href{https://osf.io/f6ztr}{osf.io/f6ztr}. 
We provide our analysis script as supplementary material.

\begin{figure*}[t]
    \centering
    \begin{subfigure}[t]{0.49\columnwidth}
        \frame{\includegraphics[width=\textwidth]{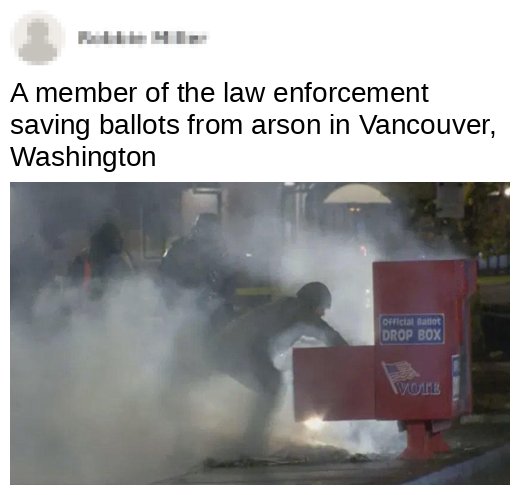}}
        \Description{A simulated social media post captioned with ``A member of the law enforcement saving ballots from arson in Vancouver, Washington''. The associated photo shows a person reaching into a ballot drop box amid fire and smoke.}
        \caption{Human/True. ``Was there an attempt to save ballots from arson in Vancouver, Washington?''\footnotemark[2]}
        \label{fig:stimuli_examples:human_true}
    \end{subfigure}
    \hfill
    \begin{subfigure}[t]{0.49\columnwidth}
        \frame{\includegraphics[width=\textwidth]{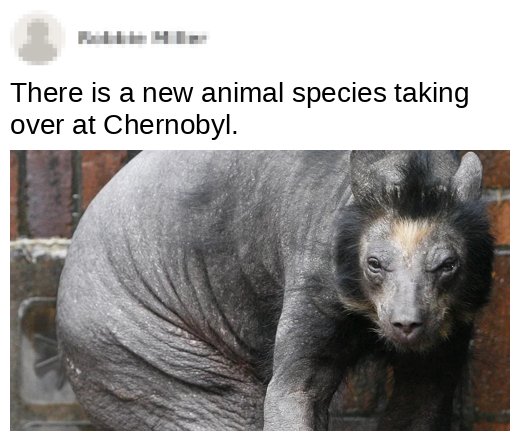}}
        \Description{A simulated social media post captioned with ``There is a new animal species taking over at Chernobyl.''. The associated photo shows a misshapen creature with leathery gray skin and a fur-covered face reminiscent of a bear.}
        \caption{Human/False. ``Has a previously unknown species been seen in Chernobyl?''\footnotemark[3]}
        \label{fig:stimuli_examples:human_false}
    \end{subfigure}
    \hfill
    \begin{subfigure}[t]{0.49\columnwidth}
        \frame{\includegraphics[width=\textwidth]{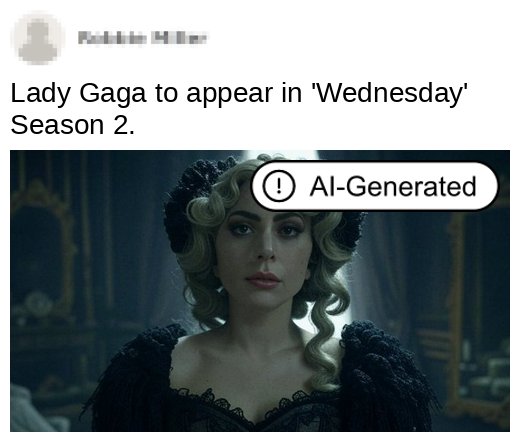}}
        \Description{A simulated social media post captioned with ``Lady Gaga to appear in 'Wednesday' Season 2.''. The associated photo shows Lady Gaga in a black robe looking at the camera. The photo is labeled as ``AI-Generated''.}
        \caption{AI/True. ``Is Lady Gaga going to appear in Wednesday Season 2?''}
        \label{fig:stimuli_examples:ai_true}
    \end{subfigure}
    \hfill
    \begin{subfigure}[t]{0.49\columnwidth}
        \frame{\includegraphics[width=\textwidth]{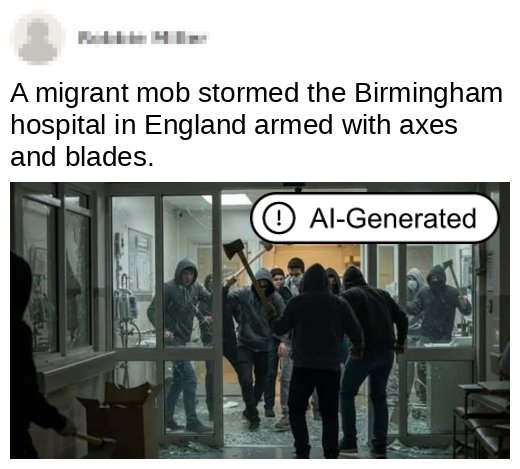}}
        \Description{A simulated social media post captioned with ``A migrant mob stormed the Birmingham hospital in England armed with axes and blades.''. The associated photo shows people in hoodies with masks and axes, smashing glass doors and storming a hallway. The photo is labeled as ``AI-Generated''.}
        \caption{AI/False. ``Did a migrant mob storm a hospital in Birmingham, England?''}
        \label{fig:stimuli_examples:ai_false}
    \end{subfigure}
    \Description{Four stimuli and claim combinations (a)-(d), each with a photo and caption, simulating a social media post. In (c) and (d), the label ``AI-Generated'' is added to the top right of the photo.}
    \caption{Example stimuli for each \emph{Image} (human, AI) $\mathbf{\times}$ \emph{Claim} (true, false) combination. The captions contain the questions that we asked participants to determine whether they believed in the posts' core claims. The complete set of stimuli and questions is provided in Appendix~\ref{app:survey:stimuli}.}
    \label{fig:stimuli_examples}
\end{figure*}

\boldparagraph{Procedure}
Our survey was conducted in English and consisted of five parts: study information and consent form, main experiment, supplementary questions, demographics, and debriefing.
Each participant was randomly assigned to one of three groups: control, labeling, and mislabeling. 

After giving their informed consent, participants were instructed about the upcoming task, which was described as ``identifying posts containing false claims that appeared on a social media platform''.
To clarify the meaning of the label, we informed participants in the treatment groups (labeling and mislabeling) that the platform uses a system to label images that might be generated using AI.
While we did not mention labels to participants in the control group (to not imply that all unlabeled posts were human-made), we informed them that posts might contain \glspl{aigi}.

In our main experiment, we employed a $2\times2\times3$ mixed factorial design with \textit{Image} (human, AI) and \textit{Claim} (true, false) as within-subjects factors and \textit{Group} (control, labeling, mislabeling) as between-subjects factor (see Table~\ref{tab:experiment_design}).
This design allowed us to disentangle the impact of both factors (\textit{Image} and \textit{Claim}).
If we had tested only human-made images with true claims and \glspl{aigi} with false claims, we would have been unable to tell whether participants' judgments were truly more accurate (in the presence of labels) or whether they simply judged a claim based on its label.
All participants saw the same 24 simulated social media posts, with six posts per condition (\textit{Image} $\times$ \textit{Claim}).
Early tests with colleagues indicated that 24 images were within our targeted timeframe and that utilizing much more images resulted in perceived monotony and fatigue.
While participants in the control group saw no AI labels at all, in the labeling group, all posts containing an \gls{aigi} were labeled.
In the mislabeling group, two out of the six posts in each condition were mislabeled.
Counterbalancing ensured that each post was mislabeled equally often.
Two additional posts served as attention checks, resulting in a total of 26~posts.
The order of posts was randomized, except for the attention checks.

For each post, participants answered two questions.
First, we asked whether they believed in the post's claim or not as a yes-no question (Q1).
Following previous work~\cite{wittenbergLabelingAIgeneratedMedia2024}, we adapted the question to each caption's core claim (see examples in Figure~\ref{fig:stimuli_examples}).
Second, we asked participants how confident they were in their assessment on a 4-point Likert scale (Q2).
Both questions were displayed next to the post (see Figure~\ref{fig:survey_screenshot} in the appendix) and appeared after a short delay, to guide the initial attention to the post.

After the main experiment, we asked participants in the treatment groups supplementary questions inspired by our focus group results.
First, we asked them to indicate whether AI labels had affected their previous ratings (Q3) and whether they noticed mislabeling (Q4--Q5).
Afterward, we were interested in their general opinion on mislabeling (Q6--Q8) and whether they would like to see AI labels on real social media platforms (Q9).

For participants in all groups, the survey concluded with demographics and a detailed debriefing.
We provide the full questionnaire in Appendix~\ref{app:survey:questionnaire}.
To estimate the exact duration and obtain feedback, we conducted three pilots on Prolific with six to 15 participants.

\begin{table*}[ht]
    \caption{Demographics of our 1\,354 valid survey participants. Our sample included 672 participants from the U.S.\ and 682 from the EU. We provide the distribution over EU countries in Appendix~\ref{app:survey_countries}. Note that not all numbers add up to the total amount, as participants had the option not to answer.}
    \Description{Table with demographics of the valid survey participants, including gender, region, education, age, and political views. The demographics are divided into EU and U.S. participants, the column labeled ‘N’ covers both regions. The column labeled ‘\%’ shows the percentages of of all participants.}
    \label{tab:survey_demographics}
    \centering
    \small
    \begin{tabular}{lrrrr@{\qquad}lrrrr}
    \toprule
    \textbf{Gender} & \textbf{EU} & \textbf{U.S.} & \textbf{N} & \textbf{\%} & \textbf{Age} & \textbf{EU} & \textbf{U.S.} & \textbf{N} & \textbf{\%} \\
        \quad Female & 340 & 334 & 674 & 49.8 & \quad 18--24 & 188 & 54 & 242 & 17.9 \\
        \quad Male & 335 & 325 & 660 & 48.7 & \quad 25--34 & 288 & 201 & 489 & 36.1  \\
        \quad Non-binary & 7 & 10 & 17 & 1.3 & \quad 35--44 & 122 & 154 & 276 & 20.4 \\
    \textbf{Education} &&&&& \quad 45--54 & 53 & 129 & 182 & 13.4 \\
        \quad 10th grade or less & 11 & 8 & 19 & 1.4 & \quad 55--64 & 22 & 97 & 119 & 8.8 \\
        \quad Secondary school & 86 & 75 & 161 & 11.9 &  \quad 65+ & 9 & 36 & 45 & 3.3 \\
    \quad Trade/technical/vocational & 29 & 11 & 40 & 3.0 & \textbf{Political views} \\
        \quad University/College w/o degree & 103 & 113 & 216 & 16.0 & \quad Very left & 98 & 168 & 266 & 19.6 \\
        \quad Associate degree & 12 & 81 & 93 & 6.9 &  \quad Left leaning & 247 & 170 & 417 & 30.8 \\
        \quad Bachelor's degree & 222 & 247 & 469 & 34.6 & \quad Center & 131 & 125 & 256 & 18.9 \\
        \quad Master's degree & 187 & 108 & 295 & 21.8 & \quad Right leaning & 121 & 129 & 250 & 18.5 \\
        \quad Professional degree & 11 & 11 & 22 & 1.6 & \quad Very right & 10 & 52 & 62 & 4.6 \\
        \quad Doctoral degree & 18 & 17 & 35 & 2.6 & 
        \quad Not interested & 61 & 23 & 84 & 6.2 \\
    \bottomrule
    \end{tabular}
\end{table*}

\boldparagraph{Stimuli}
\label{sec:survey:stimuli}
\footnotetext[2]{Image courtesy of Associated Press (AP).}
\footnotetext[3]{Image courtesy of Sebastian Willnow/DDP/AFP via Getty Images.}
All stimuli were presented as typical social media posts, consisting of a caption and a corresponding image (see Figure~\ref{fig:stimuli_examples}).
Posts were static, i.e., there were no options to like, share, or comment.
We added author names and images to make posts more realistic, but blurred them to prevent an unwanted bias. 
Labeled images contained the text ``AI-Generated'' in the top-right corner, with black font on white background.
We chose this design to inform about the use of \gls{genai} but not convey any positive or negative sentiment.
Moreover, related work~\cite{epsteinWhatLabelShould2023} shows that most people correctly associate the term ``AI-generated'' with \gls{aigc}.

Following previous studies~\cite{pennycook_implied_truth,fengExaminingImpactProvenanceenabled2023,wittenbergLabelingAIgeneratedMedia2024} we sourced social media posts from popular fact-checking sites (e.g., \href{https://snopes.com}{snopes.com}, \href{https://factcheck.afp.com}{factcheck.\newline afp.com}).
This approach ensured that our stimuli were relevant and representative of those actually circulating and avoided the risk of creating and spreading novel misinformation.
We took several measures to ensure similarity and comparability of all four conditions (\textit{Image} $\times$ \textit{Claim}).
First, we balanced the posts in all conditions regarding their topics (ranging from politics to lifestyle news). 
Second, while we kept the posts' captions as close to the original as possible, we shortened overly long ones and adjusted captions that were too emotional or did not clearly purport the post's claim, as this could have affected our results (e.g., participants needing to scroll for longer captions, but not for shorter ones).
Thirdly, for the two AI-generated conditions, we only selected posts containing realistic-looking images.
This is by no means an artificial restriction, as well-made \glspl{aigi} are nowadays almost indistinguishable from real ones.
Lastly, to reduce partisan bias, we avoided posts that were clearly left- or right-leaning. 
We provide all images, captions (original and edited), and corresponding questions in Appendix~\ref{app:survey:stimuli}.

Notably, to investigate whether AI labels simply decrease participants' belief in a claim or whether they help to make better judgments, our study design requires posts that convey a true claim through an \gls{aigi}.
Since we did not find enough appropriate posts for this condition on fact-checking sites, we took false posts with \glspl{aigi} (different from those in the AI/false condition) and adjusted the captions to make the associated claims true.
As a consequence, those claims are not directly taken from social media posts, however, the associated \glspl{aigi} are.
Given the increasing normalization of \gls{genai}, we consider it important to study how users interact with such posts.
News articles were already illustrated with \glspl{aigi} from stock image sites~\cite{oremusTheseLookPrizewinning2023,wilsonAdobeSellingFake2023}. 
Moreover, first news outlets are using \gls{genai} to compose articles, as the example of a German tabloid shows, where 11\% of all articles are written by an ``AI journalist''~\cite{nicoudBringingAI4002024,nicnewmanJournalismMediaTechnology2024}.

\boldparagraph{Recruitment}
To investigate a similar population as in our focus groups, we again sampled participants via Prolific.
We balanced our sample regarding country of residence (50\% EU and 50\% U.S.) and gender.
Moreover, we added a screener to filter for experienced participants (approval rate $>90\%$, completed surveys $>50$), that had fluent English skills and used any social media platform.
Each participant was paid \pounds 2.86 for an expected duration of 16~minutes.
The actual median completion time was 10~minutes.
Our power analysis suggested a minimum of 342 participants per group.
To account for invalid responses, we aimed for 470 participants per group, thus, 1410 in total.
In the end, we received 1\,405 completed surveys, out of which we excluded 51 participants that failed one or both attention checks.
Table~\ref{tab:survey_demographics} lists the demographics of our 1\,354 valid participants.

\boldparagraph{Metrics}
We analyzed participants' behavior by calculating their response accuracy, which is defined as the fraction of correct responses.
These include both true claims rated as true (hits) and false claims rated as false (correct rejections).
However, accuracy conflates two distinct aspects influencing the decision: \textit{sensitivity} (ability to tell apart true from false claims) and \textit{response bias} (general tendency to rate a claim as true or false).
To disentangle them, we leveraged the concept of \gls{sdt}~\cite{green1966signal}, which is well established in psychological research~\cite{andersonStatedependentAlterationFace2011,huangApplicationSignalDetection2020,LupyanLanguageCanBoost2013}.
\citet{batailler2022signal} previously showcased the value of \gls{sdt} for understanding \emph{why} people fall for misinformation.
In our setting, the sensitivity $d'$ indicates a participant's ability to correctly identify true claims (signal) against the background of misinformation (noise).
We used the classical univariate \gls{sdt} model assuming equal-variance Gaussian distributions for signal and noise.
Sensitivity is calculated as $d' = z(H) - z(FA)$, with the hit rate \textit{H} being the fraction of hits among true claims, and the false alarm rate \textit{FA} being the fraction of false claims rated as true among false claims. 
Here, $z$ denotes the inverse cumulative distribution function of the standard normal distribution 
that transforms \textit{H} and \textit{FA} into their underlying quantiles (corresponding to z-scores).
According to \gls{sdt}, the response bias $c$ is defined as $c = -1 \times \frac{z(H) + z(FA)}{2}$.
It reflects the threshold at which a participant switches their decision from false to true.
One participant might be more cautious, rather rejecting a true claim than falling for a false claim ($c > 0$), while another one might prioritize identifying all true claims, possibly accepting false claims ($c < 0$).

We tested our hypotheses using both accuracy and sensitivity to combine their advantages.  
While sensitivity is not affected by response bias, it cannot account for random effects for participants and images. Therefore, additionally fitting a \gls{glmm} with accuracy as the dependent variable allows us to generalize beyond our particular sample and stimuli set.
If the differences among participants and stimuli are random (and not systematic) and the judgment pattern is not driven by response bias, we should observe the same effects in both metrics.

\begin{table*}[t]
\centering
\small
\caption{Expected effects for our three hypotheses. For both metrics (accuracy and sensitivity), we report the statistical model we used and the effect we expect if the respective hypothesis proves correct. \textit{\contrastone} and \textit{\contrasttwo} refer to the first and second Helmert contrast of \textit{Group}, respectively.}
\Description{A table depicting our analysis approach for our three hypotheses, including the metrics used (accuracy/sensitivity), models (GLMM/ANOVA), and the expected effects that labeling as well as mislabeling should have under a given hypothesis.}
\label{tab:hypotheses_effects}

\begin{tabular}{lllll}
\toprule
\textbf{Hypothesis} & \textbf{Metric} & \textbf{Model} & \textbf{Effect of Labeling} & \textbf{Effect of Mislabeling} \\
\midrule

\multirow{2}{*}{Context-Label} 
  & Accuracy & GLMM & \textit{\contrastone} & \textcolor{gray}{no effect} \\ 
  & Sensitivity & ANOVA & \textit{\contrastone} & \textcolor{gray}{no effect} \\
\midrule
\multirow{2}{*}{Image-Label} 
  & Accuracy & GLMM & \textit{\contrastone} $\times$ \textit{Image} & \textit{\contrasttwo} $\times$ \textit{Image} \\
  & Sensitivity & ANOVA & \textit{\contrastone} $\times$ \textit{Image} & \textit{\contrasttwo} $\times$ \textit{Image} \\
\midrule
\multirow{2}{*}{Rely-on-Label} 
  & Accuracy & GLMM & \textit{\contrastone} $\times$ \textit{Image} $\times$ \textit{Claim} & \textit{\contrasttwo} $\times$ \textit{Image} $\times$ \textit{Claim} \\
  & Sensitivity & ANOVA &\textcolor{gray}{no effect} & \textcolor{gray}{no effect} \\
\bottomrule
\end{tabular}

\end{table*}

\subsection{Results}
We first outline our pre-registered hypotheses and the according analysis plan.
We then present our statistical and descriptive results.

\boldparagraph{Hypotheses}
We hypothesized that AI labels might affect participants’ judgments in two opposite ways\footnotemark[4]\footnotetext[4]{For clarity, hypotheses were renamed and restructured, but still correspond to those in the pre-registration.}.
First, the mere presence of labels might make participants focus more on the veracity of \emph{all posts} (labeled and unlabeled), leading to more accurate judgments.
This outcome would be ideal to counter misinformation.

\begin{description}
    \item[Context-Label Hypothesis:] The accuracy and sensitivity of \emph{all} posts should be higher in the treatment groups (labeling and mislabeling) compared to the control group. We expect an effect for \textit{\contrastone}. 
    Since the \emph{general presence} of labels causes the increased focus on veracity (not a post's individual label), we do not expect an effect due to mislabeling.
\end{description}
However, participants might focus more on the veracity of \emph{posts with labeled images} only, in which case only these claims would be judged more accurately.
\begin{description}
    \item[Image-Label Hypothesis:] The accuracy and sensitivity of posts with \emph{labeled} images (i.e., mostly \glspl{aigi}) should be higher in the treatment groups compared to the control group. We expect an \gls{ia} between \textit{\contrastone} and \textit{Image}. If mislabeling influences the effect, we expect an \gls{ia} between \textit{\contrasttwo} and \textit{Image}, due to more correctly judged posts with human-made images (since some are labeled) and fewer correctly judged posts with \glspl{aigi}. %
\end{description}
Alternatively, participants might simply interpret labels as an indication that a claim is false.
In this case, posts with labeled images would be judged as false more often, while posts with unlabeled images would be judged as true more often, independent of the claims' actual veracity.
\begin{description}
    \item[Rely-on-Label Hypothesis:] For posts with \glspl{aigi}, the accuracy of true claims should be lower in the treatment groups (compared to the control group), while the accuracy of false claims should be higher. In contrast, for posts with human-made images, the accuracy of true claims should be higher in the treatment groups, while the accuracy of false claims should be lower. For accuracy, this results in an \gls{ia} between \textit{\contrastone}, \textit{Image}, and \textit{Claim}. For sensitivity, we do not expect an effect since, based on our study design, both the hit and false alarm rate should change by the same amount. If mislabeling influences the effect, we similarly expect an \gls{ia} between \textit{\contrasttwo}, \textit{Image}, and \textit{Claim} for accuracy and no effect for sensitivity.
\end{description}
We summarize the expected effects in Table~\ref{tab:hypotheses_effects}.
Besides the main hypotheses on accuracy and sensitivity, we also expected that labeling (and mislabeling) might affect participants' response bias and confidence.
For both we expect an effect for \textit{\contrastone} and \textit{\contrasttwo}.

\begin{figure*}[t]
  \centering
  \begin{subfigure}[t]{0.615\columnwidth}
    \centering
    \aptLtoX[graphic=no, type=html]{
        \includegraphics[trim=0 0 2pt 0, clip=true]{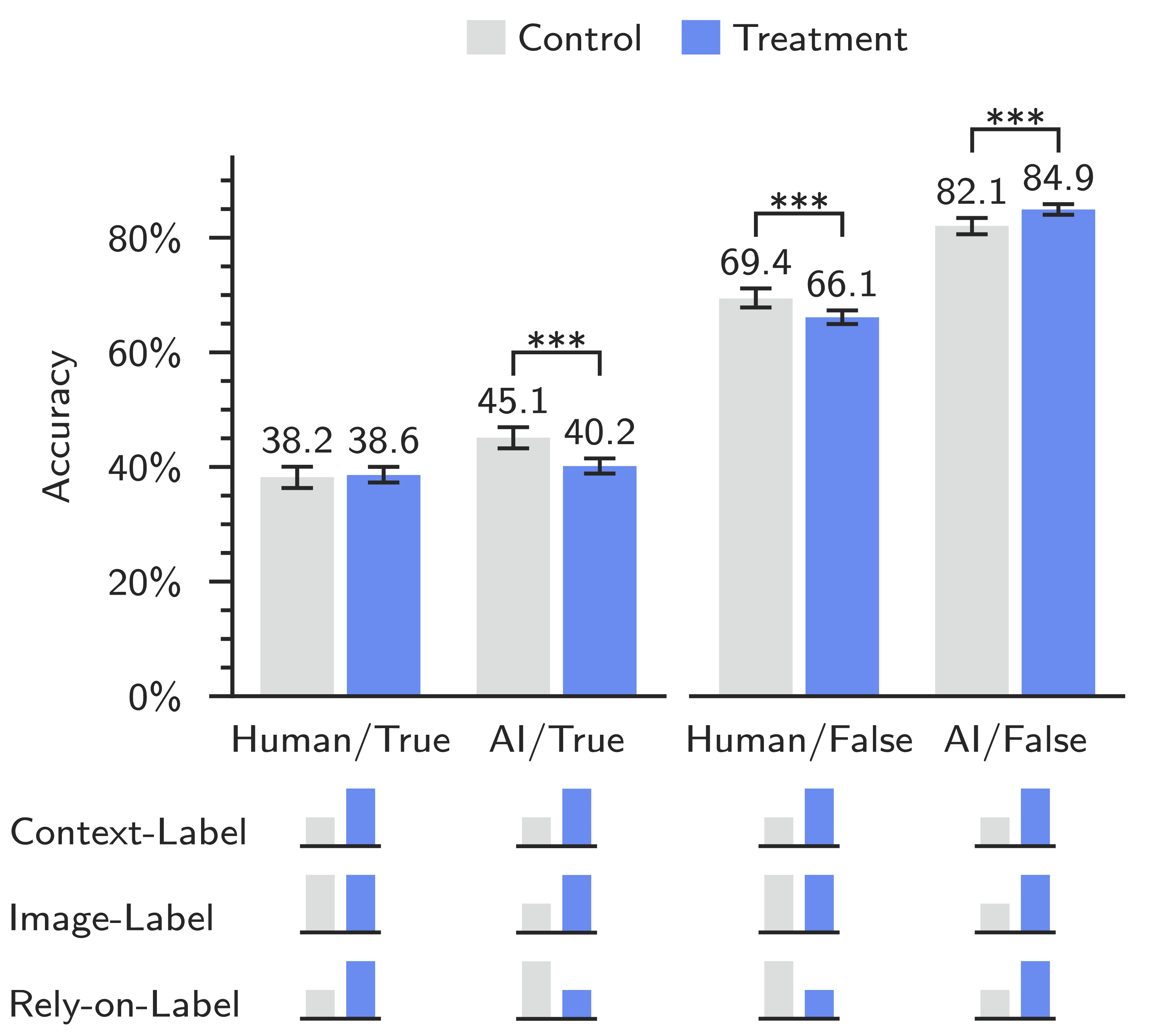}
    }{
        \includegraphics[trim=0 0 2pt 0, clip=true]{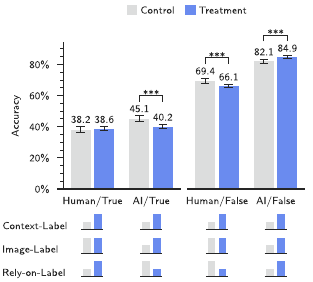}
    }
    \caption{Accuracy (control vs.\ treatment).}
    \Description{Vertical bar plot showing the accuracy of participants in the control and treatment groups. Treatment groups are both the labeling and mislabeling groups. For true claims, human-made images have an equal accuracy of about 38\% in both groups. In contrast, AI-generated images are rated more accurately by the control group (45.1\%) than the treatment groups (40.2\%) (significant). For false claims, human-made images are rated more accurately by the control group (69.4\% vs. 66.1\%) (significant), while AI-generated images are rated more accurately by the treatment groups (84.9\% vs. 82.1\%) (significant). For the context-label hypothesis, we expect an increase in accuracy from the control group to the treatment groups for all four conditions.
For the image-label hypothesis, we expect an increase in accuracy from the control group to the treatment groups for AI/true and AI/false, and no effect for human/true and human/false.
For the rely-on-label hypothesis, we expect an increase in accuracy from the control group to the treatment groups for human/true and AI/false, and a decrease for AI/true and human/false.}
    \label{fig:2a}
  \end{subfigure}
  \hfill
  \begin{subfigure}[t]{0.375\columnwidth}
    \centering
    \aptLtoX[graphic=no, type=html]{
        \includegraphics[trim=0 0 4pt 0, clip=true]{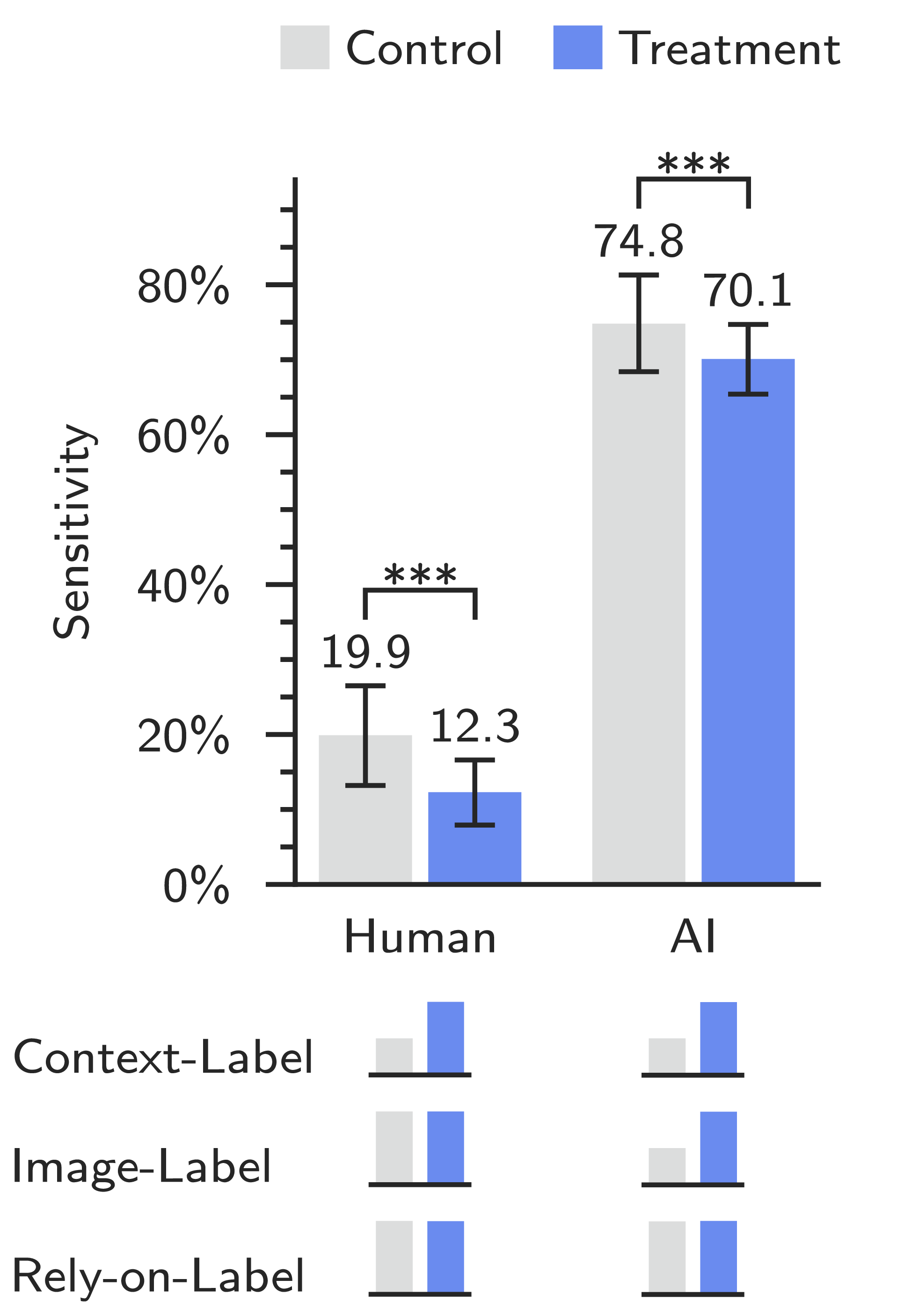}
    }{
        \includegraphics[trim=0 0 4pt 0, clip=true]{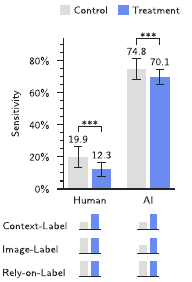}
    }
    \caption{Sensitivity (control vs.\ treatment).}
     \Description{Vertical bar plot showing the sensitivity of participants in the control and treatment groups.
 Treatment groups are both the labeling and mislabeling groups.
 For claims with human-made images the sensitivity was higher in the control group (19.9\%) compared to the treatment groups (12.3\%) (significant).
 For claims with AI-generated images the sensitivity was also  higher in the control group (74.8\%) compared to the treatment groups (70.1\%) (significant). For the context-label hypothesis, we expect an increase in sensitivity from the control group to the treatment groups for human and AI. For the image-label hypothesis, we expect an increase in sensitivity from the control group to the treatment groups for AI and no effect for human. For the rely-on-label hypothesis, we expect no effect for sensitivity for human and AI.
}
    \label{fig:3a}
  \end{subfigure}
  \hfill
  \begin{subfigure}[t]{0.615\columnwidth}
    \centering
    \aptLtoX[graphic=no, type=html]{
        \includegraphics[trim=0 0 2pt 0, clip=true]{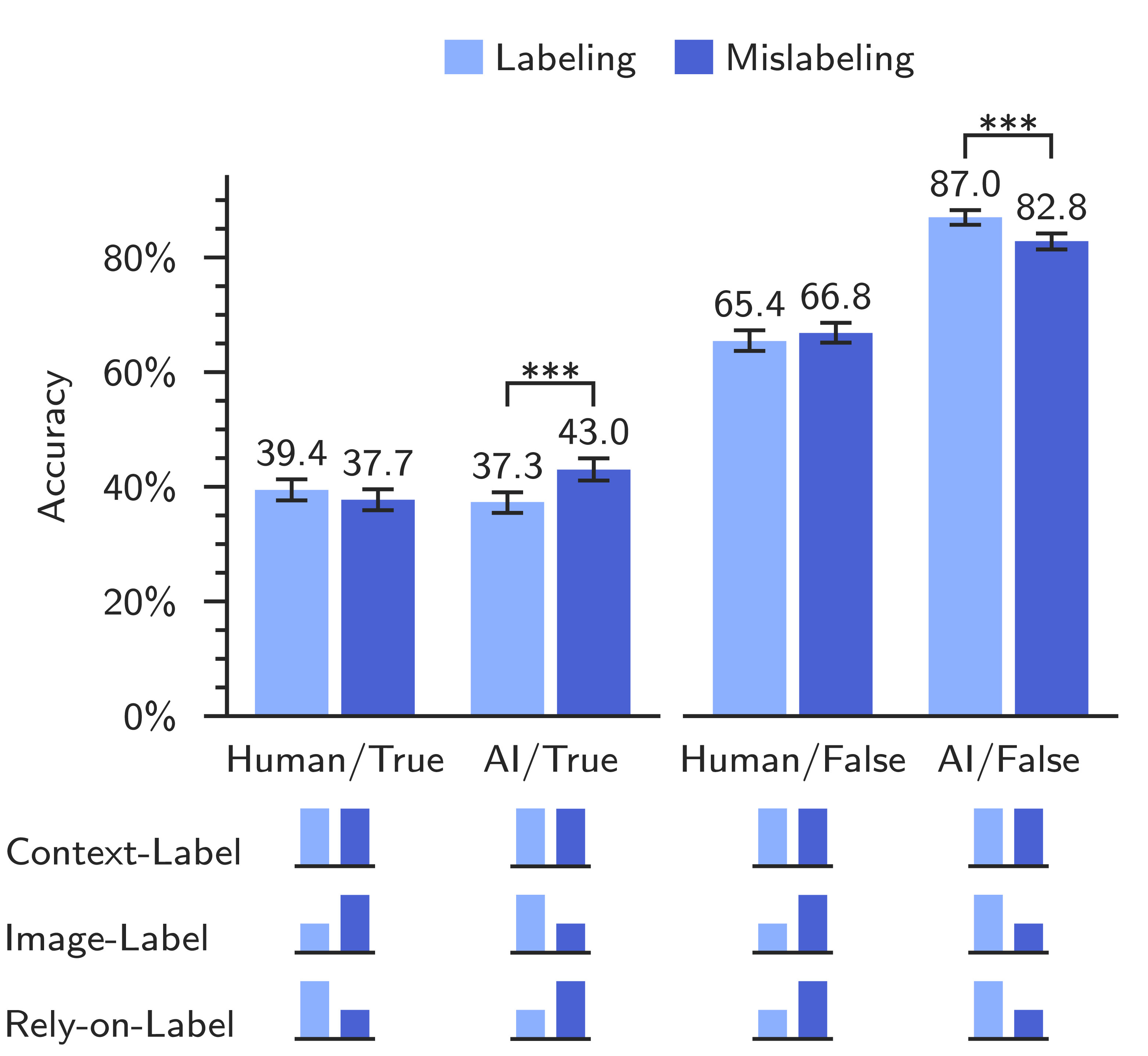}
    }{
        \includegraphics[trim=0 0 2pt 0, clip=true]{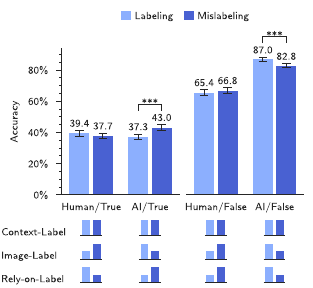}
    }
      \caption{Accuracy (labeling vs.\ mislabeling)\footnotemark[5].}
      \Description{Vertical bar plot showing the accuracy of participants in the labeling and mislabeling groups. For true claims, human-made images are rated more accurately by the labeling group (39.4\%) than the mislabeling group (37.7\%), whereas AI-generated images are rated considerably more accurately by the mislabeling group (43.0\% vs. 37.3\%) (significant). For false claims, human-made images are rated more accurately by the mislabeling group (66.8\% vs. 65.4\%), while AI-generated images are rated more accurately by the labeling group (87.0\% vs. 82.8\%) (significant).
      For the context-label hypothesis, we expect no effect on accuracy from the labeling group to the mislabeling group for all four conditions. For the image-label hypothesis, we expect an increase in accuracy from the labeling group to the mislabeling group for human/true and human/false, and a decrease for AI/true and AI/false. For the rely-on-label hypothesis, we expect an increase in accuracy from the labeling group to the mislabeling group for AI/true and human/false, and a decrease for human/true and AI/false.
}
    \label{fig:2b}
  \end{subfigure}
  \hfill
  \begin{subfigure}[t]{0.375\columnwidth}
    \centering
    \aptLtoX[graphic=no, type=html]{
        \includegraphics[trim=0 0 4pt 0, clip=true]{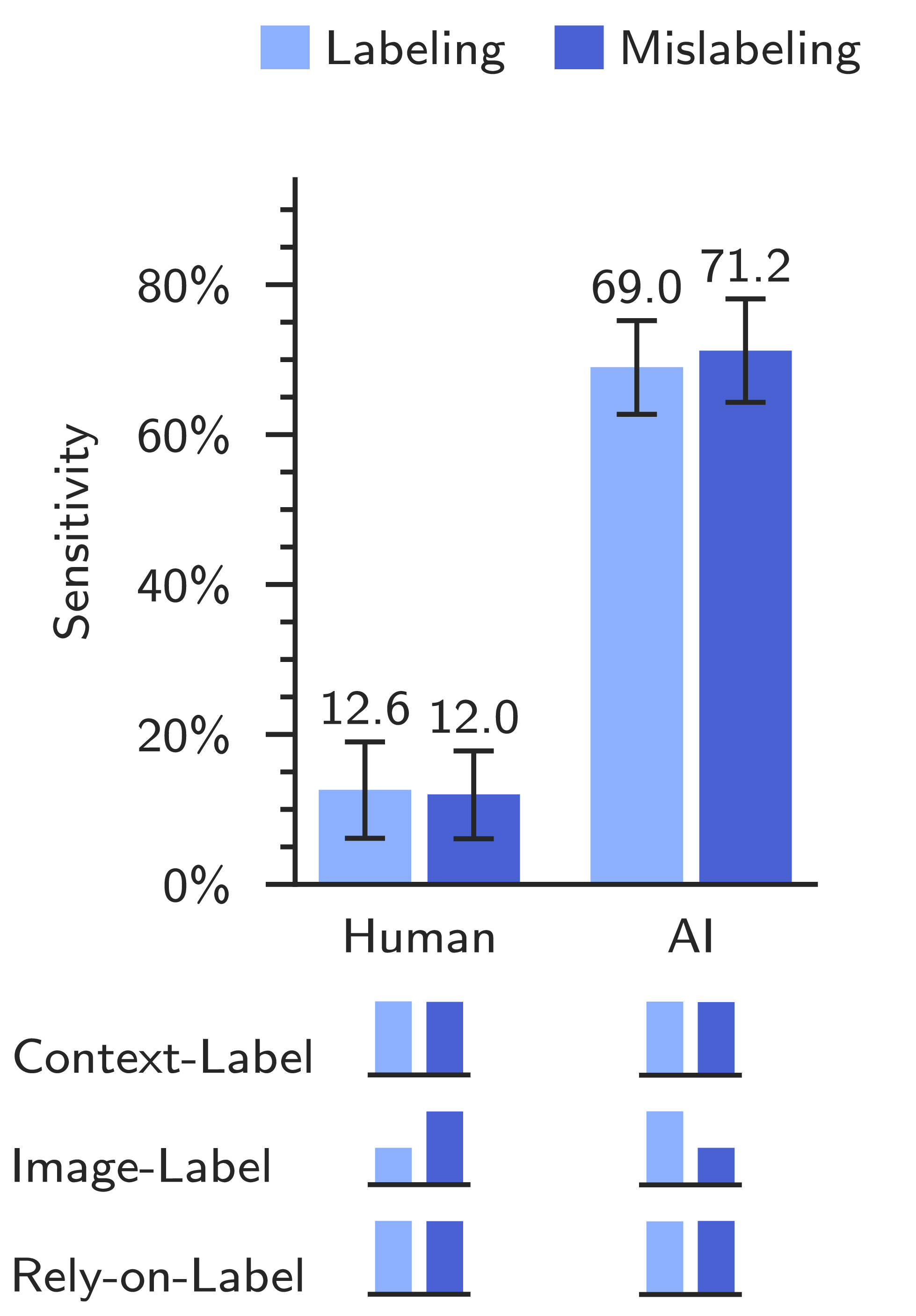}
    }{
        \includegraphics[trim=0 0 4pt 0, clip=true]{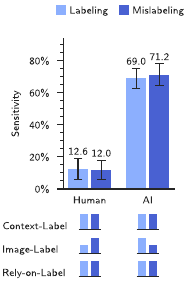}
    }
      \caption{Sensitivity (labeling vs.\ mislabeling)\footnotemark[5].}
       \Description{Vertical bar plot showing the sensitivity of participants in the labeling and mislabeling groups. Sensitivity for claims with human-made images is quite similar for both groups (12.6\% for the labeling group and 12.0\% for the mislabeling group). For claims with AI-generated images the sensitivity was lower in the control group (69.0\%) compared to the treatment groups (71.2\%). For the context-label hypothesis, we expect no effect on sensitivity from the labeling group to the mislabeling group for human and AI. For the image-label hypothesis, we expect an increase in sensitivity from the labeling group to the mislabeling group for human, and a decrease for AI. For the rely-on-label hypothesis, we expect no effect on sensitivity from the labeling group to the mislabeling group for human and AI.
}
    \label{fig:3b}
  \end{subfigure}
  \caption{Mean accuracy and sensitivity with 95\% confidence intervals. (a) and (b) compare the control group against both treatment groups (labeling and mislabeling), (c) and (d) compare the labeling against the mislabeling group. Results for accuracy ((a) and (c)) are separated by \textit{Image} (human, AI) and \textit{Claim} (true, false). Sensitivity is only separated by \textit{Image} since it is calculated from both true and false claims. ``$\mathbf{\ast\ast\ast}$'' denotes statistically significant effects ($\mathbf{p<0.001}$). Below each comparison, we illustrate the expected effects under each hypothesis 
  (\includegraphics[height=6pt]{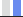}: no effect, \includegraphics[height=6pt]{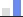}: increase, \includegraphics[height=6pt]{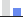}: decrease).
  }
  \Description{Four vertical bar plots displaying the mean accuracy separated by Image (human, AI) and Claim (true, false) ((a) and (c)) and mean sensitivity separated by Image (human/AI) ((b) and (d)) with 95\% confidence intervals. (a) and (b), compare the control and treatment groups while (c) and (d) compare the labeling and mislabeling group. Significant effects are highlighted. Below each plot there is a depiction of the expected effects for each of the three hypotheses. Comparing the actual with the expected effects allows for assessing the fit of the hypotheses.}
  \label{fig:bar_acc}
\end{figure*}

\boldparagraph{Analysis Plan}
We analyzed the \emph{relative differences} between groups (control, labeling, and mislabeling), i.e., we had no hypotheses for predictors only including \textit{Image} and/or \textit{Claim}.
This ensured that our results were not affected by potential material effects, since those would have influenced all groups equally.

Given our design, Helmert contrasts~\cite{granziol2025} allow for a comparison between control group vs.\ \textit{both} treatment groups (i.e., first Helmert contrast, \textit{\contrastone}), independently of the difference between the treatment groups.
This enabled us to compare the general absence of labels with the general presence of labels. 
Combining the labeling and mislabeling groups gave our results more statistical power and represented a realistic scenario, since labeling is never 100\% correct.
Importantly, with the second Helmert contrast (\textit{\contrasttwo}), we were able to find out whether the occurrence of mislabeling influences participants differently than correct labeling.

To investigate all hypotheses for accuracy (both regarding the effect of labeling and mislabeling), we fitted a \gls{glmm} with \textit{Group} (control, labeling, mislabeling), \textit{Image} (human, AI), and \textit{Claim} (true, false) as fixed effects, accuracy as the dependent variable, and planned Helmert contrasts (\textit{\contrastone} and \textit{\contrasttwo}).
The model had by-subject (i.e., participant) and by-item (i.e., image) random intercepts to account for differences between participants and posts. %
We did not interpret \glspl{me} or lower-order interactions if they were qualified by a significant higher-order interaction. 
To investigate all hypotheses regarding sensitivity, we ran a \textit{Group} $\times$ \textit{Image} mixed-model ANOVA with sensitivity as the dependent variable and planned Helmert contrasts for the \gls{me} of \textit{Group} as well as the interaction.
Analogously, we calculated a one-factorial ANOVA with \textit{Group} as independent and response bias as dependent variable, with planned Helmert contrasts for the \gls{me} of \textit{Group}.

To verify that our \gls{glmm} provides a robust model fit, we tested the underlying assumptions. The simulated residuals did not deviate significantly from uniformity (Kolmogorov–Smirnov test: $p=.47$). Neither the dispersion test ($p=.9$), the outlier test ($p=.22$), nor the Pearson chi-square test ($\chi^2=31\,561.83$, $\text{ratio}=0.97$, $p=.999$) suggested any model misspecification, presence of influential outliers, or overdispersion. Multicollinearity was assessed using variance inflation factors ($\text{VIF}<2.52$ for all predictors).
The assumptions for all other analyses were also met, except for variance homogeneity for the response bias ANOVA.
Therefore, we used heteroscedasticity-consistent covariates in this case.

If not stated otherwise, we performed all analyses according to our pre-registration.
We provide additional information on the used software in Appendix~\ref{app:survey:software}.

\begin{table*}[t]
    \Description{This table depicts the odds ratios, the 95\% confidence interval, z-value and p for each predictor (Claim, Contrast 1 (treatment vs. general labeling), Contrast 2 (labeling vs. mislabeling), Image) and their interactions.}
    \caption{Model parameters for the full GLMM with accuracy as the dependent variable and \textit{Group}, \textit{Claim}, and \textit{Image} as predictors. \textit{\contrastone} and \textit{\contrasttwo} refer to the first and second Helmert contrast of \textit{Group}, respectively.
    }
    \label{tab:accuracy_summary_all}
    \centering
    \small
    \begin{tabular}{lrrrr}
        \toprule
        \textbf{Predictor} & \textbf{Odds Ratios} & \textbf{CI (95\%)} & \textbf{z-value} & \textit{\textbf{p}} \\
        \midrule
            \rowcolor{gray!10}
        (Intercept) & 3.69 & 2.56--5.31 & 7.02 & \textbf{<0.001} \\
        \textit{Claim} & 0.17 & 0.10--0.29 & -6.68 & \textbf{<0.001} \\
            \rowcolor{gray!10}
        \textit{\contrastone} & 1.01 & 0.98--1.04 & 0.62 & 0.533 \\
        \textit{\contrasttwo} & 1.07 & 1.02--1.13 & 2.57 & \textbf{0.010} \\
            \rowcolor{gray!10}
        \textit{Image} & 0.62 & 0.43--0.90 & -2.54 & \textbf{0.011} \\
        \textit{Claim} $\times$ \textit{Image} & 1.44 & 0.86--2.41 & 1.39 & 0.164 \\
            \rowcolor{gray!10}
        \textit{Claim} $\times$ \textit{\contrastone} & 0.96 & 0.93--0.99 & -2.27 & \textbf{0.023} \\
        Image $\times$ \textit{\contrastone} & 0.94 & 0.91--0.96 & -4.77 & \textbf{<0.001} \\
            \rowcolor{gray!10}
        \textit{Claim} $\times$ \textit{\contrasttwo} & 0.90 & 0.84--0.96 & -3.33 & \textbf{0.001} \\
        Image $\times$ \textit{\contrasttwo} & 0.90 & 0.86--0.95 & -4.14 & \textbf{<0.001} \\  
       \rowcolor{gray!10}   
        \textit{Claim} $\times$ \textit{Image} $\times$ \textit{\contrastone} & 1.11 & 1.07--1.15 & 5.74 & \textbf{<0.001} \\
        \textit{Claim} $\times$ \textit{Image} $\times$ \textit{\contrasttwo} & 1.20 & 1.13--1.28 & 5.72 & \textbf{<0.001} \\
        \bottomrule
         \multicolumn{5}{l}{\scriptsize \textit{Coding of predictors: Claim (True = 1, False = 0), Image (Human = 1, AI = -1),}} \\[-0.75ex]
         \multicolumn{5}{l}{\hspace{1.7cm} \scriptsize \textit{\contrastone (Control = -2, Treatment = 1), \contrasttwo (Labeling = 1, Mislabeling = -1)}} \\
    \end{tabular}
\end{table*}

\boldparagraph{Participants Tend to Rely on Labels}
We first investigated the general effect of labeling (control vs.\ treatment groups), starting with our hypotheses regarding accuracy.
Since the three-way interaction of \textit{Claim} $\times$ \textit{Image} $\times$ \textit{\contrastone} (predicted for accuracy under the rely-on-label hypothesis) is the effect with the highest order, we first compared the full \gls{glmm} with a simpler model (R1) that excluded the three-way interaction.
This allowed us to check whether the model including the three-way interaction helps to explain our data better than a model only considering the two-way interactions and \glspl{me}.
\footnotetext[5]{For the context-label and image-label hypotheses, the expected effects only hold if mislabeling has no independent effect (e.g., it remains unnoticed).}
We used the \gls{aic} for comparison, with a lower \gls{aic} indicating a better model fit~\cite{akaike1998information}.
As the full model explained the data better than the reduced model ($\text{AIC}_\text{Full} = 36\,963.4 < \text{AIC}_\text{R1} = 36\,994.12$), we used it to further investigate our hypotheses. %
We found the three-way interaction to be significant (see Table~\ref{tab:accuracy_summary_all}) and analyzed it further with simple comparisons for each \textit{Image} $\times$ \textit{Claim} combination.
For AI/false posts, the accuracy was higher in the treatment groups ($z = 3.47, p < .001$) than in the control group.
Across all AI/false posts, the probability that participants correctly judged a post as false is 84.9\% if the image was labeled and 82.1\% if it was unlabeled.
However, the accuracy for AI/true posts was lower in the treatment groups ($z = -4.22, p < .001$).
For human/false posts accuracy was also lower in the treatment groups ($z = -3.15, p = .002$).
We found no difference for human/true posts ($z = 0.33, p = .740$) (see Figure~\ref{fig:2a}).
The observed pattern supports the \textit{rely-on-label} hypothesis: Participants appear to have used labels as an indication for a claim's veracity (except for human/true posts, where the effect did not become significant).
Consequently, our results speak against the \textit{image-label} hypothesis, according to which accuracy for posts with \glspl{aigi} should be higher in the treatment groups (due to the increased focus on veracity for labeled images) and the \textit{context-label} hypothesis, according to which accuracy should be higher for all posts in the treatment groups.\footnotemark[6]\footnotetext[6]{Beyond the three-way interactions, we found significant \glspl{me} for \textit{Claim}, \textit{Image}, and \textit{\contrasttwo} as well as two-way interactions for \textit{Image} $\times$ \textit{\contrastone}, \textit{Claim} $\times$ \textit{\contrasttwo}, and \textit{Image} $\times$ \textit{\contrasttwo}. \Glspl{me} for \textit{Image} and \textit{Claim} are not interpretable because we did not control for any material effects. 
We do not interpret the effect of \textit{\contrasttwo} and the two-way interactions as they are qualified by a significant three-way interaction.
}

To investigate whether accuracy was influenced by response bias, we next investigated our hypotheses regarding sensitivity.
Our analysis revealed an effect for \textit{\contrastone} ($t(1351) = -2.00,\ p = .046,\ d = .05$), indicating that sensitivity was higher in the control group ($M = .47,\ SD = .76,\ 95\%\ CI = (.42, .52)$) compared to both treatment groups ($M = .41,\ SD = .75,\ 95\%\ CI = (.38, .45)$), see Figure~\ref{fig:3a}.
No other effect for sensitivity reached statistical significance (all \(ps >.132\)).
In summary, differently than predicted by the \textit{context-label} and \textit{image-label} hypothesis, labeling appears to have made it harder (instead of easier) to distinguish between true and false claims.
However, the pattern can be explained by the \textit{rely-on-label} hypothesis.
A decrease in sensitivity ($d' = z(H) - z(FA)$) implies either a lower hit rate ($H$), a higher false alarm rate ($FA$), or both.
Recalling our simple comparisons, participants in the treatment groups indeed judged AI/true posts less often as true (lower $H$), AI/false posts less often as true (lower $FA$), and human/false posts more often as true (higher $FA$).
For the \textit{rely-on-label} hypothesis, we predicted no effect for sensitivity since we assumed that $H$ and $FA$ would change by the same amount.
However, our results suggest that, in total, the decrease in $H$ was not compensated by the change of $FA$, 
causing the lower sensitivity.

Finally, we directly investigated whether response bias differs between control and treatment groups.
Analyzing \textit{\contrastone} showed that response bias did not differ between the groups ($p > .253$).
Thus, our accuracy pattern cannot be explained by a difference in response bias but only by sensitivity, showing that accuracy is a valid measure for analyzing our data while accounting for random effects.
However, it is noteworthy that the overall response bias across all groups was rather conservative ($c=.47$).
Accordingly, participants were generally more likely to judge false claims as false and less likely to judge true claims as true.

\boldparagraph{Participants Still Rely on Labels in the Presence of Mislabeling}
Next, we investigated our hypotheses on the effects of mislabeling, again starting with accuracy.
We compared the full model with a simpler model (R2) that only excluded the three-way interaction of \textit{Claim} $\times$ \textit{Image} $\times$ \textit{\contrasttwo}.
Again, since the full model explained the data better than the reduced model ($\text{AIC}_\text{Full} = 36\,963.4 < \text{AIC}_\text{R2} = 36\,994.02$), meaning that the \gls{ia} between all three factors helps to explain our data, we used it to further investigate our hypotheses.
We found a significant three-way interaction between \textit{Claim}, \textit{Image}, and \textit{\contrasttwo} (see Table~\ref{tab:accuracy_summary_all}).
For the \textit{image-label} hypothesis we only predicted a two-way interaction between \textit{\contrastone} and \textit{Image}. As a three-way interaction instead qualifies the effect, we discard the \textit{image-label} hypothesis.
To investigate whether the obtained three-way interaction corresponds to the pattern predicted by the rely-on-label hypothesis, we calculated simple comparisons for each \textit{Image} $\times$ \textit{Claim} combination.
For posts with \glspl{aigi}, the accuracy for false claims was higher in the labeling group ($z = -4.27,\ p < .001$) than in the mislabeling group, while it was lower for true claims ($z=4.16,\ p < .001$).
However, for posts with human-made images, either with true claims ($z = -1.33,\ p = .18$) or false claims ($z = 1.14,\ p = .26$), accuracy did not differ between the labeling and the mislabeling group (see Figure~\ref{fig:2b}).
The observed pattern further supports the \textit{rely-on-label} hypothesis. %
For \glspl{aigi}, participants still relied on labels to judge a claim as true or false.
For human-made images, the direction of our results fits the hypothesis as well, but the effect did not become significant, potentially due to the small number of trials.

Analyzing sensitivity, neither the effect of \textit{\contrasttwo} nor the \gls{ia} for \textit{\contrasttwo} $\times$ \textit{Image} were significant ($ps > .63$), suggesting no differences in sensitivity between the groups (see Figure~\ref{fig:3b}).
This pattern is expected if people simply relied on labels, since the amount of posts where a label correctly indicates a false claim was the same in the labeling and mislabeling group.
Lastly, \textit{\contrasttwo} showed that response bias did not differ between the labeling and mislabeling group ($p > .07$), i.e., mislabeling had no effect on the response bias.

\boldparagraph{Labels Do Not Influence Confidence}
To investigate whether labeling influenced participants' confidence, we conducted \gls{lmm} analyses with Helmert contrasts for \textit{Group} as fixed effects and confidence (self-reported on a four-point Likert scale) as dependent variable (centered by the grand mean).
All models had by-subject (i.e., participant) and by-item (i.e., image) random intercepts.
We compared the full model with reduced models excluding the respective contrast of interest. 
The full model ($\text{AIC}_\text{Full} = 80\,711.69$) did not explain the data better than the reduced models ($\text{AICs} < 80\,703.5$). 
This indicates that no analyzed effect for confidence reached statistical significance (all $ps > .207$).

\boldparagraph{Analysis of Participants' Region}
Given that we recruited from the U.S.\ and the EU, we tested our full \gls{glmm} with region (U.S.\ vs.\ EU) as a moderator. 
Among the 12 effects examined in the post-hoc analysis, only the interaction of \textit{Claim} $\times$ \textit{\contrastone} ($z=-2.17,\ p=0.03$) and \textit{Claim} $\times$ \textit{\contrasttwo} ($z=-2.62,\ p =0.009$) were moderated, with a stronger effect in the U.S.\ sample.
However, both effects do not withstand Bonferroni correction (adjusted critical $p = 0.004$).

\begin{figure*}[t]
    \centering
    \begin{subfigure}[t]{\columnwidth}
        \centering
        \aptLtoX[graphic=no, type=html]{
            \includegraphics{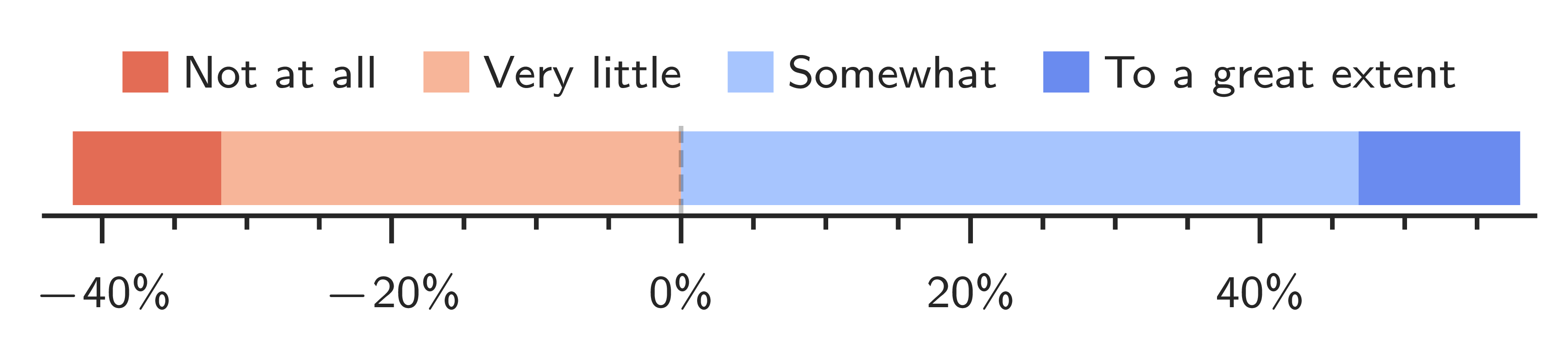}
        }{
            \includegraphics{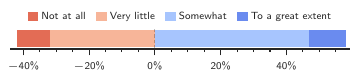}
        }
        \caption{Q3: Did you have the impression that the AI labels influenced your decisions in the previous task?}
        \Description{Participants' responses for Q3 are shown in a horizontal stacked bar plot. Slightly more than half of the people have the impression that AI label somewhat (48.82\%) or to a great extent (11.15\%) influenced their decision in the previous task. 31.77\% say it influenced them very little, 10.26\% say not at all.}
        \label{fig:q3}
    \end{subfigure}
    \hfill
    \begin{subfigure}[t]{\columnwidth}
        \centering
        \aptLtoX[graphic=no, type=html]{
            \includegraphics{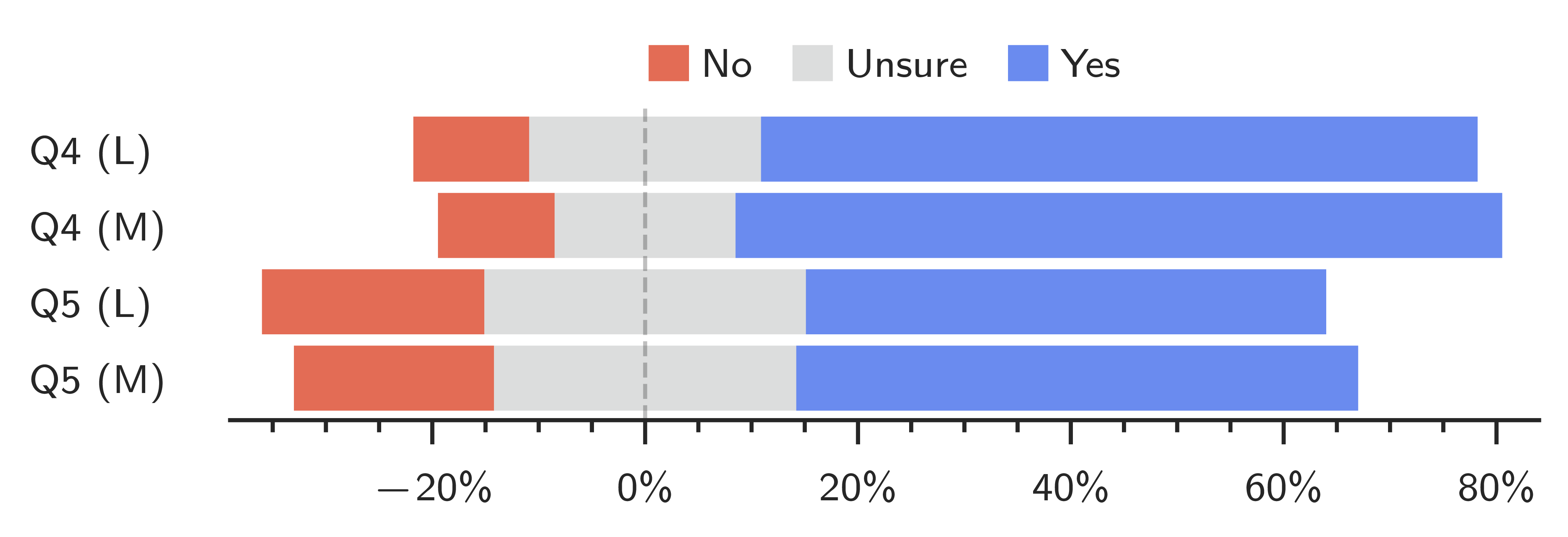}
        }{
            \includegraphics{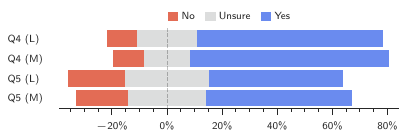}
        }
        \caption{Did you have the impression that, in the previous task... Q4:~some AI-generated images were not labeled as such? / Q5:~some human-made images were wrongfully labeled as ``AI-generated''? [L: labeling group, M: mislabeling group]}
        \Description{Participants' responses for Q4 and Q5 are shown in horizontal stacked bar plots, divided according to label and mislabel group. For Q4, in both groups (labeling and mislabeling) most people answer with yes (67.34\% and 72.04\%), some are unsure (21.78\% and 17\%), and some answer with no (10.89\% and 10.96\%). For Q5, fewer people answer yes (48.89\% and 52.8\%), more are unsure (30.22\% and 28.41\%) and answer no (20.89\% and 18.8\%). }
        \label{fig:q45}
    \end{subfigure}
    \vfill
    \begin{subfigure}[t]{\columnwidth}
        \centering
        \aptLtoX[graphic=no, type=html]{
            \includegraphics{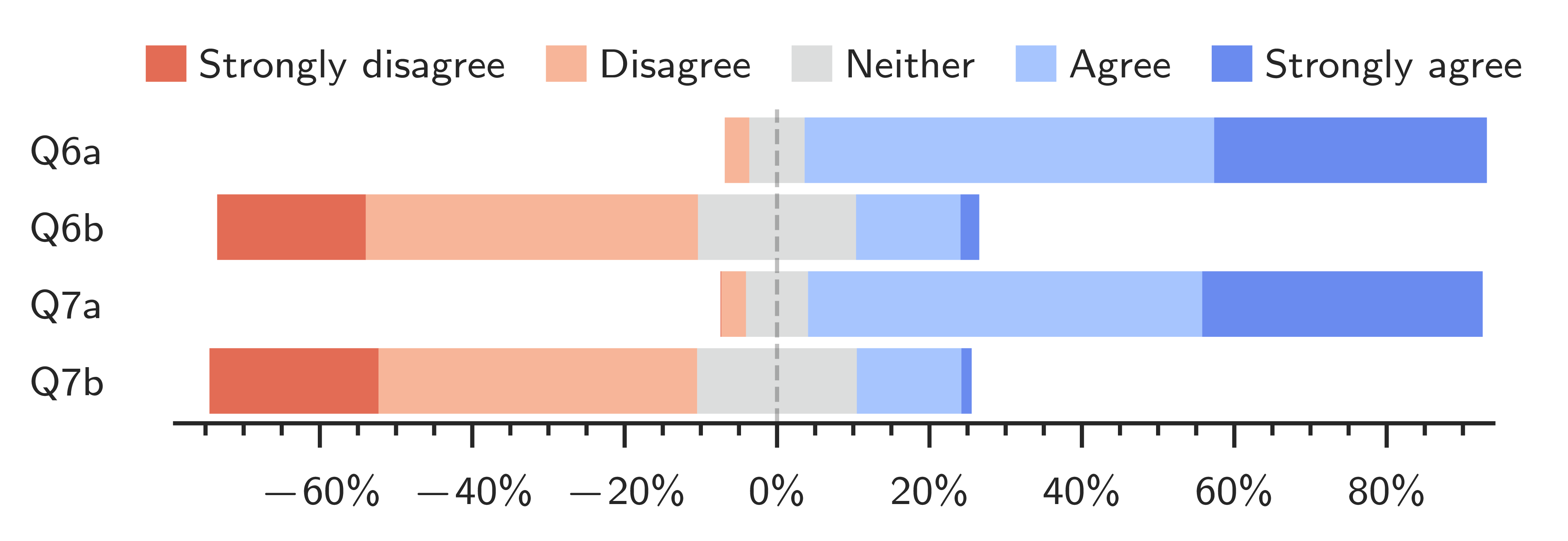}
        }{
            \includegraphics{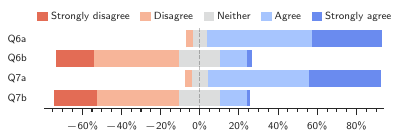}
        }
        \caption{Regarding unlabeled AI-generated images~(Q6) / wrongfully labeled human-made images~(Q7), how much do you agree with the following claims? a: Users lose trust in the labeling system if they become aware of such mislabeling. b: It is not a problem if such mislabeling only happens once in a while.}
        \Description{Participants' agreement to statements Q6a, Q6b, Q7a and Q7b from ``strongly disagree`` to ``strongly agree`` in four horizontal stacked bar plots. For both unlabeled AI-generated images (Q6) and wrongfully labeled human-made images (Q7), most participants agree (53.73\% and 51.73\%) and strongly agree (35.79\% and 36.79\%) that users lose trust in the labeling system if they become aware of such mislabeling (a). Most also say for both, that they disagree (43.59\% and 41.8\%) or strongly disagree (19.51\% and 22.19\%) that such mislabeling would not be a problem if it only happens once in a while. For both Q6b and Q7b, nearly 14\% agree with this statement.}
        \label{fig:q67}
    \end{subfigure}
    \hfill
    \begin{subfigure}[t]{\columnwidth}
        \centering
        \aptLtoX[graphic=no, type=html]{
            \includegraphics{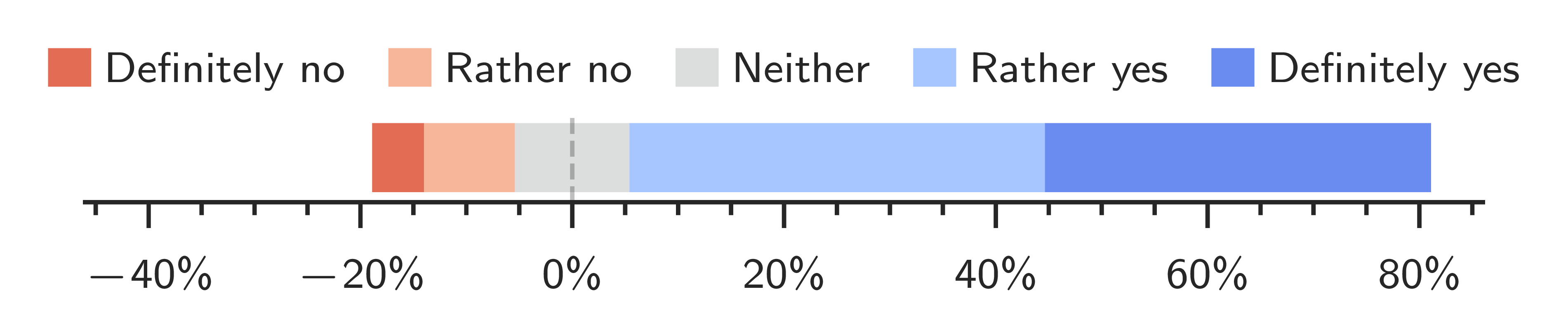}
        }{
            \includegraphics{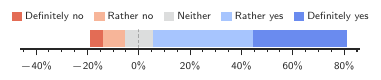}
        }
        \caption{Q9: Would you like to see AI labels (as they were presented in this study) on real-world social media platforms?}
        \Description{Participants' responses for Q9 are shown in a horizontal stacked bar plot. About 76\% of people would like to see AI labels on real-world social media platforms (39.24\% rather yes, 36.45\% definitely yes), whereas 4.91\% say definitely no, 8.58\% say rather no and 10.81\% say neither yes nor no.}
        \label{fig:q9}
    \end{subfigure}
    \caption{Participants' responses to our supplementary questions. Note that we asked these only to participants in the treatment groups, since the control group did not see any labels.}
    \Description{Four horizontal stacked bar plots, (a)-(d), showing participants' responses to the supplementary questions.}
    \label{fig:additional_questions}
\end{figure*}

\boldparagraph{Descriptive Results}
Finally, we report the responses to the supplementary questions inspired by the qualitative findings from our focus groups.
A majority of 58\% had the impression that labels had at least some effect on their judgments (Q3, see Figure~\ref{fig:q3}).
At the same time, 75.7\% would like to see AI labels on real social media platforms, while only 13.5\% were against them (Q9, see Figure~\ref{fig:q9}).
These results support the findings from our focus groups: Participants appear to find labels useful for judging a post's claim as true or false.

However, participants were relatively skeptical about the accuracy of labels (see Figure~\ref{fig:q45}).
67.3\% of the participants in the labeling group and 72\% of those in the mislabeling group stated to have noticed unlabeled \glspl{aigi} (Q4).
In contrast, only, 48.9\% (52.8\%) had the impression that some human-made images were mislabeled (Q5).
As a reminder, participants in the labeling group only saw correctly labeled images.
While the responses might be skewed due to the suggestive question, this skepticism reflects the plethora of concerns our focus group participants had about labels.

Our participants' opinion on mislabeling largely confirm our focus group findings (Q6/Q7, see Figure~\ref{fig:q67}).
Both for unlabeled \glspl{aigi} and for mislabeled human-made images, most participants expect users to lose trust, even if mislabeling occurs only rarely.
When asked directly which kind of mislabeling is worse (Q8), 47.5\% considered both errors to be equally bad, followed by unlabeled \glspl{aigi} (33.6\%) and wrongly labeled human-made images (17.4\%).

\boldparagraph{Key Findings}
Regarding \textbf{RQ2} (Effect of AI labeling on users’ perception of true/false claims with human-made/AI-generated images), 
we found that our participants relied on the presence (or absence) of labels when judging news posts as true or false.
Labels did not nudge them to focus more on the veracity of claims.
They also had no influence on the confidence with which participants made their judgments.
As a result, while labels increased the accuracy of participants' judgments for false claims with labeled \glspl{aigi}, we found significant negative side effects.
First, participants' belief in true claims accompanied by labeled \glspl{aigi} was reduced. 
Second, participants were more susceptible to misleading posts containing (unlabeled) human-made images.

Considering \textbf{RQ3} (Consequences of mislabeling and effect on users' trust),
our results indicate that users are influenced by the presence or absence of labels when judging a post's claim, regardless of whether an image is correctly labeled or mislabeled.
Thus, in a context where labels are present, they are more likely to fall for misinformation conveyed through unlabeled \glspl{aigi}.

\section{Discussion}
In this section, we discuss whether labels can help to combat AI-generated misinformation and address unintended side effects they might entail.
Moreover, we examine the implications of our findings for the practical deployment of AI labels.

\subsection{Users Want AI Labels --- But Do They Protect Against AI-Generated Misinformation?}

Our focus groups revealed that users perceive AI labels, despite their existing flaws, as a valuable tool for identifying \glspl{aigi} that might otherwise deceive them. %
We made similar observations in our survey, where over 75\% would have liked to see AI labels on social media platforms and more than 50\% felt that the labels influenced their judgments.
However, our quantitative results revealed that AI labels can only partially meet users' expectations.
While we could confirm previous work~\cite{wittenbergLabelingAIgeneratedMedia2024} that found labels to reduce the belief in misleading claims supported by \glspl{aigi}, we also found that labels could by no means safeguard against all AI-generated misinformation. %
In a similar vein, \citet{altayPeopleAreSceptical2024} discovered a small effect of AI labels. 
\citet{LabelingSyntheticContent2025} found that AI labels do not reduce user engagement on social media platforms.
Taken together, we conclude that the constructive effect of AI labels is limited and that they cannot fully prevent users from falling for deceptive \gls{aigc}.
Critically, our survey also revealed that AI labels do not work as legislators or platforms might expect.
Instead of increasing participants' awareness, nudging them to focus on a claim's veracity, users appear to rely on them to judge a claim as true or false.
Given that social media users typically spend only a few seconds per post~\cite{brunsMeasuringGainingHolding, facebookiqCapturingAttentionFeed}, they might use labels as a shortcut to judge the veracity of a claim.
\citet{fengExaminingImpactProvenanceenabled2023} made a similar observation when investigating C2PA-like provenance indicators, where participants confused the credibility of media with that of the provenance indicators.

While our research suggests that labels influence the underlying perception of a post, we currently know little about 
how labels interact with the framing effect of both image and caption.
Previous research has shown that the accompanying language influences how people perceive images~\cite{LupyanEffectsOfLanguage}.
\citet{PowellClearerPicture2015} showed that for war-related news, text and image frames are perceived differently: While the text frame influenced the participants' opinion towards a military conflict, the image frame affected their behavioral intentions, e.g., to donate.
\citet{Bingaman2021Siri} found that framing, either through text alone or the interplay between text and an image, can influence participants' support of AI.
We assume that both the visual and textual components of AI labels 
affect such framing effects, 
e.g., prior research has already shown that participants are sensitive towards different wordings of AI labels~\cite{epsteinWhatLabelShould2023, LabelingSyntheticContent2025}. 
We therefore urge future research to investigate participants perceptions not only on the label design itself but to also focus on the framing effect for the underlying post.

Users relying on labels to judge claims makes mislabeling especially problematic.
Our focus groups and survey showed that users are concerned about mislabeling, both false negatives and false positives.
Given the currently available labeling mechanisms, it is likely that many \glspl{aigi} will end up unlabeled.
If, as in our survey, users interpret the absence of a label as a sign that a post is true, the introduction of labels could \emph{increase} the risk of users falling for misleading posts.
Adversaries might intentionally circumvent the labeling system to make their misinformation appear more credible.
At the same time, users might not believe authentic content if it is wrongly labeled as AI-generated.
Our focus groups also showed that mislabeling can discredit, e.g., news agencies, and content creators.
Something similar has already happened to Meta: Their original label design, which stated that content was ``Made with AI'', was changed in a hurry to ``AI Info'' after backlash from users who considered their images to be falsely flagged~\cite{VergeInstagramsMadeWithAISwapped2024}.

Comparing the high expectations of our participants with the actual effect of AI labels and the consequences of mislabeling, we see a serious risk that users might overestimate the protection that labels can provide.
While introducing AI labels might initially reassure users, we hypothesize that, for instance, users relying on AI labels to protect their less tech-savvy loved ones will be disappointed by the outcome. 
We therefore urge legislators and platform providers not to solely rely on AI labels in the fight against AI-generated misinformation.
While they can be a complementary measure, they are no excuse to neglect or drop other safeguards, like deleting misinformation or educating users about the possible dangers of \gls{aigc}.
Notably, our survey results indicate a rather conservative response bias, meaning that participants tended to judge a claim as false more often than true.
This is in accordance with \citet{pfander2025spotting}.
In their meta-analysis of 67 papers, they find a skepticism bias, meaning that users were more prone to doubting trustworthy news than believing false news.
Therefore, we suggest not focusing solely on adding labels to \gls{aigc}, but also exploring ways to certify authentic and credible information.
As one positive example, YouTube began displaying a ``Captured with a camera'' disclosure in the video description if \gls{c2pa} metadata confirms that the content is authentic~\cite{youtubeBuildingTrustYouTube}.

\subsection{AI Labels May Have Unintended Side Effects}
Besides the intended effect of AI labels, 
we identified two problematic side effects.
These are caused by participants seemingly ``relying on'' labels to decide whether a claim is true or false.
In the following, we discuss explanations for these effects as well as their consequences.

\boldparagraph{AI Labels Make Users More Susceptible to ``Conventional'' Visual Misinformation}
Our survey showed that in the presence of labeled \glspl{aigi}, participants more often believed in misleading claims illustrated with human-made images.
Our results differ from \citet{altayPeopleAreSceptical2024}, who did not find such a spillover effect when investigating AI labels for headlines, which might be caused by a different effect of textual and visual misinformation on users~\cite{sundarSeeingBelievingVideo2021}.
One possible explanation for our findings could be an implied truth effect~\cite{pennycook_implied_truth}, paired with a misinterpretation of the label.
Users might trust the platform to label all false claims, which implies that a claim is true if an image is not labeled.
Alternatively, AI labels could distract users, by nudging them to focus on the novel dangers of \gls{aigc}.
If users are not willing or able to invest additional resources, 
they might just shift their attention from caring about ``conventional'' misinformation to AI-generated misinformation.
Despite the growing concerns about the malicious use of \gls{genai}, photos that are taken out of context or deceptively cropped are still a pressing problem~\cite{dufour2024ammeba}. 
If AI labels make users more susceptible to this kind of misinformation, the risks of them might be higher than their (current) reward.
We call upon future work to further investigate \textit{if} AI labels influence the perception of such misinformation in real-world scenarios and \textit{why} AI labels cause this change in perception.

\boldparagraph{AI Labels Make Users Believe Less in True Claims Supported by \glspl{aigi}}
As a second side effect, we found that participants' belief in true claims illustrated with labeled \glspl{aigi} was reduced in the treatment groups.
Both related work~\cite{altayPeopleAreSceptical2024,wittenbergLabelingAIgeneratedMedia2024,toffTheyCouldJust2025} and our focus groups suggest that users start to question or dislike \gls{aigc} if it is disclosed as such.
Taken together, we suspect that users' aversion towards AI might make them disbelieve claims with labeled images.
Previous research has shown that users' evaluation of content is influenced by their prior beliefs and attitudes towards a topic, e.g., a positive attitude towards an image's message causes a higher perceived image credibility~\cite{ShenFakeImages2019}.
\citet{Asher2018StrategicConversations} investigated conversations under imperfect information. In this context, a message might be interpreted selectively, so that the evaluation supports a preconceived belief about the subject, and, thus, might even reinforce this belief further. 
Transferring to labels, they increase the awareness that an image is AI-generated. Therefore, they might increase the influence that attitudes regarding \gls{genai} have on the evaluation of a claim or could even reinforce these attitudes.
Another explanation might be that participants simply find no good rationale for why real news stories should be accompanied by \glspl{aigi}.
All of this could contribute to users interpreting AI labels as an indication of misinformation, thus, leading to the observed side effect.
However, the current real-world impact on news is likely minimal, 
since most news agencies do not yet illustrate articles with realistic-looking \glspl{aigi}~\cite{heerFarAIgeneratedImages2023}.
If such an AI aversion persists, news outlets should critically evaluate if they ever want to relax their criteria for AI usage.
On the other hand, users' perception might also change with increasing exposure and normalization of \gls{aigc}. 
If the deterrent effect of AI labels is entirely built on the deterrent effect of the underlying \gls{aigc}, the results of a study like ours might turn out differently in a few years.
Therefore, it is essential that future work monitors users' reaction towards legitimate uses of \gls{aigc} as well as the effect that labeling still has on the perception of AI-generated misinformation.

\subsection{Findings in the Context of Previous Work on Trust in Technology}

Our qualitative results suggest that the successful adoption of AI labels largely depends on users' trust into their implementation and deployment.
We identify two dimensions that both have been addressed in the broader context of human trust in technology: trust in social media platforms and trust in algorithms (in our case, the labeling mechanism).
Regarding the former, \citet{Zhang2024Profiling} found that trust and distrust into social media can coexist: While users may distrust certain aspects of a platform, 
they may still have trust in other functionalities of the platform. 
In their review of 70~papers on trust in social media, \citet{ZhangWhatDoWeMean2023} identified transparency as an antecedent of trust, which was a critical requirement for our participants, too.
\citet{ShusasTrust2024} found that emerging adults trusted social media platforms to accurately utilize content labeling tools.
Moreover, they made the surprising observation that participants largely trusted automated approaches to content labeling, contradicting prior work~\cite{Waheeb2020} and also our findings.
They suspect that the increasing use of AI among younger people may increase their trust towards automated labeling systems.

While our focus groups showed that users' confidence varies from labeling mechanism to labeling mechanism, previous work by \citet{LangerLook2022} suggests that the terminology used to describe it may influence users' evaluation as well.
Testing different terms for algorithmic decision-making systems (ADMs) like ``Algorithm'', ``Automated System'', or ``Artificial Intelligence'' revealed a strong effect on participants' familiarity and tangibility.
Moreover, terminology affected their evaluation of the system's fairness and trust, which may benefit the actual use of the system.
Thus, we argue that not only AI labels themselves must be carefully designed, but also any accompanying descriptions of the underlying mechanisms.
\citet{dietvorst2015algorithm} found that users assess algorithms more harshly than humans.
After seeing an algorithm make a wrong prediction, participants lost confidence more quickly than after seeing a human forecaster make the same error.
Our results confirm that users' trust is fragile if mislabeling occurs.
Focusing on justice-related aspects in ADM systems, \citet{Binns2018ItsReducing} observed ambivalent attitudes.
While some participants considered automated decisions to be unfair per se, others saw value in the unbiased nature of algorithms.
Our focus groups leaned towards the former, having little trust in automated detection methods or fearing a possible interference by the model developers.
However, \citet{Bach03032024} found that users' trust is not fixed but can evolve with increasing user-system interactions.
They hypothesize that users trust the system more once they familiarize themselves and adjust their initial expectations.
Given that AI labels are a relatively new concept, analyzing how users' trust develops under repeated exposure is an important direction for future work.

\subsection{Implications of Our Study}

Our results indicate that expectations regarding the effect of AI labels on misinformation are currently too high.
Instead, AI labels pose the threat that they might delude into a false sense of security.
However, there currently exists no comprehensive measure to combat misinformation.
In this respect, depending on how the landscape of news (and misinformation) is evolving, a small positive effect of AI labels might already go a long way.
As of today, AI labeling is enforced by legislation worldwide and implemented by major social media platforms.
Therefore, we urge for a thoughtful utilization of AI labeling and to address concerns of users.
After all, even if labeling cannot fully protect against misinformation, it is still valued by the users.
We recommend considering the following points for deploying AI labels:

\boldparagraph{Education}
First and foremost, platforms must educate users about the merits and limits of AI labels. 
They must clarify that AI labels are not an indicator for misinformation, but only for AI origin.
Until now, AI labels were introduced more or less silently.
Many of our participants did not notice AI labeling, even on platforms that had already deployed it.
However, we argue that users must know what is asked of them when uploading their \gls{aigc}, especially for self-disclosure, as well as how to assess labels if they spot them on the platforms.
We call for further research on whether educating users about the side effects of labeling could prevent them.

\boldparagraph{Dealing with Mislabeling}
Since mislabeling can quickly erode users' trust, platforms should try to avoid it as much as possible.
Platforms exclusively relying on self-disclosure should reconsider their decision, as participants were especially critical of this mechanism, knowing malicious users would not disclose their AI usage. 
Occurrences of mislabeling should be openly and transparently addressed to reduce backlash.
Moreover, there should be mechanisms in place to appeal labeling decisions.

\boldparagraph{Transparency} %
If users have reasons to question the neutrality of the labeling system, they will likely distrust it.
To prevent this, platforms need to transparently inform about their labeling rules, which mechanisms are used, and disclose the capabilities and weaknesses of their labeling system.
This is especially relevant for detectors, as users might not comprehend their decisions, which could raise concerns about intentional biases, e.g., from training data.
Moreover, we urge future work to investigate how users' judgments are affected by the performance of the labeling system, i.e., if the risk of mislabeling is directly quantified.

\boldparagraph{Simplicity}
Participants want simple and comprehensible labeling policies.
We found evidence that users want all \glspl{aigi} to be labeled. 
This contradicts existing legislation, e.g., the EU \gls{aiact}~\cite{other/AI_Act}, which often only demands labels for contentious content.
However, more research is needed to investigate labeling policies further, e.g., our results indicated that users might expect a different approach to labeling partly-generated or edited images. 

\boldparagraph{Consistency}
Participants stressed that labeling rules need to be consistent, not only within a platform but also between platforms. %
However, the current landscape of labeling policies varies drastically.
While consensus between platforms might be hard to reach, platforms should at a minimum deploy consistent rules on their platforms, e.g., not allowing labeling exceptions.
Platforms should also avoid using different labels for different kinds of \glspl{aigi}, e.g., currently, content created with Meta's AI tools and directly shared to the platform might receive an ``Imagined with AI'' watermark, while content that is found to be generated based on ``AI signals'' will be labeled with an ``AI Info'' label~\cite{LabelAIContentInstagram}.

\boldparagraph{Responsibility}
A central concern of our participants was a potential abuse of power.
Considering users' preferences regarding who they trust to be responsible for labeling, we found indications of a wide range of opinions.
While \citet{SignalOfProvenance2025} also found diverse preferences, they focused on platforms and content creators as responsible entities.
However, our participants also considered e.g., legislation or central organizations. 
As users' trust into the labeling system is crucial for a successful adoption, we call on the research community to conduct further work to thoroughly investigate user needs and considerations regarding the responsibilities for labeling rules and enforcement, including all possible entities.

\subsection{Limitations}
\label{sec:limitations}
Both of our studies have a number of limitations.
Although Prolific has been found to produce reliable results in previous work~\cite{tangReplicationHowWell2022,peerDataQualityPlatforms2022,douglasDataQualityOnline2023}, our sample is not representative, only includes participants from the U.S.\ and EU, and might suffer from a self-selection bias.
The ATI-S~\cite{wesselATISUltrashortScale2019} and AIAS-4~\cite{grassiniDevelopmentValidationAI2023} scores of our focus group participants were relatively high.
Therefore, our participants might be more tech-savvy and more open towards new technology than the general population.
Moreover, self-reporting, social desirability bias, and group dynamics might have influenced participants' answers.
To counter these effects, we strictly adhered to the recommendations from \citet{krueger2014focus} when conducting our focus groups.
We tried to make participants comfortable and stressed that we will not judge any answers but were interested in their diverse perspectives. 
Appropriate interventions by the moderators prevented individual participants from dominating 
the discussion.
Over the course of the focus groups, we explained concepts (e.g., mislabeling) that participants might not have been aware of before.
As a result, they might have perceived \glspl{aigi} more negatively than they normally would, as this new information might have influenced or skewed their opinion.
However, these explanations were necessary to let our participants form informed opinions and allowed us to obtain meaningful insights.

To isolate the effects of our independent variables, our survey setting is artificial and does not fully correspond to a realistic interaction with social media.
Notably, we limited our evaluation to news-like posts accompanied by images, disregarding other types of content.
For posts in the AI/true condition, adjusting the originally false captions may have led to stimuli that were influenced by our own opinions and perceptions.
We also assume an equal distribution of true/false claims and human/AI images.
Finally, despite our best efforts to provide clear instructions and choose a validated label design~\cite{epsteinWhatLabelShould2023}, we cannot rule out that some participants might have considered the posts' captions to be AI-generated as well.
These limitations may have resulted in study artifacts and material effects not present in realistic social media feeds or missed interactions with other content.

\subsection{Ethical Considerations}
Our focus groups and survey were approved by the ethical review board of Saarland University under No.\ 24-09-1 and 25-01-4.
Before both studies, our participants were informed about its purpose and agreed to a consent form, also containing an option for withdrawal.
The respective consent forms can be found in Appendices~\ref{app:focus_groups:recruitment} and \ref{app:survey:recruitment}.
We minimized the collection of personally identifiable information and pseudonymized our focus group participants' names before the analysis.
We adhered to the GDPR for data collection, storage, and processing of participants' data.

To investigate the influence of AI labels on misinformation, we had to show participants misleading social media posts.
To foster discussions, we presented a few \glspl{aigi} in our focus groups.
However, we always explained their context and disclosed false claims.
For the survey, we carefully crafted our stimuli set to not upset or frighten our participants. 
While we could not fully disclose our goal to investigate AI labels when recruiting participants, we added a disclaimer that it contains misinformation that might touch on sensitive topics.
Before participants submitted the survey, they were shown a debriefing page, which was also taken into account for estimating the survey duration. 
Here, participants were clearly informed about which posts contained misinformation and/or were accompanied by an \gls{aigi}.
We also provided a link to a fact-checking article for each post.
As we carefully selected our images and all shown misinformation was strictly connected to our research, we deem the risks of our study acceptable.

\section{Conclusion}
In this work, we study the implications of labels for \glspl{aigi} in the context of misinformation using five focus groups and a pre-registered online survey with 1\,354 participants.
We found that while users considered AI labels a useful tool to distinguish real from AI-generated images, persistent concerns about labels and the underlying mechanisms 
can potentially erode their trust. %
Critically, our survey suggests that AI labels \emph{did not} help participants to judge claims more accurately.
Instead, participants appear to simply rely on the labels themselves, judging claims with labeled images more often as false and those without labels more often as true regardless of the actual veracity.
While labels could successfully decrease users' belief in misinformation containing \glspl{aigi}, an unintended side effect was that true information illustrated with a labeled \gls{aigi} was more frequently dismissed as false. 
Moreover, the presence of labels made users fall more often for ``conventional'' misinformation (without any involvement of \gls{genai}).
This overreliance on labels makes mislabeling especially problematic, as users are prone to fall for unlabeled misinformation or will not believe wrongly labeled true information.
These results underscore the need for platform providers and legislators to approach AI labeling with great caution. Policies and implementations must account for the expectations and concerns of users. 
Otherwise, users might lose trust or, even worse, the negative side effects of labels might overshadow their intended benefit for mitigating the threat posed by AI-generated misinformation.

\begin{acks}
We would like to thank all of our participants who took part in
our focus groups and survey and the anonymous reviewers for their valuable feedback.
Moreover, we thank Alexandra von Preuschen and Hoang Nguyen for their insights and for proofreading the paper.
This research was partially funded by Volkswagen\-Stiftung Niedersächsisches Vorab – ZN3695, the Deutsche Forschungsgemeinschaft (DFG, German Research Foundation) under Germany's Excellence Strategy - EXC 2092 CASA - 390781972, the Daimler and Benz Foundation
under the grant Ladenburger Kolleg, Project KonCheck, and the German Federal Ministry of Education and Research under the grants SisWiss (16KIS2330) and AIgenCY (16KIS2012).
\end{acks}

\bibliographystyle{ACM-Reference-Format}

\appendix

\section{Focus Groups}
\label{app:focus_groups}

\subsection{Prolific Description of ``Pre-Screener for Focus Groups about Labeling of AI Images''}
\label{app:focus_groups:screener_invitation}

We recruited participants using the following study description.
\begin{quote}
About this study:
The goal of this study is to investigate the labeling of AI-generated images from the end user's point of view. By collecting users' understanding, opinions and expectations of AI-labeling, we aim to pinpoint the challenges and benefits it can offer. We are particularly interested in the impact it might have on safeguarding against disinformation. Thus, our questions will focus on AI images, opinion, and expectations of labeling, and labeling mechanisms. \\

Focus Group Starting Times (a focus group takes about 90 minutes): \textit{*list of available dates*}

Important: This is only the pre-survey to recruit participants for a focus group. If you are selected, you will receive an additional invitation with the date, time, and access link for the actual focus group. The focus group will take about 90 minutes and will be compensated with £23.75. \\
Requirements to participate in the focus group:
\begin{itemize}
\item Working microphone and camera
\item Ability to participate in a Zoom call
\item The Zoom call will be recorded
\item Showing up in time to the appointment specified in the invitation
\end{itemize}

About this questionnaire:

In this questionnaire we ask for background information, demographics, and, most importantly, your availability during different time slots. Completing it should take approximately 8 minutes. Payment for the pre-survey will be released to all participants that gave valid answers. Participants that cannot participate at any of the specified dates will be screened out and receive a partial payment. 

Participation:
\begin{itemize}
    
\item  Payment for the pre-screener will be released to all participants that gave valid answers. 
\item  Participants that cannot participate at any of the specified dates will be screened out and receive a partial payment. 
\item  A publication of our results will be fully anonymous for you, at most, short anonymized quotes might be used. We will record the focus group to transcribe it for further analysis in our study. Any personal information will be de-identified in the transcripts and only appear in the form of aggregated data in our research publication later on.  
\end{itemize}
What is a focus group?:

In a focus group, various people, often with different backgrounds, come together and collectively work on or discuss a topic. We aim for our focus groups to consist of three to five participants. In the position of a moderator, we will ask various questions on the topic, which will then lead to open discussions where everyone is encouraged to express their opinion. It is not the goal of a focus group to judge or query for knowledge. During the discussion, there are no right or wrong answers. \\

For more information: https://research.teamusec.de/2024-ai-labeling/

Thank you for participating, we appreciate your valuable time and effort!
\end{quote}

\subsection{Focus Group Pre-Screening Questionnaire}
\label{app:focus_groups:screener}

We used the following questionnaire form our focus groups.

\begin{quote}
     
About this study:

The goal of this study is to investigate the labeling of AI-generated images from the end user's point of view. By collecting users' understanding, opinions and expectations of AI-labeling, we aim to pinpoint the challenges and benefits it can offer. We are particularly interested in the impact it might have on safeguarding against disinformation. Thus, our questions will focus on AI images, opinion, and expectations of labeling, and labeling mechanisms. 

About this questionnaire:

In this questionnaire we ask for background information, demographics, and, most importantly, your availability during different time slots. Completing it should take approximately 8 minutes. Payment for the pre-survey will be released to all participants that gave valid answers. Participants that cannot participate at any of the specified dates will be screened out and receive a partial payment. 

Important: This is only the pre-survey to recruit participants for a focus group. If you are selected, you will receive an additional invitation with the date, time, and access link for the actual focus group. The focus group will take about 90 minutes and will be compensated with £23.75.

Thank you for participating, we appreciate your valuable time and effort!

\end{quote}

\begin{itemize}
    \item[\textbf{Q1}] What is your Prolific ID?
Please note that this response should auto-fill with the correct ID [\textit{freetext}]
    \item[\textbf{Q2}] Please select all time slots (starting times) in which you can participate in a focus group. A focus group will take roughly 90 minutes. Important: If your time zone is not listed, please click on the respective link to convert the time slot into your time zone. The tool should automatically detect the correct time zone based on your location, if not, choose your location by clicking on "Change Your Location".
    [\textit{list of dates}, "None of the above time slots are suitable"]
    \item[\textbf{Q3}] Please confirm your selection.
Important: If you are matching our focus group criteria, we will try to assign you to one of your selected time slots and send you an invitation to the follow-up focus group as soon as possible. Registration is on a first-come, first-served basis, as we will send out more focus group invitations than necessary so that enough participants register.
["I have selected all suitable time slots, taking into account my local time zone.", "I do not want to participate in this study"]
    \item[\textbf{Q4}] In the following questionnaire, we will ask you about your interaction with technical systems. The term “technical systems” refers to apps and other software applications, as well as entire digital devices (e.g., mobile phone, computer, TV, car navigation). [Completely disagree, Largely disagree, Slightly disagree, Slightly agree, Largely agree, Completely agree, Prefer not to say]
        \begin{itemize}
            \item[\textbf{Q4a}] I like to occupy myself in greater detail with technical systems.
            \item[\textbf{Q4b}] I like testing the functions of new technical systems.
            \item[\textbf{Q4c}] It is enough for me that a technical system works; I don’t care how or why.
            \item[\textbf{Q4d}] It is enough for me to know the basic functions of a technical system.
        \end{itemize}
    \item[\textbf{Q5}] In the following, we are interested in your attitudes towards artificial intelligence (AI). AI can execute tasks that typically require human intelligence. It enables machines to sense, act, learn, and adapt in an autonomous, human-like way. AI may be part of a computer or online platform—but it can also be encountered in various other hardware devices such as robots.
    [Completely disagree, Largely disagree, Slightly disagree, Slightly agree, Largely agree, Completely agree, Prefer not to say]
        \begin{itemize}
            \item[\textbf{Q5a}] I believe that AI will improve my life.
            \item[\textbf{Q5b}] I believe that AI will improve my work.
            \item[\textbf{Q5c}] I think I will use AI technology in the future.
            \item[\textbf{Q5d}] I think AI technology is positive for humanity.
        \end{itemize}
    \item[\textbf{Q6}] Which of these social media or content creator platform do you use at least once a week? [ Facebook, Instagram, LinkedIn, Pinterest, Reddit, Snapchat, Telegram, Threads, TikTok, WeChat, X (Twitter), Youtube, "Other (please specify): , None, Prefer not to say]
    \item[\textbf{Q7}] What is your gender? [Woman, Man, Non-binary, Prefer to self-describe:, Prefer not to say]
    \item[\textbf{Q8}] What is your age in years? [\textit{freetext}]
    \item[\textbf{Q9}] What is your country of residence? [\textit{dropdown with 249 countries}, Other, Prefer not so say] 
    \item[\textbf{Q10}] In which time zone are you located? [\textit{dropdown with UTC timezones}, Prefer not to say]
    \item[\textbf{Q11}] Which of the following best describes the highest level of formal education that you have completed? [I never completed any formal education, 10th grade or less (e.g. some American high school credit, German Realschule, British GCSE), Secondary school (e.g. American high school, German Realschule or Gymnasium, Spanish or French Baccalaureate, British A-Levels), Trade, technical or vocational training, Some college/university study without earning a degree, Associate degree (A.A., A.S., etc.), Bachelor’s degree (B.A., B.S., B.Eng., etc.), Master’s degree (M.A., M.S., M.Eng., MBA, etc.), Professional degree (JD, MD, etc.), Other doctoral degrees (Ph.D., Ed.D., etc.), Other, Prefer not to say]
    
\end{itemize}

\subsection{Prolific Description of ``Focus Groups about Labeling of AI Images (\textit{[Date]})''}
\label{app:focus_groups:invitation}

The following study description was given to participants we selected to take part in a particular focus group.

\begin{quote}
This is the focus group about the labeling of AI images (following the pre-screener). The focus group will take about 90 minutes. \\
The focus group will take place on: \textit{*Date*} \\
Zoom Room: \textit{*Link*}

Please be on time for the study! Please contact us if you can no longer participate in the focus group after accepting the study. \\

Requirements to participate in the focus group:
\begin{itemize}
\item  Working microphone and camera
\item  Ability to participate in a Zoom call
\item  The Zoom call will be recorded
\item  Showing up in time to the appointment specified in the invitation
\item  Please take all necessary steps to enable undisturbed participation.
\end{itemize}

About this study:

The goal of this study is to investigate the labeling of AI-generated images from the end user's point of view. By collecting users' understanding, opinions and expectations of AI-labeling, we aim to pinpoint the challenges and benefits it can offer. We are particularly interested in the impact it might have on safeguarding against disinformation. Thus, our questions will focus on AI images, opinion, and expectations of labeling, and labeling mechanisms. 

Participation:
\begin{itemize}
\item  Payment for the focus group study will be released following your participation in the focus group.
\item  A publication of our results will be fully anonymous for you, at most, short anonymized quotes might be used. We will record the focus group to transcribe it for further analysis in our study. Any personal information will be de-identified in the transcripts and only appear in the form of aggregated data in our research publication later on. 
\end{itemize}

What is a focus group?:

In a focus group, various people, often with different backgrounds, come together and collectively work on or discuss a topic. We aim for our focus groups to consist of three to five participants. In the position of a moderator, we will ask various questions on the topic, which will then lead to open discussions where everyone is encouraged to express their opinion. It is not the goal of a focus group to judge or query for knowledge. During the discussion, there are no right or wrong answers. \\

For more information: https://research.teamusec.de/2024-ai-labeling/

Thank you for participating, we appreciate your valuable time and effort!

\end{quote}

\subsection{Study Information and Consent Form}
\label{app:focus_groups:recruitment}

Below we provide the consent form participants signed.

\begin{itemize}
    \item \textbf{This study's purpose} is to produce a scientific publication using anonymized data from the information you provide, with possible anonymous quotes from the focus group.
    \item \textbf{Eligibility} is open to individuals (1.) over the age of 18 (2.) who are active on social media platforms and  (3.) are aware of the existence of AI-generated content.
    \item A subsequent \textbf{focus group will be video recorded and transcribed} (converted to text) for analysis purposes by in-house automated transcription software or a GDPR-compliant external service.
    \item \textbf{Personal or project-related information} (e.g. your name) \textbf{will be removed} from the transcription (anony\-mized). We may only publish aggregated data or short quotes in our subsequent publication, without any traceability to you. We will delete the original record of the focus group after its transcription.
    \item \textbf{All study data will be hosted in a secure cloud or on internal servers} accessible only by project members, except in the case of external transcription.
    \item \textbf{Transcribed and anonymized data are kept for up to 10 years} in the spirit of good scientific practice, e.g. if questions about details arise later.
    \item We expect the focus groups to take up \textbf{roughly 90 minutes} of your time. We offer a \textbf{compensation of \pounds 23.75} for all participants that attended the focus group.
    \item Your \textbf{participation is voluntary}. You may stop participating at any time by closing the browser window or the program to withdraw from the survey. During the focus group, you may decide to drop out of it at any time. If you decide to withdraw your participation, we will delete your contribution to the focus group from the transcript. 
    \item The \textbf{risks to your participation} in this study are those associated with basic computer tasks, including boredom, fatigue, mild stress, or breach of confidentiality. The \textbf{benefits to you are your} compensation and the learning experience from participating in a research study. The \textbf{benefit to society} is the contribution to scientific knowledge.
    \item For any questions about this research, you may contact: 
    \begin{itemize}
        \item Sandra Höltervennhoff (Co-Project Lead, PhD student, \\ hoeltervennhoff@sec.uni-hannover.de) 
        \item Jonas Ricker (Co-Project Lead, PhD student, \\ jonas.ricker@rub.de) 
        \item Prof. Dr. rer. nat. Sascha Fahl \\ (Supervising Professor, fahl@cispa.de)
    \end{itemize}

\end{itemize}
By signing this consent form, I am affirming that...
\begin{itemize}
    \item I have read and understand the above information.
    \item I am 18+ and eligible to participate in this study.
    \item I am comfortable using the English language to participate in this study.
    \item I have chosen to participate in this study. I understand that I may stop participating at any time without penalty.
    \item I am aware that I may revoke my consent at any time by contacting the research team.
    \item I am aware that a follow-up focus group will be video recorded.
\end{itemize}

\subsection{Focus Group Guide}
\label{app:focus_groups:guide}
Here, we provide the questions we used to guide our semi-structured focus group discussions. Before the actual questions, the interviewers presented themselves and the purpose of the study, and participants were informed about how focus groups work and asked for consent regarding the use of their data. During the focus group, participants saw a slide deck showing the current topic and exemplary images.

\boldparagraph{Part A: Generative AI and Risks}
Generative AI systems are able to generate new content based on user input. A well-known example is ChatGPT, which can understand and answer questions from users. Another application is the generation of media, e.g., images, using simple descriptions in text form (e.g., the description ``A photo of a dog.''). The images are generated within a short amount of time and the users do not need to have any prior knowledge about image creation. The images generated in this way can appear very realistic and are increasingly difficult to distinguish from real media.

\begin{itemize}
    \item[\textbf{Q1.}] I will now ask everybody in turn, have you ever encountered AI-generated images?
    \begin{itemize}
        \item[\textbf{Q1a.}] What was it?
        \item[\textbf{Q1b.}] Where did you encounter it, for example, on which social media platform?
    \end{itemize}
    \item[\textbf{Q2.}] How did you recognize that the image was AI-generated?
    \begin{itemize}
        \item[\textbf{Q2a.}] How easy or hard did you find it to recognize that the image was AI-generated?
    \end{itemize}
\end{itemize}
AI-generated images can be used for many different applications, e.g., as educational content, for better illustration, or even for artistic purposes. However, AI images can also be problematic.

\begin{itemize}
    \item[\textbf{Q3.}] Can you think of any problems of AI-generated images?
\end{itemize}

\boldparagraph{Part B: Opinion and Expectations of AI Labeling}
As there is already misinformation that is created with AI today and fears are that this problem will continue to grow, efforts are being made to stop AI-generated disinformation. As one measure, various websites have started to identify and label images created with AI. Politically, this labeling is enforced, for example, in the Digital Service Act of the EU for very large online platforms.

\begin{itemize}
    \item[\textbf{Q1.}] Have you ever heard about AI labels or even encountered them yourself?
    \item[\textbf{Q2.}] What is your opinion towards such labels?
    \begin{itemize}
        \item[\textbf{Q2a.}] Do you find the idea of labeling AI-generated images helpful or not?
        \item[\textbf{Q2b.}] If you see such a label on an image, what would be your first thought/impression?
        \item[\textbf{Q2c.}] Do you think that such labels could also protect against disinformation?
    \end{itemize}
    \item[\textbf{Q3.}] Would you say that all AI-generated images should be labeled or only specific ones, like images that are misleading or could falsely appear to be authentic?
    \item[\textbf{Q4.}] Would you also label images that are edited using AI? One example is that AI filters are used to enhance the image or that the image background is adjusted, like removing a person.
    \item[\textbf{Q5.}] Who should be responsible for making such AI labeling rules and enforcing them?
\end{itemize}

\boldparagraph{Part C: Problems of AI Labeling}
\begin{itemize}
    \item[\textbf{Q1.}] How do you think that mechanisms to label AI generated images look like?
\end{itemize}
We will now present three methods of identifying AI images and would like to hear your opinion on them.
The simplest way to label AI-generated content on social media is self-disclosure. This means that when uploading something, the user is responsible to mark their content if it was created using AI.
Another technique is to automatically detect AI-generated images. These detectors typically also use AI and predict a score denoting how likely an image is AI-generated. The social media platform could apply this detector to all uploaded images and put a label on those that are found to be AI-generated.
A third option is to use metadata, which is embedded into an image when it is created. If  you use an online service to generate an image, the name of this service and some  additional information will be linked to the image file. Once you upload it, the platform can read this data and display the corresponding label.

\begin{itemize}
    \item[\textbf{Q2.}] What do you think about these approaches?
    \begin{itemize}
        \item[\textbf{Q2a.}] What do you think are advantages and disadvantages of each approach?
        \item[\textbf{Q2b.}] I will now ask everybody in turn, just your gut feeling, which of these approaches do you like the most?
        \item[\textbf{Q2c.}] Is there any approach where you would not trust the labels?
    \end{itemize}
\end{itemize}
We will now talk about the problems each of the three approaches have. With self-disclosure, people could just not indicate that they used AI to create an image, either intentionally or because they took an image from somewhere else and simply don’t know. People could also wrongfully say they used AI, reducing trust in the label.
Detectors can make wrong predictions (e.g., due to image processing or unseen generative models). This can cause false negatives (AI-generated image is not labeled) and false positives (real image is labeled as AI).
The main problem of metadata is that it can be removed intentionally (e.g., by taking a screenshot) or unintentionally (metadata is usually stripped during upload to social media platforms). Moreover, this approach only works if the providers of generative AI tools support the metadata. The approach can also be bypassed by using your own generative model.

\begin{itemize}
    \item[\textbf{Q3.}] Were you particularly surprised by any of the problems mentioned for the approaches?
    \item[\textbf{Q4.}] How do you rate the consequences if AI images are wrongfully not labeled?
    \item[\textbf{Q5.}] How do you rate the consequences if authentic images are mislabeled as AI-generated content?
    \item[\textbf{Q6.}] What do you consider worse, AI-generated images that are not labeled or authentic images that are mislabeled as AI-generated?
    \item[\textbf{Q7.}] How does mislabeling affect your opinion and trust in the label?
    \item[\textbf{Q8.}] Has your opinion towards labeling changed since the start of the focus group after hearing about concrete strategies to mark or detect AI images?
    \item[\textbf{Q9.}] Is there still anything related to the topic of AI labeling that anyone would like to share with us, maybe something that we forgot to ask?
\end{itemize}

\subsection{Codebook}
\label{app:focus_groups:codebook}

\begin{itemize}
    \item A1 Experience of AI
        \begin{itemize}
            \item A1 Social media
            \item A1 Ads
            \item A1 Creation of AI
            \item A1 News
            \item A1 Other websites
            \item A1 Seldom/ No experience
        \end{itemize}
    \item A2 Recognition of AI
        \begin{itemize}
            \item A2 Context cues
            \item A2 Recognition depends on attention
            \item A2 Recognition depends on creator
            \item A2 Recognition depends on picture
            \item A2 Recognition is easy
            \item A2 Recognition is hard
            \item A2 Recognition via label
            \item A2 Software
            \item A2 Visual recognition
        \end{itemize}
    \item A3 Problems of AI content
        \begin{itemize}
            \item A3 AI bias
            \item A3 Crime (blackmailing, deep porn, scamming etc.)
            \item A3 Deception (of skills)
            \item A3 Flooding
            \item A3 Forged evidence
            \item A3 Mis-/Disinformation
            \item A3 Unrealistic standards (beauty, good pics)
            \item A3 Availability to Everyone/Traceability
            \item A3 Bots
            \item A3 Copyright/Privacy issues
        \end{itemize}
    \item B1 Encountering of AI labels
        \begin{itemize}
            \item B1 Heard about labels (not encountered)
            \item B1 News
            \item B1 No encountering of labels
            \item B1 Social media
            \item B1 Studies
            \item B1 Used AI label
        \end{itemize}
    \item B2 Opinion towards AI labels
        \begin{itemize}
            \item B2 Helpful but limited
            \item B2 Helpful/Positive („they are great“)
            \item B2 Labeled content would be perceived negatively
            \item B2 Labeled content would be perceived positively
            \item B2 Labels could raise acceptance for using AI
            \item B2 Should be mandatory
            \item B2 Appealing system is important
            \item B2 Helpful for specific content
            \item B2 Helpful in preventing misinformation
            \item B2 Helpful in preventing scams
            \item B2 Needed in the long run
            \item B2 Spreads awareness about AI content
            \item B2 Unsure if perception would change/ Other perception
        \end{itemize}
    \item B3 AI images that should be labeled
        \begin{itemize}
            \item B3 Labeling of partly AI-generated images
                \begin{itemize}
                    \item B3 Difficult to decide on labeling rules/ Gray Area
                    \item B3 Not necessary to label minor AI manipulations (e.g. filters)
                    \item B3 Risk that labeling gets more complex
                    \item B3 All partial AI-gen images need label
                    \item B3 Depends on Content
                    \item B3 Different label
                \end{itemize}
            \item B3 Labeling of completely AI-generated images
                \begin{itemize}
                    \item B3 Difficulty to judge problems of AI images
                    \item B3 Labels for all AI-gen images
                    \item B3 Labels for contentious AI images
                    \item B3 Labels for images containing humans
                \end{itemize}
            \item B3 Decision Making/Enforcement
                \begin{itemize}
                    \item B3 Central Organization
                    \item B3 Community
                    \item B3 Consistency
                    \item B3 Creator
                    \item B3 Government/Law
                    \item B3 Other
                    \item B3 Platform
                    \item B3 Provider of AI
                \end{itemize}
        \end{itemize}
    \item B4 Problems of AI labels (overarching)
        \begin{itemize}
            \item B4 Big/complex problem, Standardization is hard
            \item B4 Mislabeling could be a problem
            \item B4 Overreliance
            \item B4 Power of Platform
        \end{itemize}
    \item C1 Known mechanisms
        \begin{itemize}
            \item C1 AI detection
            \item C1 Manual detection
            \item C1 Metadata
            \item C1 Other
            \item C1 Self-disclosure
            \item C1 Watermarks
        \end{itemize}
    \item C5 False positives/false negatives
        \begin{itemize}
            \item C5 Examples of false negatives
            \item C5 Examples of false positives
            \item C5 Evaluation
                \begin{itemize}
                    \item C5 Dependent on context/image
                    \item C5 Equally problematic
                    \item C5 No loss of trust
                    \item C5 Source/Credibility of website is important factor
                    \item C5 More problematic
                    \item C5 Not problematic
                    \item C5 Problematic
                    \item C5 Problematic in the long run
                    \item C5 Problematic in the short run
                \end{itemize}
            \item C5 Consequences
                \begin{itemize}                  
                    \item C5 Could damage reputation/ trustworthiness
                    \item C5 Disinformation / leads to questioning of facts
                    \item C5 Loss of trust if mislabeled images are obvious
                    \item C5 No consequences (sometimes)
                    \item C5 Users are getting disturbed/annoyed
                    \item C5 Users fall for scamming
                    \item C5 Enables deniability
                    \item C5 Loss of trust
                    \item C5 Loss of trust if it happens often
                \end{itemize}
            \item C5 Helper Codes: mislabeling
                \begin{itemize}
                    \item C5 Helper Codes: mislabeling: C5 false negative
                    \item C5 Helper Codes: mislabeling: C5 false positive
                    \item C5 Helper Codes: mislabeling: C5 general
                \end{itemize}
        \end{itemize}
    \item M Mechanisms
        \begin{itemize}
            \item MC HC Mechanism
                \begin{itemize}
                    \item MC HC Mechanism: HC: AI Detection
                    \item MC HC Mechanism: HC: All/Unspecified
                    \item MC HC Mechanism: HC: Metadata
                    \item MC HC Mechanism: HC: Other
                    \item MC HC Mechanism: HC: Self-Disclosure
                    \item MC HC Mechanism: HC: Watermarks
                \end{itemize}
            \item MC2 Reliability of mechanisms
                \begin{itemize}
                    \item MC2 Changes during time/ AI advances
                    \item MC2 Mechanism is reliable (for now)
                    \item\ C2 Mechanism is not reliable
                \end{itemize}
            \item MC3 Preference of mechanisms
                \begin{itemize}
                   \item MC3 Favorite Mechanisms
                    \item MC3 Combination of mechanisms is best
                    \item MC3 None
                \end{itemize}   
            \item MC4 Advantages of approach
                \begin{itemize}
                    \item MC4 Independent of user
                    \item MC4 Scalable
                    \item MC4 Easy
                    \item MC4 Independent of AI
                \end{itemize}
            \item MC42 Disadvantages of approach
                \begin{itemize}
                    \item MC42 Computing power
                    \item MC42 Lying/Misunderstanding
                    \item MC42 Manipulation/Removal
                    \item MC42 Tool Compliance
                    \item MC42 Usability
                    \item MC42 Interpretability
                    \item MC42 Other
                    \item MC42 Results dependent on training (model)
                \end{itemize}
        \end{itemize}
    \item Meta Codes
        \begin{itemize}
            \item Realization/Surprise
            \item Wish
            \item Interpretation
            \item Interesting quote
        \end{itemize}
\end{itemize}

\section{Survey}
\label{app:survey}

\subsection{Software}
\label{app:survey:software}
We used the statistical software R Version 4.4.2 for Mac \cite{R} for statistical analyses. For the \gls{glmm} and the \gls{lmm} analyses we used the \textit{lme4} package~\cite{lme4}. For the ANOVA we used the \textit{afex} \cite{afex} and \textit{emmeans} packages \cite{emmeans}.
We tested GLMM assumptions using the DHARMa package~\cite{dharma}.
We estimated covariate matrices using the \textit{sandwich} package~\cite{sandwich}.

\subsection{Prolific Description of ``Credibility of Social Media Posts''}
\label{app:survey:invitation}

We recruited participants using the following study description.

\begin{quote}
In this study we want to investigate news posts on social media. \\
The survey will show you 26 image-based news posts and will ask you to assess their credibility. Subsequently, you are asked a number of questions about the task. \\
Important: This survey is not suited for mobile devices. Please use a desktop/PC for this survey. \\

Study Disclaimer: With this study we want to investigate misinformation on social media platforms. Therefore, for example, some posts touch on news towards political figures and celebrities and include sensitive topics such as violence, catastrophes, or war. The content will not depict disturbing violence, injuries or vulnerable people. As this study contains misinformation, we will debrief after the study which content was fake or real. \\

The survey should take approximately 16 minutes to complete.

\end{quote}

\subsection{Study Information and Consent Form}
\label{app:survey:recruitment}

Below we provide the consent form participants signed.

\begin{itemize}
    \item \textbf{This study's purpose} is to produce a scientific publication using anonymized data from the information you provide.
    \item \textbf{Eligibility} is open to individuals (1.) over the age of 18 (2.) who are active on social media platforms and  (3.) are aware of the existence of AI-generated content.
    \item \textbf{All study data will be hosted in a secure cloud or on internal servers} accessible only by project members, except in the case of external transcription.
    \item \textbf{Anonymized data are kept for up to 10 years} in the spirit of good scientific practice, e.g., if questions about details arise later.
    \item We expect the survey to take up \textbf{roughly 16 minutes} of your time. \textbf{We offer a compensation of \pounds 2.86} for all participants completing the survey.
    \item \textbf{Your participation is voluntary.} You may stop participating at any time by closing the browser window or the program to withdraw from the survey. If you decide to withdraw your participation, we will not utilize your survey answers. You can also opt out of the study after completing the survey by contacting the researchers with your Prolific ID. We will then delete your responses from our dataset.
    \item The \textbf{risks to your participation} in this study involve viewing images or news that are artificial or of a sensitive nature (e.g., content related to politics or violence). The \textbf{benefits to you} are your compensation and the learning experience from participating in a research study. The \textbf{benefit to society} is the contribution to scientific knowledge.
    \item For any questions about this research, you may contact: 
      \begin{itemize}
        \item Sandra Höltervennhoff (Co-Project Lead, PhD student, \\ hoeltervennhoff@sec.uni-hannover.de) 
        \item Jonas Ricker (Co-Project Lead, PhD student, \\ jonas.ricker@rub.de) 
        \item Prof. Dr. rer. nat. Sascha Fahl \\ (Supervising Professor, fahl@cispa.de)
    \end{itemize}
\end{itemize}
By signing this consent form, I am affirming that...
\begin{itemize}
    \item I have read and understand the above information.
    \item I am 18+ and eligible to participate in this study.
    \item I am comfortable using the English language to participate in this study.
    \item I understand that I may stop participating at any time without penalty.
\end{itemize}

\subsection{Extended Demographics}
\label{app:survey_countries}

In Table~\ref{tab:survey_countries} we report the country of origin for our survey participants from the EU.

\begin{table}[htbp]
    \centering
    \small
    \caption{Country of origin for our 682 EU survey participants}
    \Description{A table depicting the country of origin for our 682 EU survey participants. Participants came from 20 different EU countries. The top three countries were Portugal (143 participants), Poland (134 participants), and Italy (111 participants).}
    \label{tab:survey_countries}
    
    \begin{tabular}{lrr}
        \toprule
        \textbf{Country} & \textbf{N} & \textbf{\%} \\
        \midrule
        Portugal & 143 & 21.0 \\
        Poland & 134 & 19.6 \\
        Italy & 111 & 16.3 \\
        Spain & 85 & 12.5 \\
        Greece & 38 & 5.6 \\
        Hungary & 31 & 4.5 \\
        France & 25 & 3.7 \\
        Netherlands & 19 & 2.8 \\
        Germany & 19 & 2.8 \\
        Czechia & 14 & 2.1 \\
        Finland & 10 & 1.5 \\
        Ireland & 10 & 1.5 \\
        Slovenia & 9 & 1.3 \\
        Sweden & 9 & 1.3 \\
        Latvia & 7 & 1.0 \\
        Austria & 5 & 0.7 \\
        Belgium & 5 & 0.7 \\
        Estonia & 4 & 0.6 \\
        Denmark & 3 & 0.4 \\
        Croatia & 1 & 0.1 \\
        \bottomrule
    \end{tabular}

\end{table}

\subsection{Questionnaire}
\label{app:survey:questionnaire}

Participants were given the following initial instructions:

\begin{quote}
Your task is to identify posts containing false claims that appeared on a social media platform. In the following, you are asked to rate the truthfulness of 26 posts. Each post consists of a short text and an image. The profile image and name of the post's author are anonymized.

\textbf{Control group:} From experience, you know that some posts on the platform contain AI-generated images. 

\textbf{Labeling/Mislabeling group:} The platform uses a system to add an ``AI-generated'' label if an image might be generated using AI.

Below are two examples of posts: [\textbf{control group:} both unlabeled, \textbf{labeling/mislabeling group:} one labeled]

You will only see the post at first, please take a look at it. Shortly after, you will see a question about an associated claim on the right. Please answer the question and indicate how confident you are.

Clicking on ``Next Page'' will start the survey.
\end{quote}

The following questions were asked for each of the 24 stimuli (see Appendix~\ref{app:survey:stimuli}), plus the two attention checks:
\begin{itemize}
    \item[\textbf{Q1.}] To the best of your knowledge, <question>? [yes, no]
    \item[\textbf{Q2.}] How confident are you in your assessment? [very unsure, unsure, sure, very sure]
\end{itemize}

The following questions were only given to participants in the labeling and mislabeling group.
\begin{itemize}
    \item[Instr.] You successfully completed the largest part of this survey! We are now interested in your perception of AI labels during the previous task.
    \item[\textbf{Q3.}] Did you have the impression that the AI labels influenced your decisions in the previous task? [not at all, very little, somewhat, to a great extent]
    \item[Instr.] The system that the platform uses to add AI labels might not always be 100\% correct. It can occur, that images are mislabeled. Mislabeling means that either, an AI-generated image is wrongfully displayed without the ``AI-generated'' label, or a human-made image is wrongfully displayed with the ``AI-generated'' label.
    \item[\textbf{Q4.}] Did you have the impression that, in the previous task, some AI-generated images were not labeled as such? [yes, no, unsure]
    \item[\textbf{Q5.}] Did you have the impression that, in the previous task, some human-made images were wrongfully labeled as ``AI-generated''? [yes, no, unsure]
    \item[Instr.] We are now interested in your opinion of mislabeling in general.
    \item[\textbf{Q6.}] Regarding unlabeled AI-generated images, how much do you agree with the following claims? [strongly disagree, disagree, neither agree nor disagree, agree, strongly agree]
    \begin{itemize}
        \item[\textbf{Q6a.}] Users lose trust in the labeling system if they become aware of such mislabeling.
        \item[\textbf{Q6b.}] It is not a problem if such mislabeling only happens once in a while.
    \end{itemize}
    \item[\textbf{Q7.}] Regarding wrongfully labeled human-made images, how much to you agree with the following claims? [strongly disagree, disagree, neither agree nor disagree, agree, strongly agree]
    \begin{itemize}
        \item[\textbf{Q7a.}] Users lose trust in the labeling system if they become aware of such mislabeling.
        \item[\textbf{Q7b.}] It is not a problem if such mislabeling only happens once in a while.
    \end{itemize}
    \item[\textbf{Q8.}] Regarding the two types of mislabeling that can occur, which one do you consider worse? [mislabeled (unlabeled) AI-generated images, wrongfully labeled human-made images, they are equally bad, none of them is a problem]
    \item[\textbf{Q9.}] Would you like to see AI labels (as they were presented in this study) on real-world social media platforms? [definitely no, rather no, neither yes nor no, rather yes, definitely yes]
\end{itemize}

The last questions were again given to participants in all groups.
\begin{itemize}
    \item[\textbf{Q10.}] Did you use any tools to rate the truthfulness of the 26 posts, e.g., a search engine or an AI? (Important: There is no right or wrong answer. Your answer does not influence your survey approval.) [yes, no, prefer not to say]
    \item[\textbf{D1.}] What is your gender? [Woman, Man, Non-binary, Prefer to self-describe, Prefer not to say]
    \item[\textbf{D2.}] What is your age? [18--24, 25--34, 35--44, 45--54, 55--64, 65+, Prefer not to say]
    \item[\textbf{D3.}] What is your country of residence? [Dropdown with 249 countries, Other, Prefer not so say]
    \item[\textbf{D4.}] Which of the following best describes the highest level of formal education that you have completed? [I never completed any formal education, 10th grade or less (e.g. some American high school credit, German Realschule, British GCSE), Secondary school (e.g. American high school, German Realschule or Gymnasium, Spanish or French Baccalaureate, British A-Levels), Trade, technical or vocational training, Some college/university study without earning a degree, Associate degree (A.A., A.S., etc.), Bachelor’s degree (B.A., B.S., B.Eng., etc.), Master’s degree (M.A., M.S., M.Eng., MBA, etc.), Professional degree (JD, MD, etc.), Other doctoral degrees (Ph.D., Ed.D., etc.), Other, Prefer not to say]
    \item[\textbf{D5.}] How would you describe your political views? [Very right, Right leaning, Center, Left leaning, Very left, Not interested in politics, Prefer not to say]
\end{itemize}

\subsection{Layout}
\begin{figure}[h]
    \centering
    \frame{\includegraphics[width=0.9\linewidth]{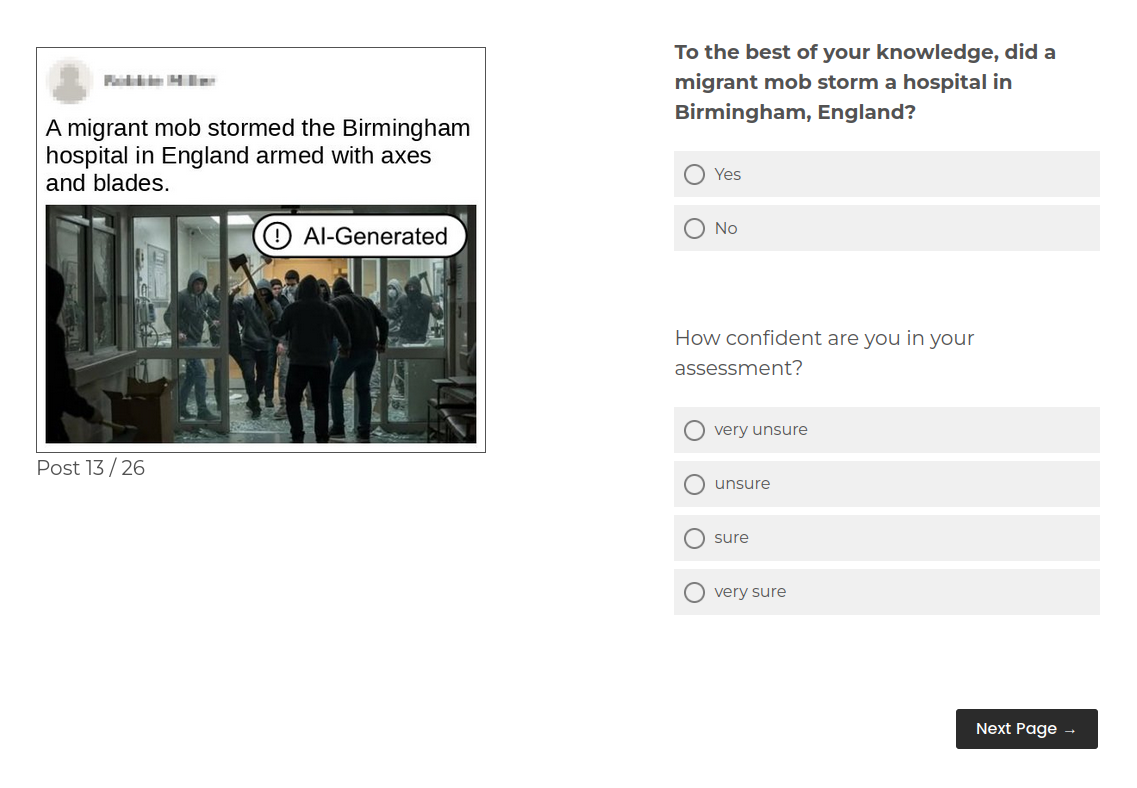}}
    \Description{The screenshot of the main experiment contains, on the left, the simulated social media post with the claim, the photo and the AI label. It is captioned ``Post 13/26''. On the right are two sets of questions and a button to access the next page. The first question is, ``To the best of your knowledge, did a migrant mob storm a hospital in Birmingham, England?'' Answer options are ``Yes'' and ``No''. The second question is ``How confident are you in your assessment`?'' with answer options ``very unsure'', ``unsure'', ``sure'' and ``very sure''.}
    \caption{Screenshot of our main experiment including the stimuli-claim combination and the corresponding questions.}
    \label{fig:survey_screenshot}
\end{figure}

In Figure~\ref{fig:survey_screenshot} we provide a screenshot of our main experiment layout, which we implemented using Qualtrics.

\subsection{Stimuli}
\label{app:survey:stimuli}
\begin{figure*}[ht]
    \centering
    \begin{subfigure}[t]{\maxstimuliwidth}
        \centering
        \frame{\includegraphics[width=\maxstimuliwidth,height=\maxstimuliheight,keepaspectratio]{assets/figures/stimuli/ballots_fire.jpeg}}
        \Description{A simulated social media post captioned with ``A member of the law enforcement saving ballots from arson in Vancouver, Washington''. The associated photo shows a person reaching into a ballot drop box amid fire and smoke.}
        \subcaption{Human/True\_1. Image courtesy of Associated Press (AP).}
    \end{subfigure}
    \hspace{30pt}
    \begin{subfigure}[t]{\maxstimuliwidth}
        \centering
        \frame{\includegraphics[width=\maxstimuliwidth,height=\maxstimuliheight,keepaspectratio]{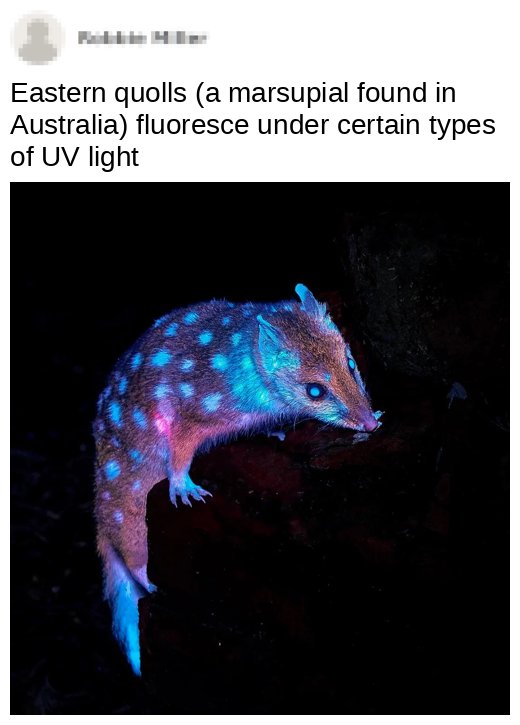}}
        \Description{A simulated social media post captioned with ``Eastern quolls (a marsupial found in Australia) fluoresce under certain types of UV light''. The associated photo shows a a bluish glowing rodent.}
        \subcaption{Human/True\_2. Image courtesy of Benjamin Alldridge.}
    \end{subfigure}
    \hspace{30pt}
    \begin{subfigure}[t]{\maxstimuliwidth}
        \centering
        \frame{\includegraphics[width=\maxstimuliwidth,height=\maxstimuliheight,keepaspectratio]{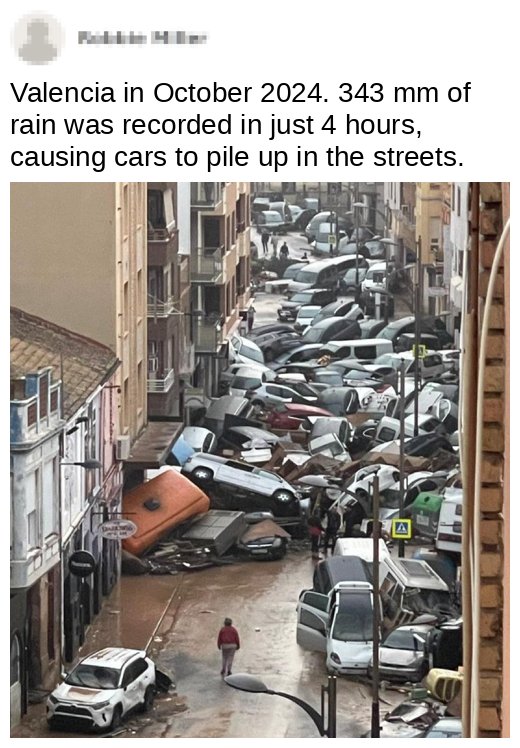}}
        \Description{A simulated social media post captioned with ``Valencia in October 2024. 343 mm of rain was recorded in just 4 hours, causing cars to pile up in the streets''. The associated photo shows a large pile of cars wedged together and pushed on top of each other.}
        \subcaption{Human/True\_3. Image courtesy of Agencia EFE.}
    \end{subfigure}
    \par\medskip
    \begin{subfigure}[t]{\maxstimuliwidth}
        \centering
        \frame{\includegraphics[width=\maxstimuliwidth,height=\maxstimuliheight,keepaspectratio]{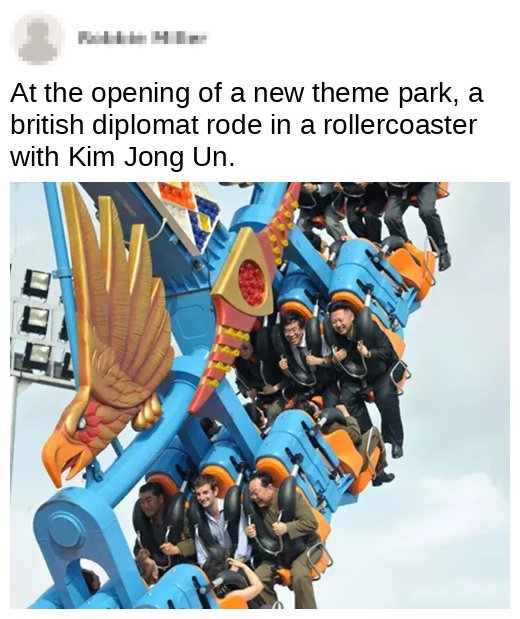}}
        \Description{A simulated social media post captioned with ``At the opening of a new theme park, a British diplomat rode in a rollercoaster with Kim Jong Un''. The associated photo shows him and other people hanging in the air in an amusement park.}
        \subcaption{Human/True\_4. Image courtesy of Korean Central News Agency.}
    \end{subfigure}
    \hspace{30pt}
    \begin{subfigure}[t]{\maxstimuliwidth}
        \centering
        \frame{\includegraphics[width=\maxstimuliwidth,height=\maxstimuliheight,keepaspectratio]{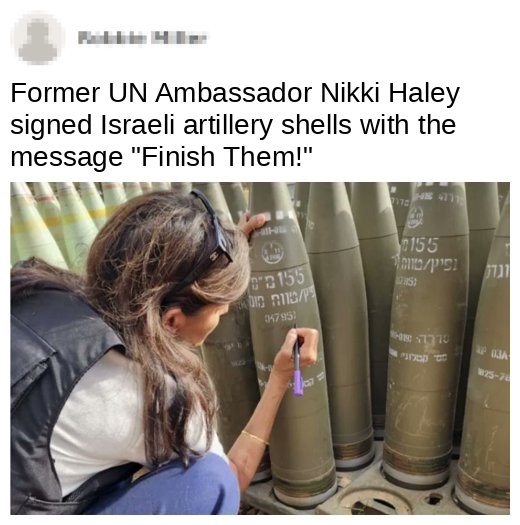}}
        \Description{A simulated social media post captioned with ``Former UN Ambassador Nikki Haley signed Israeli artillery shells with the message ``Finish Them!''''. The associated photo shows a woman writing something on a grenade.}
        \subcaption{Human/True\_5. Image courtesy of Instagram.}
    \end{subfigure}
    \hspace{30pt}
    \begin{subfigure}[t]{\maxstimuliwidth}
        \centering
        \frame{\includegraphics[width=\maxstimuliwidth,height=\maxstimuliheight,keepaspectratio]{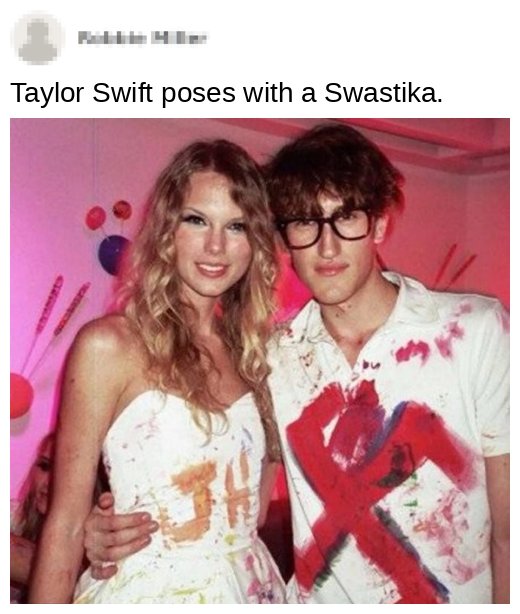}} 
        \Description{A simulated social media post captioned with ``Taylor Swift poses with Swastika''. The associated photo shows the two of them posing arm in arm and looking at the camera.}
        \subcaption{Human/True\_6. Image courtesy of TMZ.}
    \end{subfigure}
    \Description{Six stimuli and claim combinations (a)-(f) for human-made images and true claims. Each consists of a photo and an associated caption, simulating a social media post.}
    \caption{Stimuli with human-made images and true claims.}
    \label{fig:stimuli_human_true}
\end{figure*}

\begin{figure*}[ht]
    \centering
    \begin{subfigure}[t]{\maxstimuliwidth}
        \centering
        \frame{\includegraphics[width=\maxstimuliwidth,height=\maxstimuliheight,keepaspectratio]{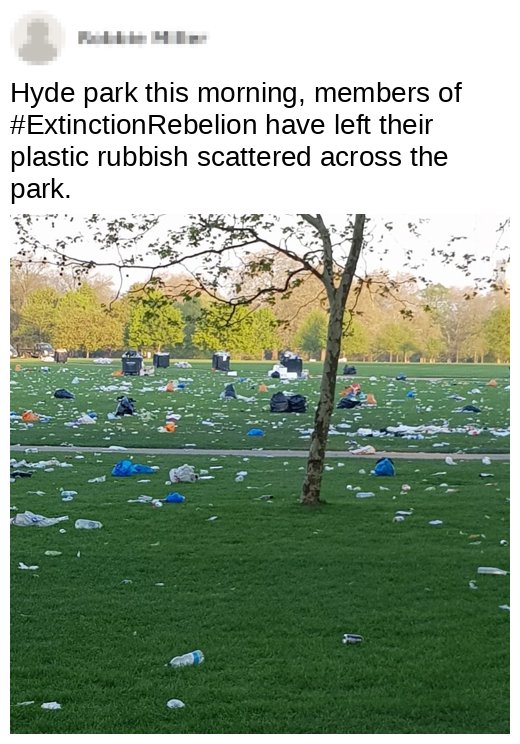}}
        \Description{A simulated social media post captioned with ``Hyde park this morning, members of \#ExtinctionRebelion have left their plastic rubbish scattered across the park''. The associated photo shows a park littered with trash. }
        \subcaption{Human/False\_1. Image courtesy of Facebook.}
    \end{subfigure}
    \hspace{30pt}
    \begin{subfigure}[t]{\maxstimuliwidth}
        \centering
        \frame{\includegraphics[width=\maxstimuliwidth,height=\maxstimuliheight,keepaspectratio]{assets/figures/stimuli/chernobyl_species.jpeg}}
        \Description{A simulated social media post captioned with ``There is a new animal species taking over at Chernobyl.''. The associated photo shows a misshapen creature with leathery gray skin and a fur-covered face reminiscent of a bear.}
        \subcaption{Human/False\_2. Image courtesy of Sebastian Willnow/DDP/AFP via Getty Images.}
    \end{subfigure}
    \hspace{30pt}
    \begin{subfigure}[t]{\maxstimuliwidth}
        \centering
        \frame{\includegraphics[width=\maxstimuliwidth,height=\maxstimuliheight,keepaspectratio]{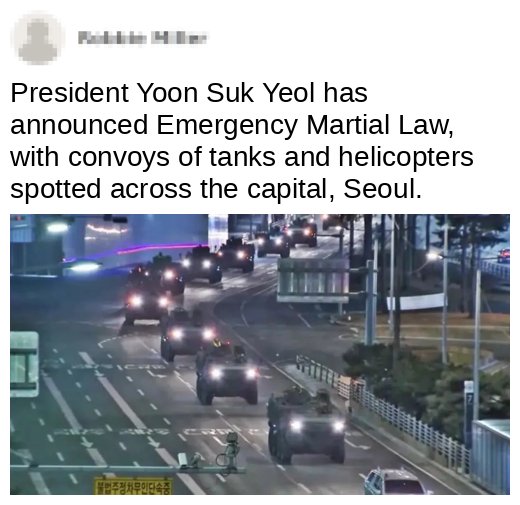}}
        \Description{A simulated social media post captioned with ``President Yoon Suk Yeol has announced Emergency Martial Law, with convoys of tanks and helicopters spotted across the capital, Seoul''. The associated photo shows a convoy of military vehicles.}
        \subcaption{Human/False\_3. Image courtesy of South Corean Defense Public Relations Agency.}
    \end{subfigure}
    \par\medskip
    \begin{subfigure}[t]{\maxstimuliwidth}
        \centering
        \frame{\includegraphics[width=\maxstimuliwidth,height=\maxstimuliheight,keepaspectratio]{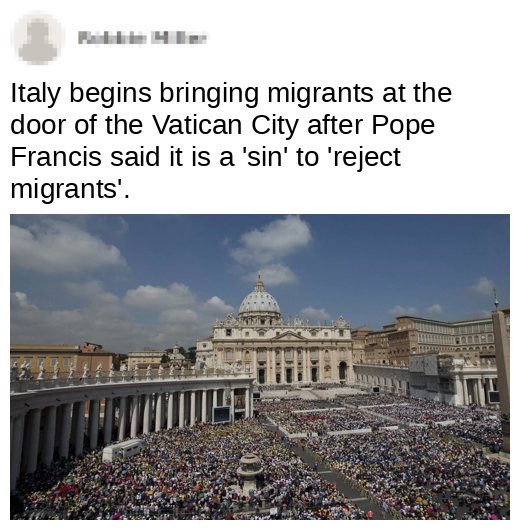}}
        \Description{A simulated social media post captioned with ``Italy begins bringing migrants to the door of the Vatican City after Pope Francis said it is a 'sin' to 'reject migrants'''. The associated photo shows St. Peter's Square filled with crowds of people.}
        \subcaption{Human/False\_4. Image courtesy of Associated Press (AP).}
    \end{subfigure}
    \hspace{30pt}
    \begin{subfigure}[t]{\maxstimuliwidth}
        \centering
        \frame{\includegraphics[width=\maxstimuliwidth,height=\maxstimuliheight,keepaspectratio]{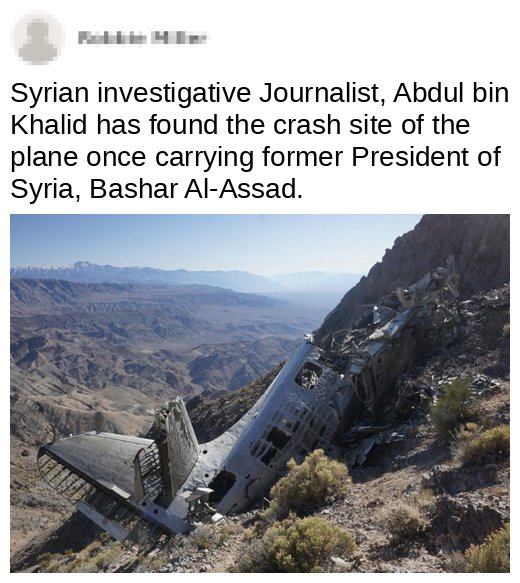}}
        \Description{A simulated social media post captioned with ``Syrian investigative Journalist, Abdul bin Khalid has found the crash site of the plane once carrying former president of Syria, Bashar Al-Assad''. The associated photo shows parts of a badly damaged, presumably crashed aircraft in the hills.}
        \subcaption{Human/False\_5. Image courtesy of Caters News Agency Ltd.}
    \end{subfigure}
    \hspace{30pt}
    \begin{subfigure}[t]{\maxstimuliwidth}
        \centering
        \frame{\includegraphics[width=\maxstimuliwidth,height=\maxstimuliheight,keepaspectratio]{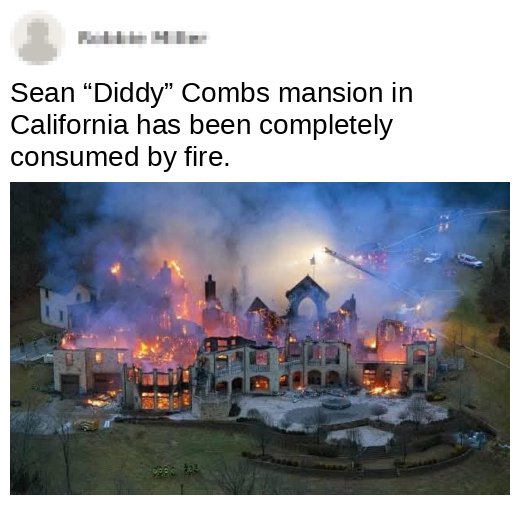}} 
        \Description{A simulated social media post captioned with ``Sean ``Diddy'' Combs mansion in California has been completely consumed by fire''. The associated photo shows a mansion on fire.}
        \subcaption{Human/False\_6. Image courtesy of The Enquirer/Joseph Fuqua II.}
    \end{subfigure}
    \Description{Six stimuli and claim combinations (a)-(f) for human-made images and false claims. Each consists of a photo and an associated caption, simulating a social media post.}
    \caption{Stimuli with human-made images and false claims.}
    \label{fig:stimuli_human_false}
\end{figure*}

\begin{figure*}[ht]
    \centering
    \begin{subfigure}[t]{\maxstimuliwidth}
        \centering
        \frame{\includegraphics[width=\maxstimuliwidth,height=\maxstimuliheight,keepaspectratio]{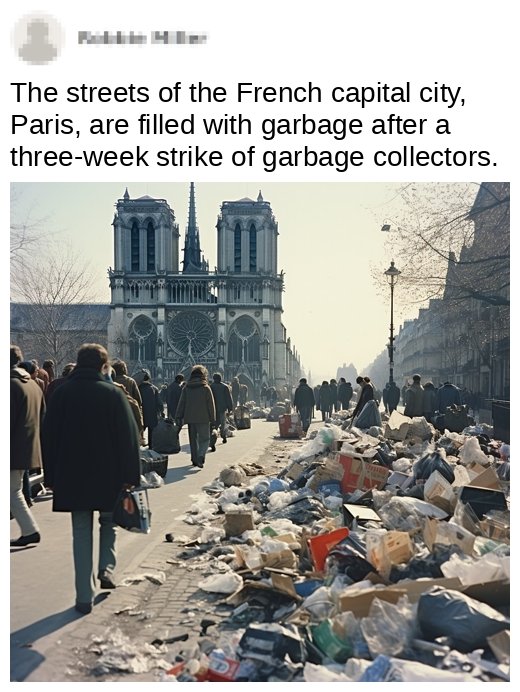}}
        \Description{A simulated social media post captioned with ``The streets of the French capital city, Paris, are filled with garbage after a three-week strike of garbage collectors''. The associated photo shows a street with large piles of trash.}
        \subcaption{AI/True\_1.}
    \end{subfigure}
    \hspace{30pt}
    \begin{subfigure}[t]{\maxstimuliwidth}
        \centering
        \frame{\includegraphics[width=\maxstimuliwidth,height=\maxstimuliheight,keepaspectratio]{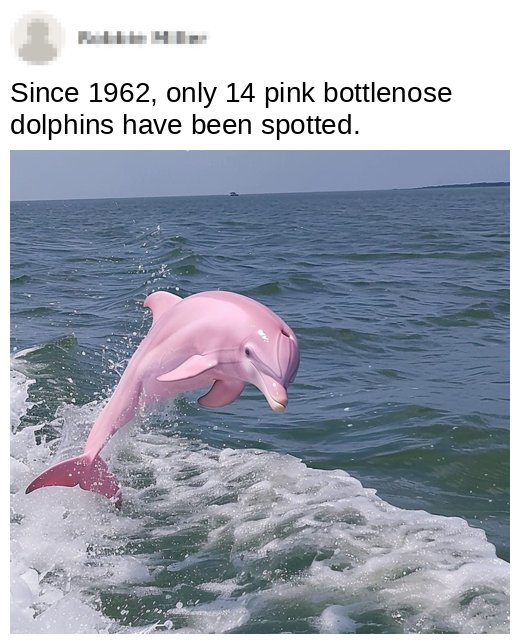}}
        \Description{A simulated social media post captioned with ``Since 1962, only 14 pink bottlenose dolphins have been spotted''. The associated photo shows a bright pink dolphin jumping out of the sea.}
        \subcaption{AI/True\_2.}
    \end{subfigure}
    \hspace{30pt}
    \begin{subfigure}[t]{\maxstimuliwidth}
        \centering
        \frame{\includegraphics[width=\maxstimuliwidth,height=\maxstimuliheight,keepaspectratio]{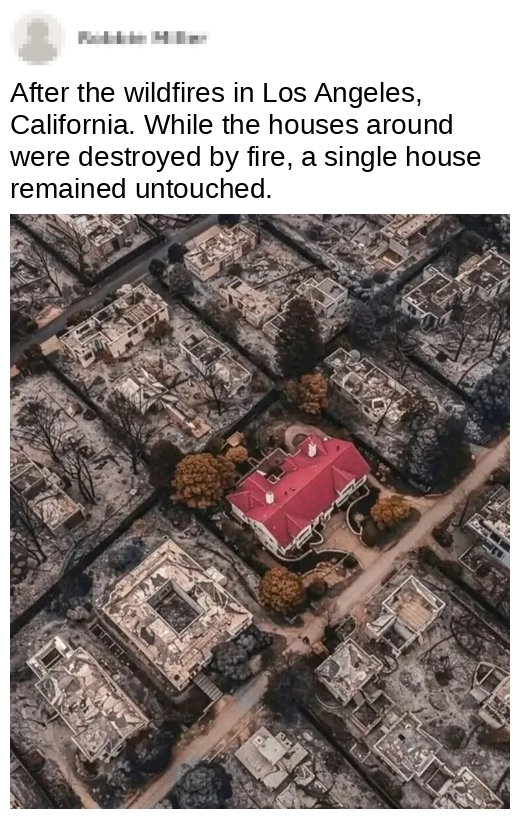}}
        \Description{A simulated social media post captioned with ``After the wildfires in Los Angeles, California. While the houses around were destroyed by fire, a single house remained untouched''. The associated photo shows an aerial view of a suburb. All properties are devastated and covered in ash, only one house remains undamaged with its garden intact.}
        \subcaption{AI/True\_3.}
    \end{subfigure}
    \par\medskip
    \begin{subfigure}[t]{\maxstimuliwidth}
        \centering
        \frame{\includegraphics[width=\maxstimuliwidth,height=\maxstimuliheight,keepaspectratio]{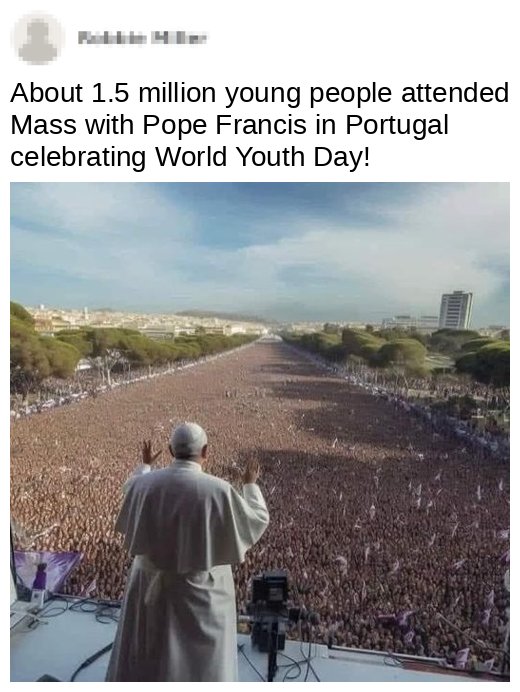}}
        \Description{A simulated social media post captioned with ``About 1.5 million young people attended Mass with Pope Francis in Portugal celebrating World Youth Day!'' The associated photo shows the back view of a pope with a huge crowd in the background.}
        \subcaption{AI/True\_4.}
    \end{subfigure}
    \hspace{30pt}
    \begin{subfigure}[t]{\maxstimuliwidth}
        \centering
        \frame{\includegraphics[width=\maxstimuliwidth,height=\maxstimuliheight,keepaspectratio]{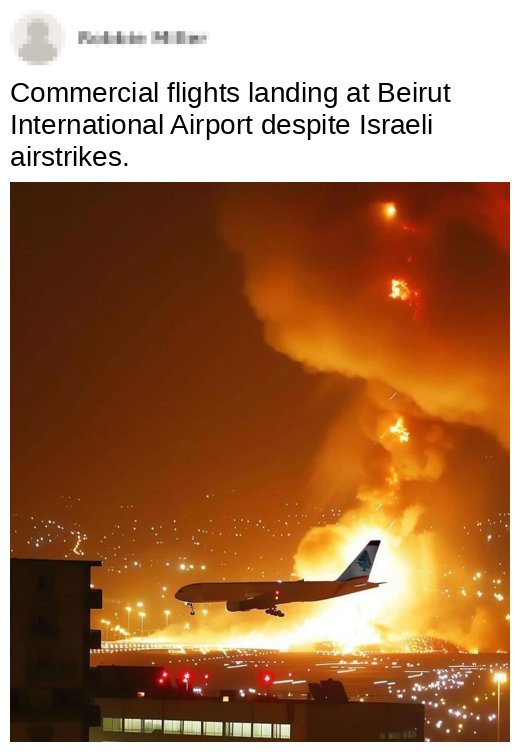}}
        \Description{A simulated social media post captioned with ``Commercial flights landing at Beirut International Airport despite Israeli airstrikes''. The associated photo shows an airplane against a backdrop of destruction with columns of flame rising up to the sky.}
        \subcaption{AI/True\_5.}
    \end{subfigure}
    \hspace{30pt}
    \begin{subfigure}[t]{\maxstimuliwidth}
        \centering
        \frame{\includegraphics[width=\maxstimuliwidth,height=\maxstimuliheight,keepaspectratio]{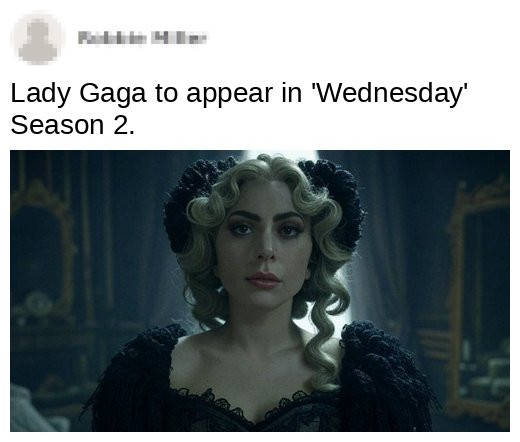}} 
        \Description{A simulated social media post captioned with ``Lady Gaga to appear in 'Wednesday' Season 2.''. The associated photo shows Lady Gaga in a black robe looking at the camera.}
        \subcaption{AI/True\_6.}
    \end{subfigure}
    \Description{Six stimuli and claim combinations (a)-(f) for AI-generated images and true claims. Each consists of a photo and an associated caption, simulating a social media post.}
    \caption{Stimuli with AI-generated images and true claims.}
    \label{fig:stimuli_ai_true}
\end{figure*}

\begin{figure*}[ht]
    \centering
    \begin{subfigure}[t]{\maxstimuliwidth}
        \centering
        \frame{\includegraphics[width=\maxstimuliwidth,height=\maxstimuliheight,keepaspectratio]{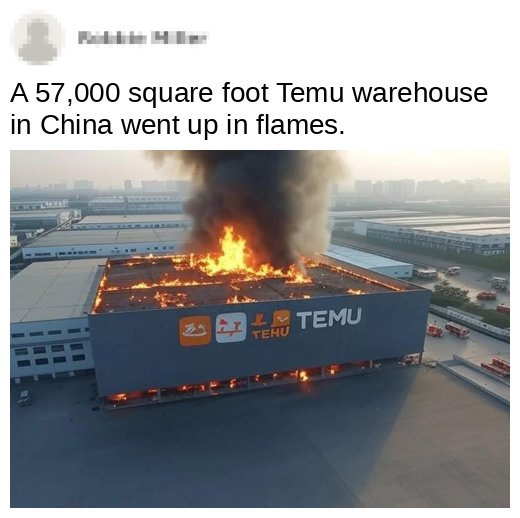}}
        \Description{A simulated social media post captioned with ``A 57,000 square foot Temu warehouse in China went up in flames''. The associated photo shows a Temu building with flames licking out of its roof.}
        \subcaption{AI/False\_1.}
    \end{subfigure}
    \hspace{30pt}
    \begin{subfigure}[t]{\maxstimuliwidth}
        \centering
        \frame{\includegraphics[width=\maxstimuliwidth,height=\maxstimuliheight,keepaspectratio]{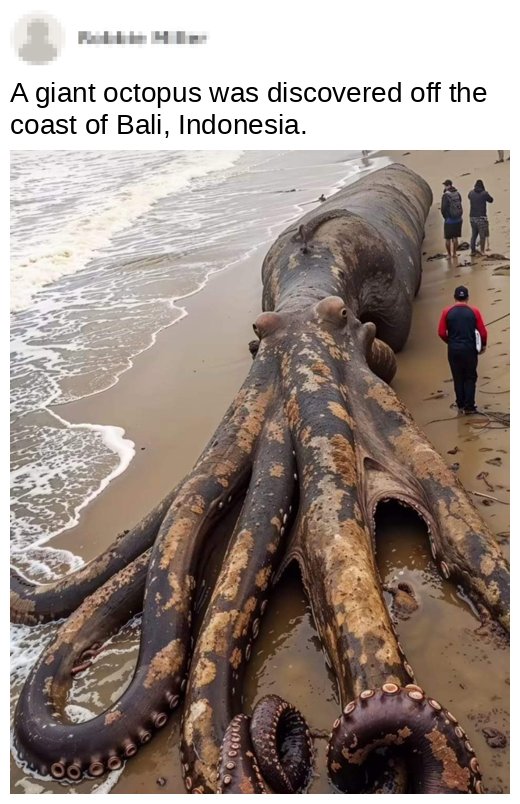}}
        \Description{A simulated social media post captioned with ``A giant octopus was discovered off the coast of Bali, Indonesia''. The associated photo shows an octopus over 20 meters long lying on the beach, with people walking past it.}
        \subcaption{AI/False\_2.}
    \end{subfigure}
    \hspace{30pt}
    \begin{subfigure}[t]{\maxstimuliwidth}
        \centering
        \frame{\includegraphics[width=\maxstimuliwidth,height=\maxstimuliheight,keepaspectratio]{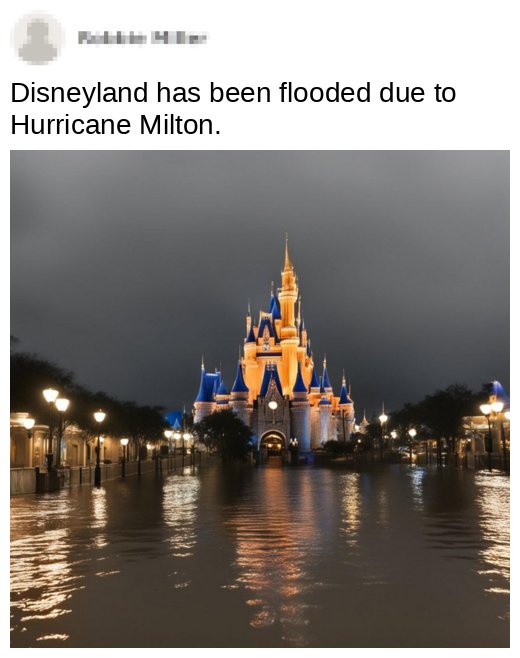}}
        \Description{A simulated social media post captioned with ``Disneyland has been flooded due to Hurricane Milton''. The associated photo shows a lake from which the Disney building rises.}
        \subcaption{AI/False\_3.}
    \end{subfigure}
    \par\medskip
    \begin{subfigure}[t]{\maxstimuliwidth}
        \centering
        \frame{\includegraphics[width=\maxstimuliwidth,height=\maxstimuliheight,keepaspectratio]{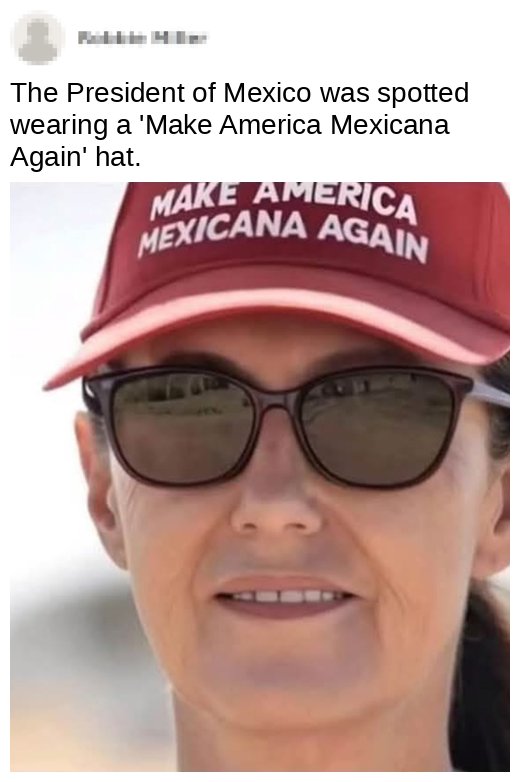}}
        \Description{A simulated social media post captioned with ``The President of Mexico was spotted wearing a 'Make America Mexicana Again' hat''. The associated photo shows her wearing sun glasses and the named hat.}
        \subcaption{AI/False\_4.}
    \end{subfigure}
    \hspace{30pt}
    \begin{subfigure}[t]{\maxstimuliwidth}
        \centering
        \frame{\includegraphics[width=\maxstimuliwidth,height=\maxstimuliheight,keepaspectratio]{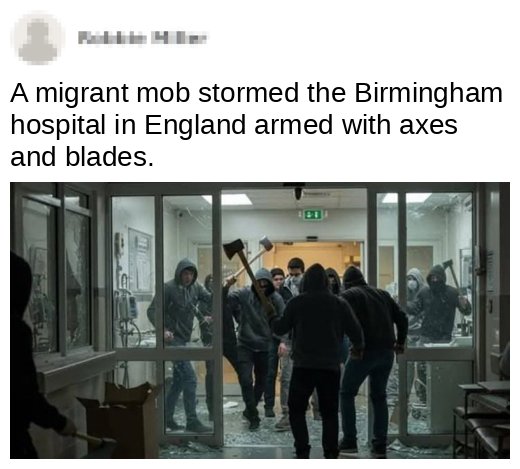}}
        \Description{A simulated social media post captioned with ``A migrant mob stormed the Birmingham hospital in England armed with axes and blades.''. The associated photo shows people in hoodies with masks and axes, smashing glass doors and storming a hallway.}
        \subcaption{AI/False\_5.}
    \end{subfigure}
    \hspace{30pt}
    \begin{subfigure}[t]{\maxstimuliwidth}
        \centering
        \frame{\includegraphics[width=\maxstimuliwidth,height=\maxstimuliheight,keepaspectratio]{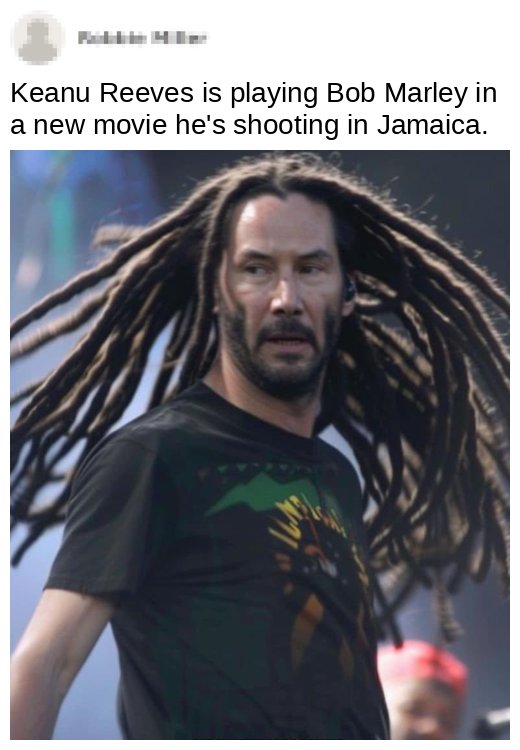}} 
        \Description{A simulated social media post captioned with ``Keanu Reeves is playing Bob Marley in a new movie he's shooting in Jamaica''. The associated photo shows him with chest-length dreadlocks in an action pose.}
        \subcaption{AI/False\_6.}
    \end{subfigure}
    \Description{Six stimuli and claim combinations (a)-(f) for AI-generated images and false claims. Each consists of a photo and an associated caption, simulating a social media post.}
    \caption{Stimuli with AI-generated images and false claims.}
    \label{fig:stimuli_ai_false}
\end{figure*}

\begin{figure*}[ht]
    \centering
    \begin{subfigure}[t]{\maxstimuliwidth}
        \centering
        \frame{\includegraphics[width=\maxstimuliwidth,height=\maxstimuliheight,keepaspectratio]{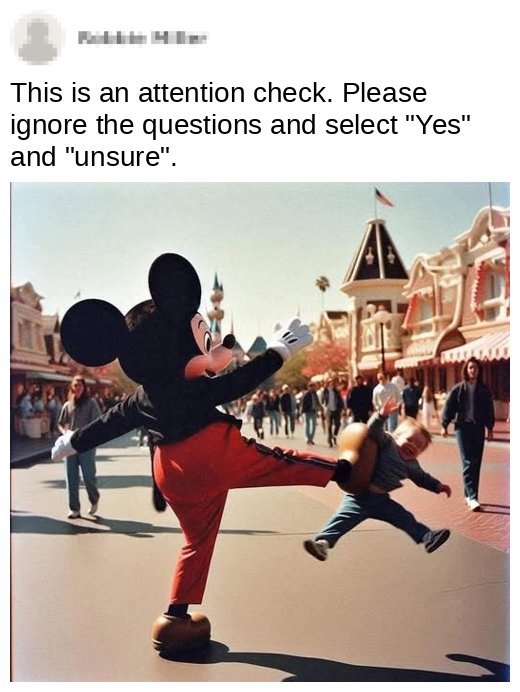}}
        \Description{An attention check looking like the simulated social media post. The caption says ``This is an attention check. Please ignore the question and select ``Yes'' and ``unsure''''. The associated photo shows an actor dressed as Mickey Mouse in a theme park kicking a child away.}
        \subcaption{First attention check, shown after the tenth post.}
    \end{subfigure}
    \hspace{30pt}
    \begin{subfigure}[t]{\maxstimuliwidth}
        \centering
        \frame{\includegraphics[width=\maxstimuliwidth,height=\maxstimuliheight,keepaspectratio]{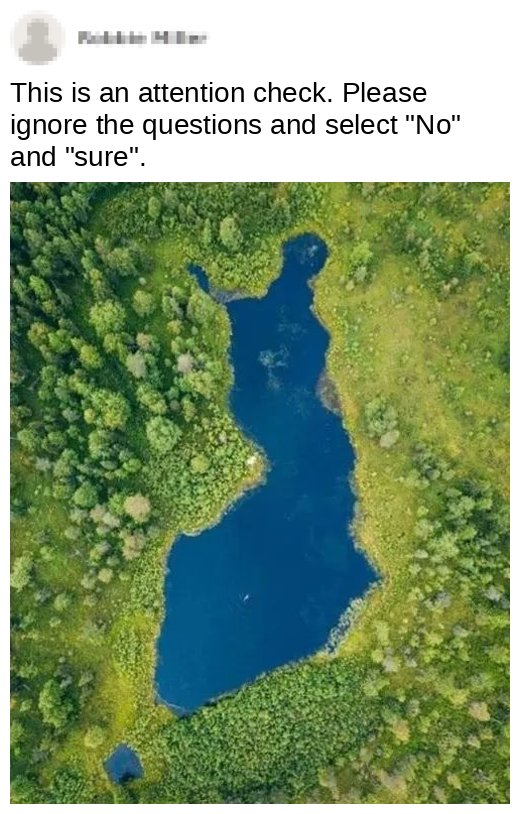}}
        \Description{An attention check looking like the simulated social media post. The caption says ``This is an attention check. Please ignore the question and select ``No'' and ``sure''''. The associated photo shows a lake from a bird's eye view.}
        \subcaption{Second attention check, shown after the 20th post. Image courtesy of Simo Räsänen via Getty Images.}
    \end{subfigure}
    \Description{Two attention checks (a)-(b). Each consists of a photo and an associated caption. This makes them look like simulated social media posts, although in this case the text does not match the image shown.}
    \caption{Attention checks.}
    \label{fig:stimuli:attention_checks}
\end{figure*}

We provide the stimuli used in our survey, separated by subset (human/true, human/false, AI/true, and AI/false) in Figures~\ref{fig:stimuli_human_true} to \ref{fig:stimuli_ai_false}. 
Table~\ref{tab:stimuli} lists the original and adjusted captions as well as the corresponding questions. We also provide links to fact-checking articles. 
Figure~\ref{fig:stimuli:attention_checks} depicts our attention checks.

\begin{table*}
    \centering
    \caption{Overview of our stimuli's original caption, modified caption (used in our survey), corresponding question, and the link to a fact check. Note that in our survey, each question was introduced by ``To the best of your knowledge, ...?''}
    \Description{A table listing all the social media posts that we used for our survey. The table includes the original caption of the post, the modified caption for the survey, the question we asked about each caption, and a link to a fact checking side for the post. The social media posts are divided into their four categories. First all human/true posts are listed, then human/false, AI/true, AI/false, and lastly the two posts that were used for the attention checks.}
    \label{tab:stimuli}
    \tiny
    \begin{tabular}{@{}p{37pt}p{140pt}p{100pt}p{90pt}p{100pt}@{}}
        \toprule
         \textbf{ID} & \textbf{Original Caption} & \textbf{Modified Caption} & \textbf{Question} & \textbf{Fact Check URL} \\
         \midrule
         Human\_True\_1 & Saving ballots from arson in Vancouver, Washington this morning & A member of the law enforcement saving ballots from arson in Vancouver, Washington & Was there an attempt to save ballots from arson in Vancouver, Washington? & \href{https://www.snopes.com/fact-check/ballots-saved-fire-vancouver-washington/}{https://www.snopes.com/fact-check/ballots-saved-fire-vancouver-washington/} \\
         Human\_True\_2 & Eastern quolls (Dasyurus viverrinus) fluoresce under certain types of UV light	& Eastern quolls (a marsupial found in Australia) fluoresce under certain types of UV light & Do eastern quolls glow under certain types of UV light?& \href{https://www.snopes.com/fact-check/fluorescent-marsupial/}{https://www.snopes.com/fact-check/fluorescent-marsupial/} \\
         Human\_True\_3 & Valencia this morning. A jaw-dropping 343 mm of rain was recorded in just 4 hours yesterday, between 4:30pm and 8:30pm. & Valencia in October 2024. 343 mm of rain was recorded in just 4 hours, causing cars to pile up in the streets. & Did cars pile up in the streets due to heavy rainfall in Valencia? & \href{https://www.snopes.com/fact-check/valencia-spain-flooding-photo/}{https://www.snopes.com/fact-check/valencia-spain-flooding-photo/} \\
         Human\_True\_4 & Pyongyang’s diplomatic community was invited to the opening of the Rungna People’s Pleasure Ground. This included the Chargé d’Affaires of the British Embassy, who accepted the invitation to attend. & At the opening of a new theme park, a british diplomat rode in a rollercoaster with Kim Jong Un. & Did Kim Jong Un ride a rollercoaster with a British diplomat? & \url{https://www.motherjones.com/politics/2012/08/kim-jong-un-amusement-park-photo-british-diplomat/} \\
         Human\_True\_5 & Former UN Ambassador Nikki Haley signed Israeli artillery shells with the message "Finish Them!" Conflict is no place for stunts. Conflict has rules. Civilians must be protected. & Former UN Ambassador Nikki Haley signed Israeli artillery shells with the message "Finish Them!" & Did former UN Ambassador Nikki Haley sign Israeli artillery shells with the message "Finish Them!" & \href{https://www.snopes.com/fact-check/haley-bombs-finish-them/}{https://www.snopes.com/fact-check/haley-bombs-finish-them/} \\
         Human\_True\_6 & If you're getting annoyed by a Taylor Swift fan just upload this photo and watch them vanish. & Taylor Swift poses with a Swastika. & Did Taylor Swift pose with a man in a swastika-adorned shirt? & \href{https://www.snopes.com/fact-check/taylor-swift-swastika-picture/}{https://www.snopes.com/fact-check/taylor-swift-swastika-picture/} \\
         \midrule
         Human\_False\_1 & Hyde park this morning, the eco worriers \#ExtinctionRebelion have left their plastic rubbish scattered across the park. so much care and concern for the earth is quite touching really!! & Hyde park this morning, members of \#ExtinctionRebelion have left their plastic rubbish scattered across the park. & Did members of Extinction Rebellion left Hyde Park full of garbage? & \url{https://www.snopes.com/fact-check/protesters-hyde-park-rubbish/} \\
         Human\_False\_2 & There Are a New Animal Species Taking Over at Chernobyl.	& There is a new animal species taking over at Chernobyl. & Has a previously unknown species been seen in Chernobyl? & \url{https://www.snopes.com/fact-check/chernobyl-animals/} \\
         Human\_False\_3 & MARTIAL LAW DECLARED IN SOUTH KOREA. President Yoon Suk Yeol has announced Emergency Martial Law, with convoys of tanks and helicopters spotted across the capital, Seoul. & President Yoon Suk Yeol has announced Emergency Martial Law, with convoys of tanks and helicopters spotted across the capital, Seoul. & Did military convoys drive through Seoul after martial law was declared? & \href{https://www.snopes.com/fact-check/martial-law-south-korea-photo/}{https://www.snopes.com/fact-check/martial-law-south-korea-photo/} \\
         Human\_False\_4 & JUST IN: Italy begins dumping migrants at the door of the Vatican City after Pope Francis said it is a 'sin' to 'reject migrants'. & Italy begins bringing migrants at the door of the Vatican City after Pope Francis said it is a 'sin' to 'reject migrants'. & Did Italy bring migrants to Vatican city? & \href{https://www.reuters.com/fact-check/italy-did-not-transfer-crowd-migrants-vatican-november-2024-12-11/}{https://www.reuters.com/fact-check/italy-did-not-transfer-crowd-migrants-vatican-november-2024-12-11/} \\
         Human\_False\_5 & Syrian investigative Journalist, Abdul bin Khalid has found the crash site of the plane once carrying former President of Syria, Bashar Al-Assad. & Syrian investigative Journalist, Abdul bin Khalid has found the crash site of the plane once carrying former President of Syria, Bashar Al-Assad. & Did Bashar Al-Assad crash with a plane?& \href{https://www.dw.com/en/fact-check-fakes-surrounding-assads-escape-to-moscow/a-71016174}{https://www.dw.com/en/fact-check-fakes-surrounding-assads-escape-to-moscow/a-71016174} \\
         Human\_False\_6 & P Diddy's mansion in California has been completely consumed by fire. & Sean “Diddy” Combs mansion in California has been completely consumed by fire. & Has Sean “Diddy” Combs mansion in California been consumed by fire? & \href{https://www.reuters.com/fact-check/photo-2014-fire-mislabeled-combs-la-mansion-2025-2025-02-06/}{https://www.reuters.com/fact-check/photo-2014-fire-mislabeled-combs-la-mansion-2025-2025-02-06/} \\
         \midrule
         AI\_True\_1 & This is what the French capital city, Paris, looks like. The dream city... now turned into this in reality & The streets of the French capital city, Paris, are filled with garbage after a three-week strike of garbage collectors. & Did Parisian garbage collectors went on strike, causing uncollected garbage littering the streets? & \url{https://factcheck.afp.com/doc.afp.com.33QV2QL} \\
         AI\_True\_2 & Rare pink Dolphin spotted in Bohol & Since 1962, only 14 pink bottlenose dolphins have been spotted. & Is there a species of dolphins that is pink? & \url{https://factcheck.afp.com/doc.afp.com.34ZE9BT} \\
         AI\_True\_3 & This is the home of a Christian in Los Angeles, California. While the houses around him were destroyed by fire, his house remained untouched. God's promise in Psalm 91:1-6 was fulfilled. You can't imagine how much he cried for joy, knowing he was protected by God. Truly, God is his refuge. & After the wildfires in Los Angeles, California. While the houses around were destroyed by fire, a single house remained untouched. & Did a single house remain untouched while the houses around it were destroyed during the LA wildfires? & \url{https://www.snopes.com/news/2025/01/14/la-fires-home-god-saved/} \\
         AI\_True\_4 & LOOK: Picture of about two million young people that attended Mass with Pope Francis in Lisbon! I'm Catholic For Life! \#WorldYouthDay2023 & About 1.5 million young people attended Mass with Pope Francis in Portugal celebrating World Youth Day! & Did 1.5 million people attend mass with Pope Francis in Portugal celebrating World Youth Day? & \url{https://factcheck.afp.com/doc.afp.com.33R24HY} \\
         AI\_True\_5 & This is Beirut tonight this is not self-defense. & Commercial flights landing at Beirut International Airport despite Israeli airstrikes.	& Did the airport in Beirut still operate despite airstrikes? & \href{https://www.reuters.com/fact-check/images-aircraft-landings-into-flaming-beirut-airport-are-ai-generated-2024-10-29/}{https://www.reuters.com/fact-check/images-aircraft-landings-into-flaming-beirut-airport-are-ai-generated-2024-10-29/} \\
         AI\_True\_6 & First Look at Lady Gaga in WEDNESDAY Season 2! & Lady Gaga to appear in 'Wednesday' Season 2. & Is Lady Gaga going to appear in Wednesday Season 2? & \url{https://www.comingsoon.net/guides/news/1894416-wednesday-season-2-lady-gaga-first-look-image-real-fake-ai} \\
         \midrule
         AI\_False\_1 & A 57,000 square foot Temu warehouse in China went up in flames today. The total loss of inventory has been estimated to be as high as \$56.19 USD. & A 57,000 square foot Temu warehouse in China went up in flames. & Did a Temu warehouse in China go up in flames? & \href{https://www.snopes.com/fact-check/temu-warehouse-fire-china/}{https://www.snopes.com/fact-check/temu-warehouse-fire-china/} \\
         AI\_False\_2 & A giant octopus was discovered off the coast of Bali, Indonesia.	& A giant octopus was discovered off the coast of Bali, Indonesia. & Was a giant octopus discovered off the coast of Bali, Indonesia? & \href{https://www.snopes.com/fact-check/giant-octopus-indonesian-coast/}{https://www.snopes.com/fact-check/giant-octopus-indonesian-coast/} \\
         AI\_False\_3 & It is reported that Disneyland has been flooded due to Hurricane Milton.	& Disneyland has been flooded due to Hurricane Milton. & Has Disneyland been flooded due to Hurricane Milton? & \url{https://factcheck.afp.com/doc.afp.com.36JU2AM} \\
         AI\_False\_4 & This is hysterical. The President of Mexico was spotted wearing a 'Make America Mexicana Again.' & The President of Mexico was spotted wearing a 'Make America Mexicana Again' hat. & Was the President of Mexico spotted wearing a 'Make America Mexicana Again' hat? & \href{https://leadstories.com/hoax-alert/2025/02/fact-check-mexican-president-claudia-sheinbaum-did-not-wear-make-america-mexicana-again-hat-it-was-made-using-xs-ai-tool.html}{https://leadstories.com/hoax-alert/2025/02/fact-check-mexican-president-claudia-sheinbaum-did-not-wear-make-america-mexicana-again-hat-it-was-made-using-xs-ai-tool.html} \\
         AI\_False\_5 & Image released of the migrant mob that stormed a Birmingham hospital armed with axes and blades.	& A migrant mob stormed the Birmingham hospital in England armed with axes and blades.	& Did a migrant mob storm a hospital in Birmingham, England? & \href{https://www.reuters.com/fact-check/image-armed-hospital-ambush-is-ai-not-evidence-disorder-uk-2025-02-21/}{https://www.reuters.com/fact-check/image-armed-hospital-ambush-is-ai-not-evidence-disorder-uk-2025-02-21/} \\
         AI\_False\_6 & Keanu Reeves is playing Bob Marley in a new movie he's shooting in Jamaica. & Keanu Reeves is playing Bob Marley in a new movie he's shooting in Jamaica. & Is Keanu Reeves playing Bob Marley in a new movie? & \href{https://www.snopes.com/fact-check/keanu-reeves-dreads/}{https://www.snopes.com/fact-check/keanu-reeves-dreads/} \\
         \midrule
         Attention\_Check\_1 & Mickey Mouse Drop-Kicks Toddler at Disneyland After Being Called 'Annoying' & This is an attention check. Please ignore the question and select ``Yes'' and ``unsure''. & To the best of your knowledge, did a person in a Mickey Mouse costume kick a toddler at Disneyland? & \href{https://www.snopes.com/fact-check/mickey-mouse-toddler-disneyland/}{https://www.snopes.com/fact-check/mickey-mouse-toddler-disneyland/} \\
         Attention\_Check\_2 & There is a lake in Finland, that looks like Finland. & This is an attention check. Please ignore the question and select ``No'' and ``sure''. & To the best of your knowledge, is there a lake in Finland that looks like Finland? & \href{https://www.snopes.com/fact-check/finland-shaped-lake-in-finland/}{https://www.snopes.com/fact-check/finland-shaped-lake-in-finland/}
  \\
         \bottomrule
    \end{tabular}
\end{table*}

\end{document}